\documentclass[prx,twocolumn,amsmath,amssymb,eqsecnum]{revtex4-2}
\usepackage{graphicx,amsmath,relsize,epstopdf,color,mathtools,bm,newtxtext,newtxmath,braket,rotating}
\usepackage[hyphenbreaks]{breakurl}
\usepackage[colorlinks=true,linkcolor=blue,citecolor=blue,urlcolor =blue]{hyperref}
\usepackage[normalem]{ulem}

\usepackage{booktabs}

\usepackage[table,xcdraw]{xcolor}
\newcommand{\eq}[1]{\begin{equation}\begin{aligned}#1\end{aligned}\end{equation}}
\newcommand{\expct}[1]{\left\langle#1\right\rangle}
\newcommand{\Cov}{\mathop{\mathrm{Cov}} \nolimits}

\newcommand{\RE}{\mathop{\mathrm{Re}} \nolimits}

\newcommand{\Tr}{\mathop{\mathrm{Tr}} \nolimits}

\newcommand{\eu}{\mathrm{e}}
\newcommand{\iu}{\mathrm{i}}

\newcommand{\rhop}{\rho_{\mathrm{p}}}
\newcommand{\rhoc}{\rho_{\mathrm{c}}}

\usepackage{soul,xcolor}

\usepackage{dsfont}

\begin{document}
	
\setstcolor{red}

\title{Evading noise in multiparameter quantum metrology with indefinite causal order}

\begin{abstract}
    Quantum theory allows the traversing of multiple channels in a superposition of different orders. When the order in which the channels are traversed is controlled by an auxiliary quantum system, various unknown parameters of the channels can be estimated by measuring only the control system, even when the state of the probe alone would be insensitive. Moreover, increasing the dimension of the control system increases the number of simultaneously estimable parameters, which has important metrological ramifications. We demonstrate this capability for simultaneously estimating both unitary and noise parameters, including multiple parameters from the same unitary such as rotation angles and axes and from noise channels such as depolarization, dephasing, and amplitude damping in arbitrary dimensions. We identify regimes of unlimited advantages, taking the form of $p^2$ smaller variances in estimation when the noise probability is $1-p$, for both single and multiparameter estimation when using our schemes relative to any comparable scheme whose causal order is definite.
\end{abstract}

\author{Aaron Z. Goldberg}
\affiliation{National Research Council of Canada, 100 Sussex Drive, Ottawa, Ontario K1N 5A2, Canada}
\affiliation{Department of Physics, University of Ottawa, Advanced Research Complex, 25 Templeton Street, Ottawa, Ontario K1N 6N5, Canada}

\author{L. L. S\'anchez-Soto}
\affiliation{Max-Planck-Institut für die Physik des Lichts, 91058 Erlangen, Germany}
\affiliation{Departamento de \'Optica, Facultad de F\'{\i}sica, Universidad Complutense, 28040 Madrid, Spain}

\author{Khabat Heshami}
\affiliation{National Research Council of Canada, 100 Sussex Drive, Ottawa, Ontario K1N 5A2, Canada}
\affiliation{Department of Physics, University of Ottawa, Advanced Research Complex, 25 Templeton Street, Ottawa, Ontario K1N 6N5, Canada}
\maketitle

\section{Introduction}
All measurements comprise four steps: initializing a probe or receiver, letting the probe interact with some system whose properties are to be measured, performing a measurement on the probe by which to extract data, and estimating the unknown parameter based on the data~\cite{Escheretal2011}. Classical estimation theory dictates how to optimize the fourth step, quantum estimation theory the third, and judicious changes in the first can lead to remarkable advantages when using probes with particular quantum properties; the interaction step, in contradistinction, is typically taken to be immutable. Introducing indefinite causal order (ICO) provides a paradigm for changing the interaction step of a measurement protocol, thereby offering a further avenue for quantum advantages, which can now be exploited to great avail.

Quantum estimation theory establishes the potential advantages of quantum probe states and quantum measurement techniques for estimating parameters in a variety of physical processes~\cite{Caves1981,Dowling1998,Giovannettietal2004,Berryetal2009,Tayloretal2013,Tsangetal2016,Liuetal2020}. This power has been demonstrated in remarkable experiments~\cite{Mitchelletal2004,LIGO2011,Whittakeretal2017,Youetal2021,Qinetal2023} and is expected to lead to practical, quantum-enhanced technologies in the near future~\cite{RaymerMonroe2019,Sussmanetal2019,Yamamotoetal2019,OIDA2020,KnightWalmsley2019}. The regime of multiparameter estimation is especially rich~\cite{Matsumoto2002,Paris2009,TothApellaniz2014,Szczykulskaetal2016,Braunetal2018,Liuetal2019,SidhuKok2020,Albarellietal2020,Polinoetal2020,DemkowiczDobrzanskietal2020,Liuetal2022,Goldbergetal2021singular}, with questions about incompatible observables~\cite{Zhu2015,Ragyetal2016,Heinosaarietal2016,Albarellietal2019,BelliardoGiovannetti2021}, nuisance parameters~\cite{Suzuki2020,Suzukietal2020}, and tradeoffs between parameters~\cite{Goldbergetal2021intrinsic} rising to the fore, which is prominent because many practical measurement scenarios involve the simultaneous estimation of multiple parameters~\cite{Koschorrecketal2011,Humphreysetal2013,BaumgratzDatta2016,Rehaceketal2017,Proctoretal2018,Rubioetal2020,Goldbergetal2021rotationspublished,Fidereretal2021}. It is to this multiparameter scenario that we apply ICO, in order to coax more practical advantages from quantum systems.

The idea of ICO stems from studies of causal structures in quantum gravity and quantum computation \cite{Hardy2007,Chiribellaetal2013} and has since burgeoned into a pervasive research field. Incorporating ICO in particular tasks leads to enhancements relative to \textit{quantum} advantages in computation~\cite{Colnaghietal2012,Morimae2014,Araujoetal2014,Taddeietal2021,Wechsetal2021}, communication~\cite{Chiribella2012,Feixetal2015,Guerinetal2016,DelSantoDakic2018,Ebleretal2018,Procopioetal2019,Chiribellaetal2021}, cooling~\cite{FelceVedral2020,Caoetal2022,GoldbergHeshami2023HBAC,Nieetal2022arxiv}, work extraction~\cite{Guhaetal2020,Guhaetal2022arxiv,Simonovetal2022}, and sensing~\cite{Mukhopadhyayetal2018arxiv,Frey2019,Zhaoetal2020,ChapeauBlondeau2021ICO,Liuetal2023}, many of which have been experimentally realized~\cite{Procopioetal2015,Rubinoetal2017,Rubinoetal2022,Goswamietal2018,Weietal2019,Massaetal2019,Goswamietal2020,Guoetal2020,Felceetal2021arxiv,Rubinoetal2021,Yinetal2022researchsquare}, along with more foundational ramifications~\cite{Oreshkovetal2012,Araujoetal2015,BaumelerWolf2016,OreshkovGiarmatzi2016,Zychetal2019,Oreshkov2019,Dimicetal2020,Milzetal2021,Barrettetal2021,PurvesShort2021,VilasiniRenato2022arxiv}. Moreover, ICO can sometimes be used to inspire protocols with \textit{definite} causal order (DCO) that outperform previously known methods \cite{ChapeauBlondeau2021ICOinspired}. The dramatic improvements possible in particular metrological tasks \cite{Zhaoetal2020}, as well as the ability to sense hitherto-hidden parameters \cite{companion2023depol}, motivate our current study.

Quantum noise, in general, ruins many proposed quantum advantages \cite{DemkowiczDobrzanskietal2009,DemkowiczDobrzanskietal2012,Crowleyetal2014,Giananietal2021,BaiAn2023arxiv}, yet it is prevalent in all realistic scenarios across quantum information, making a quantum advantage in the presence of noise even more impressive.  We recently showed that ICO confers dramatic advantages for estimating the phase of a unitary in arbitrary dimensions in the presence of depolarization noise, offering $\mathcal{O}(p^2)$ smaller variances than any scheme with DCO when the depolarization strength is large and $p$ is thereby small~\cite{companion2023depol}. These results can even be obtained using maximally mixed probe states that are completely insensitive to such parameters when evolved with a DCO.

We here demonstrate how to apply ICO to a wide variety of estimation problems and find dramatic sensitivity advantages in the presence of noise for both single and multiparameter estimations. After first providing a background on ICO, its implementation using a ``quantum switch,'' and a background on the quantum Fisher information (QFI) paradigm in Sec.~\ref{sec:ICO switch background}, we showcase our recent results for estimation of an arbitrarily large-dimensional unitary's phase in the presence of depolarization noise in Sec.~\ref{sec:depol companion}, followed by new advantages for dephasing (Sec.~\ref{sec:dephase single}) and amplitude damping (Sec.~\ref{sec:amp damp single}) noise in arbitrary dimensions. The advantages, in terms of how much smaller the estimators' variances are for our ICO scheme relative to the \textit{best possible} scheme with DCO, are on the order of: $\mathcal{O}\left[\left(\frac{p_Ap_B}{p_A+p_B}\right)^2\right]$ for depolarization, where small $p_O$ means that channel $O$ is very noisy; $\mathcal{O}\left[\left(\tfrac{1}{2}-p_A\right)^2\left(\tfrac{1}{2}-p_B\right)^2\right]$ for dephasing along a particular axis, where small $|\tfrac{1}{2}-p_O|$ means that the dephasing (or spin-flip) channel is very noisy; and $\mathcal{O}\left[\frac{p_A p_B}{\left(\sqrt{p_A}+\sqrt{p_B}\right)^2+c\left(\sqrt{p_A}-\sqrt{p_B}\right)^2}\right]$ for amplitude damping noise with some dimension-dependent constant $c$, where small $p_O$ again means that channel $O$ is very noisy. All schemes have the amount of information decrease as the amount of noise increases, but ICO is more resilient to noise and therefore more efficient in terms of the number of times the unitary must be probed in the large-noise limit, as attested to by its advantageous scaling in noise parameters. Formally, \textit{infinite} advantages are thus possible when either $p_A=0$ or $p_B=0$ for depolarization and amplitude damping, as well as when either or both of $p_A=\tfrac{1}{2}$ and $p_B=\tfrac{1}{2}$ for dephasing. By \emph{infinite advantages} we herein mean that ICO confers the ability to measure something that would be impossible without ICO. ICO is beneficial for metrology in and around these limits of when schemes with DCO fail or begin to fail due to being overwhelmed by noise.

In Sec.~\ref{sec:multiparam}, we take the opportunity to show how ICO can be used, not just in the presence of noise, but to characterize properties of the noise itself by developing the theory of ICO for multiparameter estimation. Noise characterization is paramount for developing quantum devices and quantum networks, in both practical and adversarial scenarios. We show there how ICO can be used to simultaneously measure parameters from the noisy channels and the unitary operator being applied, investigating all three noise scenarios in turn. The crucial upgrade required to be sensitive to more parameters is to increase the dimension of the control system governing the causal order of operations, which requires the ability to consider all orders of the unitary and noise channels.  For simultaneously estimating the unitary's phase as well as the strengths of the two noise channels, both depolarization and dephasing again offer formally infinite advantages for ICO when one of the noise channels is completely depolarizing or completely dephasing. Depolarization channels with complete control of the order of the channels even allow estimation of the unitary's phase when \textit{both} channels are completely depolarizing ($p_A=p_B=0$) and, for amplitude damping channels, we qualitatively show ICO's advantage to rapidly grow with decreasing $p_O$. The amplitude damping channel can also be used with a higher-dimensional control to simultaneously estimate the unitary's phase and rotation axis in addition to noise parameters. 

The aforementioned sections allow ICO to change the order in which the noise and unitary channels are applied on a probe state. In contrast, in all other studies of noisy metrology with ICO, multiple copies of the noisy unitary are applied in an indefinite order, with the causal relationship between the noise and unitary fixed in each channel \cite{Mukhopadhyayetal2018arxiv,ChapeauBlondeau2021ICO,ChiribellaZhao2022arxiv,ChapeauBlondeau2022,Delgado2022,Liuetal2023}. Even when those studies find advantages for ICO, they tend to be small, as the crucial component of our work is controlling the very order in which the noise and unitaries are applied, even with a single copy of the unitary. For completeness, we show in Sec.~\ref{sec:multiparam copied channels} how multiple copies of the same unitary subject to the same depolarization channel, with a fixed relationship between the noise and unitary, can be augmented with ICO to simultaneously measure the unitary and noise parameters; this extends the lines of previous work to multiparameter estimation and arbitrary dimensional probe systems. We also discuss advantages in scaling for these related scenarios, showing how to decrease the variance in estimating a unitary's phase decreases by $\mathcal{O}(p^{D-1})$ for $D$ identical copies of the noisy unitary channel.

Finally, we observe in Sec.~\ref{sec:independent from probe} that, when the probe state is a qubit and is sent through arbitrary numbers of channels in arbitrary numbers of orders controlled by a control state with arbitrary dimensions, the measurements on the control are independent from the chosen probe state for a general class of channels. This, to our knowledge, is the second foray of ICO into the context of multiparameter metrology~\cite{Delgado2022} and our widespread results indicate that ICO will remain a stalwart in this field.

\section{Background}
\label{sec:ICO switch background}
\subsection{Primer on indefinite causal order using quantum switches}
When two independent quantum channels $\mathcal{E}^{(A)}$ and $\mathcal{E}^{(B)}$ act sequentially on a quantum state $\rhop$ with a DCO,  the total evolution is governed by the sequential application
\begin{equation}
\rhop \mapsto \mathcal{E}^{(B)}\circ\mathcal{E}^{(A)} (\rhop ) \quad \mathrm{or} \quad \rhop\mapsto \mathcal{E}^{(A)}\circ\mathcal{E}^{(B)}  (\rhop ) \, .
\end{equation} 
We use $\rhop$ to denote the \textit{probe} state. A quantum switch breaks from this paradigm by allowing an external quantum system to control the order in which two or more channels act on $\rhop$. Such a device is sufficient for achieving a number of advantages in a variety of tasks and has been realized experimentally in groundbreaking experiments.

To wit, suppose the sequence is governed by the state of an auxiliary system, termed  \textit{control} state $\rhoc$. When $\rhoc$ is in some state $\ket{0}\bra{0}$, the probe evolves following $\mathcal{E}^{(B)}\circ\mathcal{E}^{(A)}\left(\rhop\right)$, while $\rhoc=\ket{1}\bra{1}$ dictates the evolution $ \mathcal{E}^{(A)}\circ\mathcal{E}^{(B)} ( \rhop )$. What, then, occurs when $\rhoc$ is prepared in a superposition of $\ket{0}$ and $\ket{1}$? This is the realm of ICO.

The dynamics are easiest to picture with unitary operations $\mathcal{E}^{(O)} ( \bullet )=U^{(O)} ( \bullet ) U^{(O)\dagger}$. The total evolution is encapsulated by the unitary operator
\begin{equation} 
\mathcal{U}=\ket{0}\bra{0}\otimes U^{(B)} U^{(A)}+\ket{1}\bra{1}\otimes U^{(A)} U^{(B)}
\label{eq:unitary from ICO}
\end{equation}
acting on the joint state $\rhoc \otimes \rhop$, which can be immediately verified for its action when $\rhoc$ is in state $\ket{0}$ or $\ket{1}$. This leads to cross terms in the joint dynamics when the control is prepared in some superposition state $\psi_0\ket{0}+\psi_1\ket{1}$:
\begin{align}
\mathcal{U}\rhoc & \otimes\rhop \mathcal{U}^\dagger=  | \psi_0 |^2 \ket{0}\bra{0}\otimes \mathcal{E}^{(B)} [ \mathcal{E}^{(A)} ( \rhop ) ] \nonumber \\
& + | \psi_1 |^2 \ket{1}\bra{1}\otimes \mathcal{E}^{(A)} [ \mathcal{E}^{(B)} ( \rhop ) ] \nonumber \\
&
+\psi_0\psi_1^* \ket{0}\bra{1}\otimes U^{(B)} U^{(A)}\rhop U^{(B)\dagger} U^{(A)\dagger} \nonumber \\
& + \left(\psi_0\psi_1^* \ket{0}\bra{1}\otimes U^{(B)} U^{(A)}\rhop U^{(B)\dagger} U^{(A)\dagger} \right)^\dagger;
\end{align} 
the final two terms represent novel interference effects that have found a number of applications. A natural assumption throughout this paper is that none of the channels [here: neither $U^{(A)}$ nor $U^{(B)}$] change on timescales relevant to the amount of time it takes $\rhop$ to traverse them. {Even though each unitary $U^{(A)}$ and $U^{(B)}$ appears twice in $\mathcal{U}$, the quantum switch ensures that each channel is only probed once. This can be seen by considering auxiliary flag' degrees of freedom in quantum states $\ket{0}_{\mathrm{F}A}$ and $\ket{0}_{\mathrm{F}B}$ that transform as $\ket{n}_{\mathrm{F}O}\mapsto \ket{n+1}_{\mathrm{F}O}$ whenever $U^{(O)}$ is applied to the system; the flag degrees of freedom factor out after the application of the switch and are uniquely in the states $\ket{1}_{\mathrm{F}A}$ and $\ket{1}_{\mathrm{F}B}$.}

Similar dynamics result from quantum channels that are not unitary. For example, we can consider maps characterized by Kraus operators:
\begin{equation}
\mathcal{E}^{(O)} ( \bullet ) = \sum_l K_l^{(O)}\left(\bullet\right) K_l^{(O)\,\dagger},
\qquad 
\sum_l K_l^{(O)\,\dagger}K_l^{(O)}=\openone \, .
\end{equation}  
The total evolution is then governed by a quantum channel with Kraus operators of the form~\cite{Chiribellaetal2013,Goswamietal2020,Guoetal2020}
\begin{equation}
\mathcal{K}_{ij}=\ket{0}\bra{0}\otimes K^{(B)}_i K^{(A)}_j+\ket{1}\bra{1}\otimes K^{(A)}_j K^{(B)}_i 
\label{eq:Kraus two channels}
\end{equation} 
acting on the joint state $\rhoc \otimes \rhop$, which can again be immediately verified for its action when $\rhoc$ is in state $\ket{0}$ or $\ket{1}$ and requires no correlations between the Kraus operators for distinct modes. This again leads to interference terms in the dynamics, which again can be seen when the control is prepared in the superposition state
$\psi_0\ket{0}+\psi_1\ket{1}$:
\begin{align}
\sum_{i,j}\mathcal{K}_{ij}  \rhoc & \otimes\rhop \mathcal{K}_{ij}^\dagger=  |\psi_0 |^2 \ket{0}\bra{0}\otimes \mathcal{E}^{(B)} [\mathcal{E}^{(A)} ( \rhop  ) ] \nonumber \\
&+ | \psi_1 |^2 \ket{1}\bra{1}\otimes \mathcal{E}^{(A)}[ \mathcal{E}^{(B)} (\rhop  )] \nonumber \\
& +\psi_0\psi_1^\ast \ket{0}\bra{1}\otimes \sum_{i,j} K^{(B)}_i K^{(A)}_j \rhop  K^{(B)\,\dagger}_i K^{(A)\,\dagger}_j \nonumber \\
& + \left(\psi_0\psi_1^* \ket{0}\bra{1}\otimes \sum_{i,j} K^{(B)}_i K^{(A)}_j\rhop  K^{(B)\,\dagger}_i K^{(A)\,\dagger}_j \right)^\dagger \, .
\label{eq:two superposed Kraus full action}
\end{align} 


The quantum-channel evolutions under ICO may be deduced by interpretting Kraus operators as remnants from unitary operations on an enlarged Hilbert space that have had the auxiliary degrees of freedom traced out. We can always 
consider the Kraus operators $K_i^{(O)}$ to represent the actions of unitary operators $U^{(O,O^\prime)}$ acting on $\rhoc\otimes\ket{0}_O\bra{0}$ via
\begin{equation}
K_{i}^{(O)}=  _{O^\prime} \!\! \bra{i}U^{(O,O^\prime)}\ket{0}_{O^\prime}.
\label{eq:Kraus from unitary}
\end{equation} 
Assuming each of the sequential operations to possess their own auxiliary modes, 
we can enlarge the unitary operators
of Eq. \eqref{eq:unitary from ICO} to become
\begin{equation}
\mathcal{U}=\ket{0}\bra{0}\otimes U^{(B,B^\prime)} U^{(A,A^\prime)}+\ket{1}\bra{1}\otimes U^{(A,A^\prime)}U^{(B,B^\prime)}.
\end{equation} 
Tracing out the auxiliary modes from the evolution 
\begin{align} 
&\rhoc\otimes\rhop\otimes\ket{0}_{A^\prime} \bra{0}\otimes\ket{0}_{B^\prime} \bra{0} \nonumber \\
&\mapsto \mathcal{U}\rhoc\otimes\rhop\otimes\ket{0}_{A^\prime} \bra{0}\otimes\ket{0}_{B^\prime} \bra{0}\mathcal{U}^\dagger  
\end{align} immediately yields the Kraus operators given by Eqs. \eqref{eq:Kraus two channels} and \eqref{eq:Kraus from unitary}. 
Notwithstanding this interpretation, only Kraus operators of the form of Eq.~\eqref{eq:Kraus two channels} reduce to the unitary $\mathcal{U}$ in the limit of a single Kraus operator, because a quantum switch is a superoperator that must act in the same manner regardless of the process in question \cite{Chiribellaetal2013}. In fact, any alternative Kraus-operator decompositions for the individual channels $\mathcal{E}^{(O)}$ will lead to the same overall dynamics when the alternative Kraus operators are fed into Eq.~\eqref{eq:Kraus two channels}. As such, given only the two respective descriptions of channels $\mathcal{E}^{(A)}$ and $\mathcal{E}^{(B)}$, a quantum switch is guaranteed to lead to evolution with Kraus operators from Eq.~\eqref{eq:Kraus two channels} without requiring any control of the details of the channels or correlations between $A$ and $B$.

This form of the Kraus operators arising from superpositions of sequences of operations holds true when there are arbitrary numbers of operations whose orders of application are being superposed. By increasing the dimension $D$ of the control system, we can increase the number of possible orderings. If we label the Kraus operators from each channel $A_j$ in the sequence by $K_i^{A_j}$, the control system can enable $D$ different permutations of the channels $A_j$, leading to Kraus operators of the form
\begin{equation}
\mathcal{K}_{i_1,i_2,\cdots,i_3}=\sum_{j=0}^{D-1}\ket{j}\bra{j}\otimes K_{i_{\pi_j(0)}}^{(A_{\pi_j(0)})}K_{i_{\pi_j(1)}}^{(A_{\pi_j(1)})}\cdots K_{i_{\pi_j(D-1)}}^{(A_{\pi_j(D-1)})}\, ,
\end{equation}
where we have denoted by $\pi_j(k)$ the $k$th element of the $j$th permutation of $(0,1,\cdots,D-1)$ and assumed there to be $D$ channels without loss of generality \footnote{If there are fewer channels, some of the Kraus operators can be set to an appropriate multiple of the identity and, if there are more channels, some of them can be concatenated into one channel [$ABC$ and $CAB$ to $(AB)C$ and $C(AB)$] or applied with a definite causal order ($ABC$ and $CBA$ always have operation $B$ at the same time). }. Any time the control state is prepared in a superposition $\sum_j \psi_j\ket{j}$, interference terms with $j_1\neq j_2$ will arise that can lead to unique effects in
\begin{equation}
\sum_{i_1\cdots i_{D}}\mathcal{K}_{i_1\cdots i_{D}}\rhop\otimes\rhoc \mathcal{K}_{i_1\cdots i_{D}}^\dagger=\sum_{j_1 j_2}\psi_{j_1}\psi_{j_2}^* \ket{j_1}\bra{j_2}\otimes \mathcal{R}_{j_1 j_2} \, .
\end{equation} 
Here, 
\begin{align} 
\mathcal{R}_{j_1 j_2}&= \nonumber \\
\sum_{i_1,\cdots,i_3}&
\left ( K_{i_{\pi_{j_1}(0)}}^{(A_{\pi_{j_1}(0)})}\cdots K_{i_{\pi_{j_1}(D-1)}}^{(A_{\pi_{j_1}(D-1)})} \right )
\rhop
\left(K_{i_{\pi_{j_2}(0)}}^{(A_{\pi_{j_2}(0)})}\cdots K_{i_{\pi_{j_2}(D-1)}}^{(A_{\pi_{j_2}(D-1)})}\right)^\dagger.
\label{eq:probe state conditioned on control basis}
\end{align}

In our work, as is often the case, we will solely use properties of the control system to learn about the interactions of the probe. The control evolves to 
\begin{equation}
\rhoc^\prime=
\sum_{j_1,j_2}\psi_{j_1}\psi_{j_2}^*\ket{j_1}\bra{j_2} R_{j_1 j_2}(\rhop),
\label{eq:rhoc prime in terms of S}
\end{equation} 
where we have defined the traces
\begin{equation}
R_{j_1 j_2}=\Tr(\mathcal{R}_{j_1 j_2}).
\label{eq:Rij control}
\end{equation}
Trace-preserving channels lead to $R_{j j}(\rhop)=1$; the interference terms with $R_{j_1 j_2}(\rhop)<1$ lead to entanglement between the control and the probe systems that can be used to estimate properties of the channels by measuring only the control. For consistency, we note that the case of two identical channels with a single Kraus operator $K^{(A)}=K^{(B)}=U$ simply has $R_{j_1j_2}=\Tr(\rhop U^\dagger U)=1$, such that the control only changes state when the channels $A$ and $B$ are nonunitary or not identical.

\subsection{Quantum Fisher information}
Suppose one has a set of parameters $\boldsymbol{\theta}$ to estimate. Given access to a probability distribution $P(x|\boldsymbol{\theta})$ for some measurement with outcomes labelled by $x$, the Cram\'er-Rao bound dictates that the covariances between any estimators $\hat{\theta}_i$ of the parameters will locally be lower bounded by the inverse of the Fisher information (FI) matrix
\begin{equation}
\Cov(\hat{\theta}_i,\hat{\theta}_j)\geq \left(\mathsf{\mathbf{F}}^{-1}_x(\boldsymbol{\theta})\right)_{ij} \, ,
\label{eq:CRB classical}
\end{equation}
where the latter has components
\begin{equation}
\left[\mathsf{\mathbf{F}}_x(\boldsymbol{\theta})\right]_{ij}=\sum_x P(x|\boldsymbol{\theta}) \frac{\partial \ln P(x|\boldsymbol{\theta})}{\partial \theta_i}\frac{\partial \ln P(x|\boldsymbol{\theta})}{\partial \theta_j} \, .
\label{eq:classical FIM}
\end{equation} 
Analogous expressions can be found for continuous measurement outcomes $x$ with integrals replacing the sums. The quantum Fisher information (QFI) matrix provides the ultimate upper bound for $\mathsf{\mathbf{F}}$ for any given probe state and underlying values of $\boldsymbol{\theta}$, thereby providing the ultimate lower limit for the covariance matrix. Given a probe state that has evolved to depend on the parameters, $\rho_{\boldsymbol{\theta}}$, one can always define the symmetric logarithmic derivatives 
\begin{equation}
\frac{\partial \rho_{\boldsymbol{\theta}}}{\partial \theta_i}=\frac{\rho_{\boldsymbol{\theta}} L_{i}+ L_{i}\rho_{\boldsymbol{\theta}}}{2}
\end{equation}
to provide a matrix analog of the derivatives in Eq. \eqref{eq:classical FIM}, where $L_{i}$ may depend on $\rho_{\boldsymbol{\theta}}$ and $\boldsymbol{\theta}$ and is always Hermitian. Then, the QFI matrix is defined componentwise as \cite{Paris2009}
\begin{equation}
\left[\mathsf{\mathbf{Q}}_{\rho_{\boldsymbol{\theta}}}(\boldsymbol{\theta})\right]_{ij}= \tfrac{1}{2} \Tr (\rho_{\boldsymbol{\theta}}  \{ L_i , L_j \} ) \, ,
\end{equation} 
where $\{ \cdot \, , \cdot \}$ stands for the anticommutator $\{ A, B \} = A B + BA$.  
The matrix inequality 
\begin{equation}
\mathsf{\mathbf{Q}}_{\rho_{\boldsymbol{\theta}}}(\boldsymbol{\theta})\succeq \mathsf{\mathbf{F}}_x(\boldsymbol{\theta})
\end{equation}
always holds in the sense that $\mathsf{\mathbf{Q}}-\mathsf{\mathbf{F}}$ is always positive semidefinite. Remarkably, the most general probability distribution $P(x|\boldsymbol{\theta})=\Tr(\Pi_x \rho_{\boldsymbol{\theta}})$ for a positive operator-valued measure (POVM) with elements $\{\Pi_x\}$ can always be optimized in the asymptotic limit, {in the sense
that, for any positive-definite weight matrix $\mathsf{\mathbf{W}}$, there exists an optimal POVM such that} 
\begin{equation}
\label{eq:falt}
\Tr[\mathsf{\mathbf{W}} \Cov(\hat{\boldsymbol{\theta}},\hat{\boldsymbol{\theta}})] {=} f\,  \Tr[\mathsf{\mathbf{W}}\mathsf{\mathbf{Q}}_{\rho_{\boldsymbol{\theta}}}^{-1}(\boldsymbol{\theta})] \, .
\end{equation} 
{where $f=1$ for single-parameter estimation and $1 \leq f \leq 2$ for multiparameter estimation and where the equality holds after many repeated optimal measurements \cite{Carolloetal2019,Tsangetal2020}}. This connects the ultimate lower bounds on the covariances of the estimators to the ultimate measurement scheme for any probe state; the optimal overall protocol then involves optimizing the QFI matrix over all probe states. We are generous throughout with this factor of $f$: we use the QFI matrix for all schemes with fixed causal order, even though the results attainable will be smaller by a factor of $f$, and provide fixed measurement schemes for all of our new protocols with ICO, which can be directly fed into the Cram\'er-Rao bound of Eq.~\eqref{eq:classical FIM}. This means that our quoted results hereafter for ICO may actually outperform schemes with a fixed causal order by an extra factor of $f$. Because the QFI matrix is additive when the same measurement process is repeated, we henceforth consider a single trial when comparing QFI values for different protocols.

\subsection{Estimating the phase of a unitary}

Consider any finite-dimensional unitary operator $U$, which can always be considered as an element of SU$(N)$ for some positive integer $N$ without loss of generality. If the generators of SU$(N)$ are labelled by $\mathbf{G}=(G_1,G_2,\cdots)$, then we can define $G_{\mathbf{n}}=\mathbf{n}\cdot\mathbf{G}$ for some unit vector $\mathbf{n}$ and always express the unitary as
\begin{equation}
U(\theta,\mathbf{n})=\exp(i  \theta G_{\mathbf{n}}).
\end{equation} 
Estimating the phase $\theta$ is a basic problem with broad applications due to the ubiquity of unitary operations; we name interferometry, magnetometry, and imaging as examples. To fix the resources used in the estimation, we choose a particular irreducible representation of the Lie group with dimension $d$, equivalent to fixing the number of particles in or energy used by a probe state. Such a fixed irreducible representation has some eigenstates $\ket{\pm\mathbf{n}}$ of $G_{\mathbf{n}}$ with some maximal and minimal eigenvalues $\lambda_{\pm}$. Then, the best possible quantum strategy with DCO for estimating $\theta$ involves preparing the pure superposition state~\cite{Giovannettietal2006}
\begin{equation}
\ket{\psi_{\mathrm{opt}}}= \frac{1}{\sqrt{2}} (\ket{\mathbf{n}}+\ket{-\mathbf{n}}) 
\end{equation}  
and allowing it to evolve to $U\ket{\psi}$. In the contexts of interferometry with light and atoms, e.g., imaging or magnetometry, such states are often known as NOON~\cite{Dowling2008} or GHZ~\cite{Greenbergeretal1990} states, respectively. The QFI in this case $\mathsf{Q}_{\psi_{\mathrm{opt}}}(\theta)=(\lambda_+-\lambda_-)^2$ informs us that the best possible estimate $\hat{\theta}$ for the angle $\theta$ will have its variance be lower bounded as
\begin{equation}
\Delta^2\hat{\theta}\geq\frac{1}{\mathsf{Q}_{\psi_{\mathrm{opt}}}(\theta)}=\frac{1}{(\lambda_+-\lambda_-)^2}.
\end{equation} 
The QFI is additive for repeated measurements and so here and henceforth we consider the QFI \textit{per trial} (i.e., per state probing the parameter of interest in the asymptotic limit{; we will always use one probe state per application of the unitary channel}). The worst possible scheme, in contrast, uses a probe that remains unchanged by $U$, such as the pure states $\ket{\pm\mathbf{n}}$ or the maximally mixed state $\openone/d$.

\section{Advantageous unitary estimation in the presence of noise using ICO}
\label{sec:single param}

\begin{figure}
    \centering
    \includegraphics[width=\columnwidth]{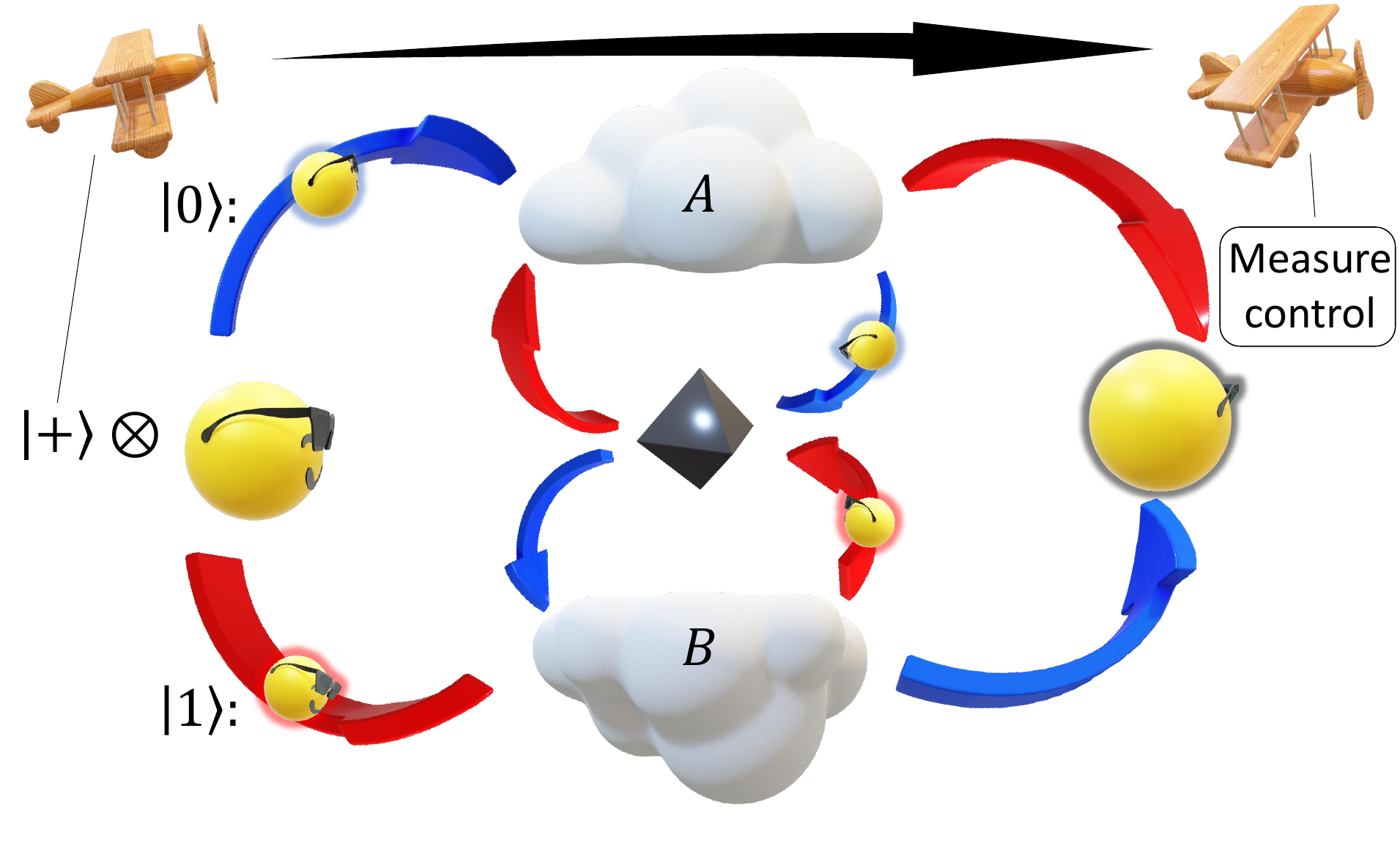}
    \caption{Schematic for estimating a unitary in the presence of noise. The probe, a maximally mixed state insensitive to unitaries (sunglasses clad), is sent through  noise channels (clouds) before and after probing a unitary (diamond). When the control is in state $\ket{0}$ ($\ket{1}$), the probe follows the blue (red) path whose first noise channel is $A$ ($B$). A superposition-state control leads to ICO, by way of which a final measurement on the control alone (carried by airplane to circumvent the unitary and noise channels) can learn about the unitary with dramatic advantages over any causally ordered scheme. This scheme can be generalized to higher-dimensional quantum switches that control more than two different orders of operations amongst the three channels here and to more than three channels.}
    \label{fig:general one unitary two noise}
\end{figure}

In the presence of noise, the QFI tends to decrease, except for some fortuitous situations in which it remains constant. We here consider a general schematic, depicted in  Fig.~\ref{fig:general one unitary two noise}, in which some noise affects a probe system both before and after it experiences the unitary transformation, which is a generic scenario where we simply supply different labels for the noise experienced by a probe on either side of a unitary. The order in which the probe traverses the noise and unitary channels can be controlled by a quantum system, again depicted in Fig.~ \ref{fig:general one unitary two noise}, such that measuring the control qubit alone allows one to learn about the unitary with a dramatic advantage over any causally ordered scheme. Experimental demonstrations of such quantum control of the order of traversing noise channels have already succeeded \cite{Goswamietal2020,Guoetal2020}, making the application of this idea to metrology practicable. We here showcase these advantages for three different types of noise: depolarization, dephasing, and amplitude damping, acting on probes of arbitrarily large dimensions so as to allow for arbitrary unitaries to be estimated.

\subsection{Depolarization noise}
\label{sec:depol companion}
We first recapitulate the ICO-driven advantage in estimating the phase of any unitary operation in the presence of strong depolarization noise presented in \cite{companion2023depol}.

Depolarization noise strongly reduces the ability to estimate $\theta$. Depolarization adds white noise to the state such that, with some probability $1-p$, one loses all information about the original state and becomes insensitive to all unitary parameters:
\begin{equation}
\mathcal{E}_{\mathrm{depol}}(\rho)=p\rho+(1-p)\frac{\openone}{d} \, .
\label{eq:depol channel}
\end{equation}

If depolarization occurs either before or after a pure probe state undergoes the unitary transformation, the QFI diminishes as~\footnote{We take this opportunity to correct a typo in the factor of two that does not appear in the earlier sources \cite{Toth2012,Hyllusetal2012} but propagates in review articles~\cite{TothApellaniz2014,SidhuKok2020}.}
\begin{equation}
\mathsf{Q} [ \mathcal{E}_{\mathrm{depol}}(\ket{\psi}\bra{\psi});\theta ]=\frac{p^2}{p+\frac{1-p}{d/2}}\mathsf{Q}(\ket{\psi}\bra{\psi};\theta) \, ,
\end{equation} 
where we employ the alternate notation $\mathsf{Q}_\rho(\theta)\leftrightarrow \mathsf{Q}(\rho;\theta)$ when convenient. Convexity of the QFI $\mathsf{Q}(\sum_i p_i \rho_i;\theta)\leq \sum_i p_i \mathsf{Q}(\rho_i;\theta)$ and flatness of the depolarization channel $\mathcal{E}_{\mathrm{depol}}(\sum_i p_i \rho_i)= \sum_i p_i \mathcal{E}_{\mathrm{depol}}(\rho_i)$ lead to the following inequality for all states undergoing depolarization, even including probe states entangled with ancillary quantum systems and joint measurements on the entangled systems:
\begin{equation}
\mathsf{Q} [ \mathcal{E}_{\mathrm{depol}}(\rho);\theta ] \leq \frac{p^2}{p+\frac{1-p}{d/2}} (\lambda_+-\lambda_-)^2 \, .
\label{eq:QFI depol no ICO}
\end{equation} 
The resulting minimum variance for any estimate of $\theta$ grows as $\mathcal{O}(p^{-2})$ whenever the depolarization probability is close to unity (i.e., $1-p\sim 1$).

It comes as no surprise that depolarizing a probe state both before and after it undergoes a unitary evolution worsens estimates of the unitary's phase. If the two depolarizations have strengths $p_A$ and $p_B$, the resulting minimum variance grows as $\Delta^2\hat{\theta}=\mathcal{O}(p_A^{-2}p_B^{-2})$. Yet, placing these two depolarizations in a coherently controlled superposition of their causal orders will significantly decrease the estimator variance.

To apply ICO to depolarizing channels, we need a Kraus-operator representation of $\mathcal{E}_{\mathrm{depol}}$. This can be furnished by defining $d^2+1$ operators: $d^2$ two-index operators that provide white noise for a $d$-element orthonormal basis $\{\ket{n}\}$ by completely mixing up all information,
\begin{equation}
K_{kl}(p)=\sqrt{\frac{1-p}{d}}\ket{k}\bra{l},
\end{equation} 
and the identity operator $K_{\openone}(p)=\sqrt{p}\mathds{1}$ that leaves states unchanged.

We use a single application of the unitary channel and two different depolarizing channels, with the orders $\mathcal{E}_{\mathrm{depol}}^{(A)}$-then-$U$-then-$\mathcal{E}_{\mathrm{depol}}^{(B)}$ when the control is in state $\ket{0}$ and $\mathcal{E}_{\mathrm{depol}}^{(B)}$-then-$U$-then-$\mathcal{E}_{\mathrm{depol}}^{(A)}$ when the control is in state $\ket{1}$. Defining $K^{(A)}_{kl}$ and $K^{(B)}_{mn}$ with $p_A$ and $p_B$, respectively, we can compute using Eq.~\eqref{eq:Rij control}
\begin{align}
        R_{01} & = \sum_{ijkl} \Tr\left(K^{(B)}_{ij} U K^{(A)}_{kl}\rho K^{(B)\dagger}_{ij} U^\dagger K^{(A)\dagger}_{kl}\right) \nonumber \\
        & = \frac{p_A(1-p_B)}{d}\Tr(U^\dagger)\expct{U}+\frac{p_B(1-p_A)}{d}\Tr(U)\expct{U^\dagger} \nonumber \\
        &+\frac{(1-p_A)(1-p_B)}{d^2}+p_Ap_B \, ,
\end{align}
where expectation values $\expct{\cdot}$ are taken with respect to the initial probe state $\rhop$. Choosing the least remarkable probe state $\rhop=\openone/d$, which is maximally mixed, possesses the least quantum mechanical properties, and should be insensitive to unitaries because $U\rhop U^\dagger=\rhop$, allows one to directly learn about
\begin{equation}
u= | \Tr(U) |^2 = \left|\sum_i e^{i  \lambda_i \theta} \right|^2 \, .
\end{equation}
Since the eigenvalues $\{ \lambda_i\}$ of the generators of SU$(N)$  can be readily calculated, this provides a direct window into estimating the unitary's phase $\theta$.

How well can this be done? Defining the $\ket{\pm}=(\ket{0}\pm\ket{1})/\sqrt{2}$ basis, starting with the control state in $\rhoc=\ket{+}\bra{+}$, then measuring the control state $\rhoc^\prime$ in the $\ket{\pm}$basis provides an FI equal to the QFI for this state of
\begin{align}
&\mathsf{Q}_{\mathrm{ICO}}(\theta)= \nonumber \\
&\frac{(p_A+p_B-2 p_A p_B)^2 \left(\frac{\partial u}{\partial \theta}\right)^2}{ d^4- [ (1-p_A) (1-p_B)+(p_A+p_B-2 p_A p_B)u +d^2p_Ap_B ]^2} \nonumber \\
& \approx 
\frac{(p_A+p_B)^2 }{ d^4-1}\left(\frac{\partial u}{\partial \theta}\right)^2+\mathcal{O}(p_A^3,p_B^3,p_Ap_B^2,p_A^2p_B ) \, .
\label{eq:depolarization Q ICO}
\end{align} 
{Because this scales with the second power of the noise and not the fourth, we learn that, for any rotation angle other than $\theta=0$ or $\theta=\pi$, there always exists a noise threshold above which ICO has an advantage over DCO. }

Per the quantum Cram\'er-Rao bound, the minimum uncertainty on any estimator of $\theta$ is given by the inverse of the QFI, showing that this outperforms the best QFI for sensing $\theta$ with DCO by a factor on the order of $\mathcal{O}(p_A^2,p_B^2,p_Ap_B)$:
\begin{equation}
\min_{\mathrm{ICO}}\Delta^2\hat{\theta}=\mathcal{O}\left(\frac{1}{(p_A+p_B)^2}\right)\ll \min_{\mathrm{DCO}} \Delta^2\hat{\theta}=\mathcal{O}\left(\frac{1}{(p_Ap_B)^2}\right)\, , 
\end{equation} 
which provides an essentially unlimited advantage as the depolarization noise increases and $p_A$ and $p_B$ decrease to zero. {Even though the expression $\partial u/\partial \theta$ appears in this expression, we have computed the FI in terms of the probability distribution $p_{\pm}(\theta)=\langle \pm|\rhoc|\pm\rangle$ and so need not worry about accidentally choosing the optimal estimator for $u(\theta)$ instead of the optimal estimator for $\theta$.}

\begin{figure}
    \centering
    \includegraphics[width=\columnwidth]{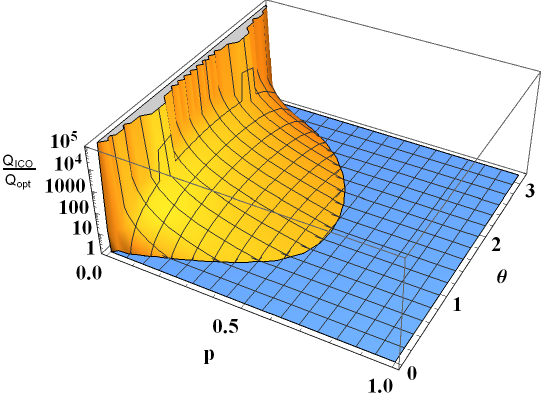}
    \caption{Advantage of ICO over the best DCO strategy for estimating a rotation angle of a qubit. Here, $\mathsf{Q}_{\mathrm{ICO}}$ is given by Eq.~\eqref{eq:depolarization Q ICO} with $d=2$, $p_A=p_B\equiv p$, and $u=4\cos^2 (\theta/2)$; whereas, $\mathsf{Q}_{\mathrm{opt}}$ is given by the upper bound in Eq.~\eqref{eq:QFI depol no ICO} with $\lambda_\pm=\pm\tfrac{1}{2}$ and two applications of the depolarizing channel of strength $p$ (i.e., $\mathsf{Q}_{\mathrm{opt}}= p^4/[p^2+ 2 (1-p^2)/d]$). Plotted is the  increase in QFI versus rotation angle and depolarization noise, which may be interpreted as how many more times a DCO scheme must be performed to obtain the same precision as an ICO scheme. If we had chosen a different dimension, the dependence on $\theta$ would have changed. For $\theta$ near $\pi/2$, the advantage persists even with noise level $p>1/2$, while it grows rapidly and boundlessly with shrinking $p$. In the blue region of larger $p$ and thus smaller noise, schemes with ICO can be much worse than schemes with definite causal order and so the former should be avoided.}
    \label{fig:qubit advantage depol}
\end{figure}

We plot the relative advantage for $d=2$ in Fig.~\ref{fig:qubit advantage depol} with $p_A=p_B\equiv p$; when $p_A$ and $p_B$ are different for a given total $p_A+p_B$ or a given fixed $p_Ap_B$, the ICO-driven advantage is even greater. These results require only a binary measurement on a single quantum system, as opposed to a generic measurement on a large-dimensional probe state or a joint measurement on the entangled control-probe state. Similar results can be obtained when the control system begins in any equal-magnitude superposition of $\ket{0}$ and $\ket{1}$ so as to maximize the effect of $R_{01}$ on $\rhoc^\prime$; different relative phase choices for the initial control state lead to different optimal measurement bases.

The same result cannot be achieved by entangling the control and probe systems, letting the probe system evolve through one causal sequence, then measuring the final state of the control system. This is because the overall quantum channel that the probe experiences, formed by iterations of the map from Eq. \eqref{eq:depol channel}, preserves the trace of the probe state. As such, tracing out the probe state before or after it evolves does not affect the control state: $\rhoc=\rhoc^\prime$. If either $p_A=0$ or $p_B=0$ with DCO, no scheme will be able to estimate $U$, even with access to ancillary entangled systems. The interference between the different causal sequences is essential (i.e., it is necessary, but not always sufficient) for probing the properties of the noisy channels.

\subsection{Dephasing noise}
\label{sec:dephase single}
Now suppose an alternate noise source, in which the state is subject to dephasing or spin-flip noise. We start by considering a qubit probe state subject to dephasing along a particular axis $\mathbf{u}$ with some probability $1-p$, characterized by the Pauli matrices $\sigma_{\mathbf{u}}=\mathbf{u}\cdot\boldsymbol{\sigma}$:
\begin{equation}
\mathcal{E}_{\mathrm{dephase}}(\rho)=p\rho+(1-p)\sigma_{\mathbf{u}}\rho\sigma_{\mathbf{u}} \, .
\end{equation} 
Again considering applications of this channel in a definite order, such as $\mathcal{E}_{\mathrm{dephase}}^{(A)}$ prior to $U$ and $\mathcal{E}_{\mathrm{dephase}}^{(B)}$ afterward, the QFI for a probe state decreases as a function of $p_A$ and $p_B$. In this situation, the noise channel does not commute with the unitary operation and so the resulting state is not symmetric with respect to $p_A\leftrightarrow p_B$.

Without loss of generality, we fix the axis of the unitary $\mathbf{n}$ to be the $z$ axis of the coordinate system used to define the computational basis for the probe qubit. Then, an optimal probe state without dephasing is $\ket{\psi}=\left(\ket{0}+\ket{1}\right)/\sqrt{2}$. With dephasing, the state evolves to a convex mixture of the four linearly dependent pure states $U\ket{\psi}$, $U\sigma_{\mathbf{u}}\ket{\psi}$, $\sigma_{\mathbf{u}}U\ket{\psi}$, and $\sigma_{\mathbf{u}}U\sigma_{\mathbf{u}}\ket{\psi}$. We can calculate the QFI in terms of the eigenvalues $\varrho_k$ and eigenvectors $\ket{\psi_k}$ of $\rho$ through~\cite{Paris2009}
\begin{equation}
\mathsf{Q}_\rho(\theta)=\sum_k\frac{1}{\varrho_k}\left(\frac{\partial \varrho_k}{\partial \theta}\right)^2+2\sum_{k\neq l}\frac{\left(\varrho_k-\varrho_l\right)^2}{\varrho_k+\varrho_l}\left|\bra{\psi_k} \frac{\partial}{\partial \theta}\ket{\psi_l}\right|^2.
\label{eq:QFI from evals evecs}
\end{equation}

The axis along which the dephasing acts significantly affects the results. With $\mathbf{u}$ along the $y$-axis, for example, the QFI vanishes at $p_A=1/2$, because the state becomes maximally mixed for all values of $\theta$ and $p_B$. With $\mathbf{u}$  along the $x$-axis, the state becomes independent from $p_A$, with the QFI remaining independent from both $p_A$ and $p_B$. As a final example, for $\mathbf{u}$ along the $z$-axis, the QFI vanishes when either $p_A=1/2$ or $p_B=1/2$, taking the form $(\tfrac{1}{2}-p_A)^2(\tfrac{1}{2}-p_B)^2$. Some of these can be dramatically outperformed by schemes with ICO and some cannot.

To introduce ICO, we need only the Kraus operators $K^{O)}_{\openone}=\sqrt{p_O}\openone$ and $K^{O)}_{\mathbf{u}}=\sqrt{1-p_O}\sigma_{\mathbf{u}}$. As before, we calculate using Eq.~ \eqref{eq:Rij control}
\begin{align}
R_{01}&=\sum_{ij}\Tr\left(K^{(B)}_i U K^{(A)}_j\rho K^{(B)\dagger}_i U^\dagger K^{(A)\dagger}_j\right) \nonumber \\
&=p_A(1-p_B)s+p_B(1-p_A)s^\ast \nonumber \\ 
& + (1-p_A)(1-p_B)+p_A p_B \,  ,
\end{align} 
where we have defined
\begin{equation}
s(\theta,\mathbf{u};\rhop)=\expct{\sigma_{\mathbf{u}} U^\dagger \sigma_{\mathbf{u}} U}\, .
\end{equation}

How does the amount of information about $\theta$ in $\rhoc^\prime$ compare to the QFI for probe states with DCO? Whereas, the QFI often vanishes for probe states with DCO when $p_A=p_B=1/2$, $R_{01}$ and therefore $\rhoc^\prime$ retains information about $\theta$ at $p_A=p_B=1/2$, so long as $s$ depends on $\theta$. One can compute $\sigma_{\mathbf{u}} U^\dagger \sigma_{\mathbf{u}} U$ to find it independent from the azimuthal angle of $\mathbf{u}$. The quantity $s$ depends on $\theta$ so long as $|u_z|\neq 1$; thus, for \textit{any} probe state $\rhop$, even a maximally mixed probe state, one can learn about $\theta$ using ICO.  When $\mathbf{u}$ was along the $y$-axis and $p_A=p_B=1/2$, we saw that the QFI was zero for DCO; whereas, here it is
\begin{equation}
\mathsf{Q}_{\mathrm{ICO}}(\theta)= \frac{1}{3-2 \RE (s) -\RE^2( s)}\left [ \frac{\partial \RE (s)}{\partial \theta}\right ]^2 \, ,
\end{equation} 
with $s=\expct{U^{\dagger 2}}$. This constitutes an \emph{infinite advantage} (an infinite increase in the QFI ratio) in this particular scenario: even a maximally mixed probe state, with $s=\cos\theta$, can provide nonzero information about $\theta$ in a situation in which definite causal order provides zero information about $\theta$.

The treatment here can be repeated for arbitrary dimensions by replacing $\sigma_{\mathbf{u}}$ by some other unitary operation. For some channel
\begin{equation}
\mathcal{E}_{\mathrm{dephase}}(\rho)=p\rho+(1-p) V\rho V^\dagger \, ,
\end{equation}
 the final result for $R_{01}$ remains the same, now with 
\begin{equation}
s=\expct{V^\dagger U^\dagger VU} \, .
\end{equation} 
So long as $U$ and $V$ do not commute, this provides information about $\theta$ to the control state $\rhoc^\prime$ that can be measured. In fact, the term $s=\expct{V^\dagger U^\dagger VU}$ is also connected to ``out-of-time-ordered correlators'' that characterize quantum information scrambling and, through it, the celebrated Kirkwood-Dirac distribution~\cite{YungerHalpernetal2018} {(see Ref.~\cite{Gaoetal2022arxiv} for further applications of the quantum switch for measuring noncommutativity)}. 

Whether or not there is an \textit{advantage} from ICO depends on whether or not schemes with DCO lose all information from such dephasing. Considering a spin system, with $U$ performing an SU(2) rotation of a spin-$J$ particle, all of the previous calculations hold true with $\theta\mapsto 2J\theta$ for generalized dephasing $V$ enacting a $\pi$ rotation about some axis $\mathbf{u}$. This means that one can again attain an infinite advantage in estimating $\theta$ using ICO for dephasing along the $y$ axis, even using a maximally mixed probe state, relative to the optimal quantum strategy of using a pure superposition of extremal eigenstates of $U$ (i.e, NOON- or GHZ-type states of the correct orientation).

Next, considering more general dephasing operators $V$, we can speculate on a large class of ICO-driven advantages. In large dimensions and for all but pathological cases of $U$ and $V$, the four pure states $U\ket{\psi}$, $UV\ket{\psi}$, $VU\ket{\psi}$, and $VUV\ket{\psi}$ are linearly independent, even though they were dependent for the qubit case. If they are all orthogonal, the four eigenvalues of $\rho$ after evolving through the unitary and pair of dephasing channels are $p_A p_B$, $p_A (1-p_B)$, $p_B (1-p_A)$, and $(1-p_A) (1-p_B)$; these cause the QFI to identically vanish at $p_A=p_B=1$ in Eq.~\eqref{eq:QFI from evals evecs}. Since the ICO-evolved state $\rhoc^\prime$ continues to depend on $\theta$ through $s$, these could present another array of ICO-driven advantages for estimation of unitaries in the presence of noise.

\subsection{Amplitude damping noise}
\label{sec:amp damp single}

Now consider a final type of noise, in which the probe has a propensity to relax from some excited state to its ground state. This well-known amplitude damping channel~\cite{NielsenChuang2000} acting on a qubit
\begin{equation}
\mathcal{E}_{\mathrm{amp.\,damp.}}\left[
\begin{pmatrix}
    \rho_{00}& \rho_{01}\\
    \rho_{10}& \rho_{11}
\end{pmatrix}
\right]=\left[
\begin{pmatrix}
    \rho_{00}+(1-p)\rho_{11}& \rho_{01}\sqrt{p}\\
    \sqrt{p}\rho_{10}& p\rho_{11}
\end{pmatrix}
\right]
\end{equation} 
is characterized by the two Kraus operators 
\begin{equation}
K_0=\begin{pmatrix}
    1&0\\0&\sqrt{p}
\end{pmatrix}\, , \qquad 
K_1=\begin{pmatrix}
    0&\sqrt{1-p}\\0&0
\end{pmatrix} \, ,
\end{equation} 
that cause a system to relax toward state $\ket{0}$ when $p$ gets closer to $0$. Again considering a unitary along the $z$ axis and the optimal probe state $\ket{+}$, the QFI for measuring the unitary's phase with amplitude damping both before and after application of the unitary degrades to
\begin{equation}
\mathsf{Q}_{\ket{+}}(\theta)=p_A p_B \, .
\label{eq:QFI damp DCO}
\end{equation}

Considering the general case of fixed causal order where the initial probe state is arbitrary, the evolved state is
\begin{equation}
    \mathcal{E}(\rho)=\left(
\begin{array}{cc}
 1-p_A p_B (1-\rho_{00}) & \eu^{\iu \theta} \sqrt{p_A p_B} \rho_{01} \\
 \eu^{-\iu \theta} \sqrt{p_A p_B} \rho_{10} & p_A p_B (1-\rho_{00}) \\
\end{array}
\right).
\end{equation} 
This has a QFI of $\mathsf{Q}=4p_Ap_B |\rho_{01}|^2$, so maximal information about $\theta$ is obtained when $|\rho_{01}|$ is maximized, confirming our intuition that $\ket{+}$ remains an optimal probe state in the presence of noise.

With a quantum switch controlling the orders of applications of two amplitude damping channels on a maximally mixed state, we can readily compute
\begin{equation}
R_{01}=\frac{1}{2} [ 1-\eu^{\iu \theta} \left(p_A-1\right) \sqrt{p_B}-\eu^{-\iu \theta} \sqrt{p_A} \left(p_B-1\right)+p_A p_B ] \, .
\label{eq:R01 amp damp qubit}
\end{equation}The QFI for the evolved control state is a bit involved, though it is simply given by Eq. \eqref{eq:QFI from evals evecs} and the eigensystem of a $2\times 2$ matrix, so we write the results in the relevant limit of neglecting terms of order $\mathcal{O}(p_A^{3/2},p_B^{3/2},p_Ap_B^{1/2},p_A^{1/2}p_B)$:
\begin{equation}
\mathsf{Q}_{\mathrm{ICO}}(\theta)\approx \frac{(\sqrt{p_A}-\sqrt{p_B} )^2}{4}+\frac{\sin^2\theta}{12} (p_A+p_B+14\sqrt{p_Ap_B} ) \, .
\end{equation} 
This again provides an unlimited benefit in terms of QFI ratio or minimum uncertainty ratio relative to all schemes with DCO in the limit of small $p_A$ and $p_B$ and a formally infinite advantage when either $p_A$ or $p_B$ vanishes (this can also be seen because schemes with DCO leave the probe state independent from $\theta$ when $p_A$ or $p_B$ vanishes). Notably, this advantage can be attained for any unitary, even when $\theta=0$, with the exception of $p_A=p_B$ when $\theta=0$.

Another type of amplitude damping channel has $\mathcal{E}^{(A)}$ sending a system toward state $\ket{0}$ and $\mathcal{E}^{(B)}$ toward $\ket{1}$; different relaxation tendencies occur on different sides of the unitary. Mathematically, this happens when $\mathcal{E}^{(A)}$ has the Kraus operators from before and $\mathcal{E}^{(B)}$ has Kraus operators
\begin{equation}
K_0^{(B)}=\begin{pmatrix}
    \sqrt{p}&0\\0&1
\end{pmatrix} \, , \qquad
K_1^{(B)}=\begin{pmatrix}
    0&0\\\sqrt{1-p}&0
\end{pmatrix} \, .
\end{equation} 
In this case, the QFI for DCO schemes with unitary $U$ about the $z$-axis and optimal probe state $\ket{+}$ again takes the form $\mathsf{Q}_{\ket{+}}(\theta)=p_A p_B$ and is optimal among DCO schemes, as can again be recognized from the DCO QFI $\mathsf{Q}=4p_Ap_B |\rho_{01}|^2$. In contrast, the small-$p$ limit of the QFI for schemes with ICO is again $\mathsf{Q}_{\mathrm{ICO}}(\theta)\approx (\sqrt{p_A}-\sqrt{p_B})^2/4$.

Amplitude damping toward some state $\ket{0}$ can be extended to amplitude damping occurring identically on $n=\log_2 d$ qubits in parallel. In the case of ICO, $R_{01}$ simply gets modified as $R_{01}\mapsto R_{01}^n$, retaining the dependence on $\sqrt{p_A}$ and $\sqrt{p_B}$ to first order as
\begin{equation}
R_{01}\approx\frac{1+\eu^{\iu \theta} n\sqrt{p_B}+\eu^{-\iu \theta} n\sqrt{p_A} }{2^n} \, .
\end{equation} 
The QFI then becomes, to lowest order in $p_A$ and $p_B$,
\begin{align}
\mathsf{Q}_{\mathrm{ICO}}(\theta) & =\frac{n^2\sin^2\theta}{4^n-1}\left(\sqrt{p_A}+\sqrt{p_B}\right)^2 \nonumber \\
& +\frac{n^2\cos^2\theta}{4^n}\left(\sqrt{p_A}-\sqrt{p_B}\right)^2 \, .
\end{align} 
This worsens with $n$ but retains the same scaling with $p_A$ and $p_B$ as for our depolarization case in Sec.~ \ref{sec:depol companion}. The optimal DCO scheme now involves entangled qubits in a GHZ state $\left(\ket{0}^{\otimes n}+\ket{1}^{\otimes n}\right)/\sqrt{2}$, but amplitude damping diminishes its QFI by $p_A^n p_B^n$:
\begin{equation}
\mathsf{Q}_{\mathrm{GHZ}}(\theta)=2\frac{p_A^n p_B^n}{1+ (1-p_A p_B )^n+p_A^n p_B^n} \, .
\end{equation} 
A better DCO scheme might be to use $n$ qubits in parallel, each in the $\ket{+}$ state, which would allow the QFI so simply scale with $n$ instead of diminishing exponentially. Still, such DCO schemes are limited to $\mathsf{Q}\sim \mathcal{O}(p_A p_B)$, while ICO allows $\mathsf{Q}_{\mathrm{ICO}}\sim \mathcal{O}(p_A, p_B,\sqrt{p_A p_B})$ for any number of qubits $n$ \footnote{See Ref.~\cite{Liuetal2023} for a comparison of ICO strategies for the amplitude damping channel when given access to two (or more) copies of $U$ each followed by an amplitude damping channel, where one can control the orders of the two noisy channels but cannot control the order of the noise and the unitary within each channel.}. We again see a general advantage for ICO over DCO schemes for estimation of a unitary in the presence of noise, even using maximally mixed states as inputs, with the advantage growing with the amount of noise and diminishing with the dimension of the probe system. 

\section{Multiparameter estimation}
\label{sec:multiparam}

One notices above that the control state depends not only on the parameters of the unitary being estimated but also on the strength of the noise channels. As such, one can imagine using the same protocol to estimate the noise levels instead of the unitary's parameters. One cannot estimate both simultaneously, because the control state only depends on a single parameter, $R_{01}$, through which both $\theta$ and $p$ are to be estimated. It then follows that higher dimensional control states that depend on more parameters may be used to simultaneously estimate multiple parameters of the quantum channels, as we presently show.

\subsection{Estimation of depolarization noise and unitary phase}
We now step into the world of multiparameter estimation. For a measurement only of the control to yield information about more than one parameter, it must have more than one functional dependence on those parameters. In the examples above, we only had access to the parameter $R_{01}$, which was often real, notably in the case of depolarization channels with maximally mixed probe states. Here we show how using control systems with larger dimensions for the quantum switch allows one to simultaneously estimate both depolarization noise strength and the unitary channel's phase.

Referring again to Fig. \ref{fig:general one unitary two noise}, there are three total channels: two depolarizations and one unitary. The six different orders of traversing these channels can be combined to give different functional dependencies on $p_A$, $p_B$, and $u(\theta)$ such that the noise and unitary parameters can be simlutaneously estimated. We will not require all six orders to determine only three parameters, so we choose the three orders $\mathcal{E}^{(A)}\circ\mathcal{E}^{(B)}\circ U$, $U\circ \mathcal{E}^{(A)}\circ\mathcal{E}^{(B)}$, and $\mathcal{E}^{(B)}\circ U\circ\mathcal{E}^{(A)}$ when the control is in state $\ket{0}$, $\ket{1}$, and $\ket{2}$, respectively. We then calculate using Eqs.~\eqref{eq:probe state conditioned on control basis} and \eqref{eq:Rij control} for the three orders:
\begin{align}
R_{01}&=p_A p_B+\frac{1-p_A p_B}{d}\expct{U}\Tr(U^\dagger) ,\nonumber \\
R_{02}&=p_A +\frac{(1-p_A)(1-p_B)}{d^2}+\frac{p_B(1-p_A)}{d}\expct{U}\Tr(U^\dagger) , \nonumber \\
R_{12}&=p_B +\frac{(1-p_A)(1-p_B)}{d^2}+\frac{p_A(1-p_B)}{d}[\expct{U}\Tr(U^\dagger)]^* \, .
\end{align} 
These three different functional dependencies on $p_A$, $p_B$, and $\expct{U}\Tr(U^\dagger)$ allow all three to simultaneously be estimated from the evolved control state $\rhoc^\prime$, even though there was only one copy of each channel being probed. This again holds even if the probe state is maximally mixed and therefore insensitive to each of $p_A$, $p_B$, and $U$ for DCO  schemes. When the probe is maximally mixed, we again find the dependence on $\theta$ for the ICO scheme through $u(\theta)=|\Tr(U)|^2=d\expct{U}\Tr(U^\dagger)$. Moreover, dependence on $\theta$ is maintained even when $p_A=p_B=0$, showing that the ability to completely control the order of the depolarization and unitary channels (not restricted to the unitary always occurring between the two depolarizations) leads to sensitivity that is ``even more impossible'' with DCO.

Measuring the control state in the $(\ket{i}\pm\ket{j})/\sqrt{2}$ basis directly yields the real part of $R_{ij}$, where $R_{ij}$ is automatically real when the probe is maximally mixed. This can be facilitated by a POVM with six elements $(\ket{i}\pm\ket{j})(\bra{i}\pm\bra{j})/4$, $i<j$, where the extra factor of two is required for normalization. The six probabilities are 
\begin{equation}
P_{ij\pm}=\frac{1\pm R_{ij}}{6},\qquad i<j \, ,
\end{equation} 
and we can use them to calculate the QFI matrix. Incredibly, this matrix is nonzero and invertible even when $p_A=p_B=0$, where \textit{no} DCO strategy could ever determine $u$. We write the FI  matrix in the $\boldsymbol{\theta}=(\theta,p_A,p_B)$ basis and record the result for $p_A=p_B=0$, with the full expression given in the Supplemental Material:
\begin{equation}
\mathsf{\mathbf{F}}_{\mathrm{ICO}}(\boldsymbol{\theta})=\left(
\begin{array}{ccc}
 \frac{\left(\partial u/\partial\theta\right)^2}{3 d^4-3 u^2} & 0 & 0 \\
 0 & \frac{d^4-2 d^2+(u-2) u+2}{3 \left(d^4-1\right)} & \frac{2 (u-1)}{3 \left(d^2+1\right)} \\
 0 & \frac{2 (u-1)}{3 \left(d^2+1\right)} & \frac{d^4-2 d^2+(u-2) u+2}{3 \left(d^4-1\right)} \\
\end{array}
\right) \, .
\end{equation} 
We can call this a \emph{formally infinite} advantage due to the unlimited increase in Fisher information in simultaneously estimating three parameters using ICO, even using a maximally mixed probe for the latter, valid for unitaries and depolarization channels in arbitrary dimensions $d$. Of course, there might exist another measurement strategy that coaxes even more information from $\rhoc^\prime$, as we have not computed the QFI matrix for this state, but we are satisfied with an infinite increase in QFI relative to any strategy with DCO, especially because our FI matrix is attainable through a straightforward measurement procedure. Moreover, since we have chosen a fixed POVM to obtain these results, we need not worry about factors of $f$ from Eq.~\eqref{eq:how tight is multiparam QFI} and are guaranteed to saturate the classical Cram\'er-Rao bound for the minimum uncertainties of each parameter [Eq.~\eqref{eq:CRB classical}] in the asymptotic limit.

We can also inspect the large-$d$ limit, which makes it more difficult to estimate $\theta$ as seen before. Even in this limit, each of $p_A$ and $p_B$ can be estimated without much trouble, given the constant term in $[d^4-2 d^2+(u-2) u+2]/[3 (d^4-1)] = (1-2/d^2)/3+\mathcal{O}(d^{-4})$. This prompts us to calculate the $d\to\infty$ limit of $\mathsf{\mathbf{F}}$ for arbitrary $p_A$ and $p_B$, perhaps having in mind a measurement with macroscopic probe systems. Different functional forms $u(\theta)$ will behave differently in the limit of large $d$, so we inspect only the $(p_A,p_B)$ submatrix of $\mathsf{\mathbf{F}}$ to show how it allows $p_A$ and $p_B$ to simultaneously be estimated
\begin{equation}
\lim_{d\to\infty}\mathsf{\mathbf{F}}_{\mathrm{ICO}}(p_A,p_B)=\left(
\begin{array}{cc}
 \frac{\left(1-2 p_A^2\right) p_B^2+1}{3 \left(p_A^2-1\right) \left(p_A^2 p_B^2-1\right)} & \frac{p_A p_B}{3-3 p_A^2 p_B^2} \\
 \frac{p_A p_B}{3-3 p_A^2 p_B^2} & \frac{\left(1-2 p_B^2\right) p_A^2+1}{3 \left(p_B^2-1\right) \left(p_A^2 p_B^2-1\right)} \\
\end{array}
\right).
\end{equation}

Just as for single-parameter estimation, probe states other than the maximally mixed state can be used to investigate other properties of $U$. There is still only one functional dependence on the unitary's parameters through $\expct{U}\Tr(U^\dagger)$, so this parameter could simultaneously be estimated alongside $p_A$ and $p_B$ if one desires. The real part of $\expct{U}\Tr(U^\dagger)$ would be accessible through the POVM described in this section, while a more general POVM might gain access to the imaginary part at the same time.

To conclude this section, we note that one could have chosen other combinations of orders to estimate these three parameters. We tabulate in the Supplemental Material all of the matrix elements $R_{ij}$ that would arise from all 36 interference terms of the six possible orders of traversing the two depolarization channels and one unitary channel. One could use higher-dimensional control states to gain redundant information about the parameters of interest because these 36 elements have more than three functional dependencies on $p_A$, $p_B$, and $u(\theta)$. These interference terms are responsible for ICO's advantages in metrology.

\subsection{Estimating dephasing noise and unitary phase}
Consider the same three orders as above but replacing the depolarizing channels with dephasing channels in the direction $\mathbf{u}$ pointing anywhere along the equator for qubit rotations about the $z$ axis. The three off-diagonal elements of $\rhoc^\prime$ can be calculated using
\begin{align}
R_{01}&=(p_A+p_B-2 p_A p_B) \cos \theta+ 2 p_A p_B-p_A-p_B+1  , \nonumber\\
R_{02}&=p_A (1-\cos \theta)+\cos \theta , \nonumber\\
R_{12}&=p_B (1-\cos \theta)+\cos \theta \, .
\end{align} 
These three linearly independent functions then facilitate the simultaneous estimation of $\theta$ and the noise strengths $p_A$ and $p_B$ using the six-element POVM with projections of the control state onto the $(\ket{i}\pm\ket{j})/\sqrt{2}$ basis.

The FI matrix is again a complicated expression, so we here record the result for $p_A=p_B=1/2$, where DCO strategies cannot be used to estimate $\theta$. In this limit, even the inverse is not too unwieldy, becoming
\begin{widetext}
\begin{equation}
\mathsf{\mathbf{F}}^{-1}_{\mathrm{ICO}}(\theta,p_A,p_B)=\left(
\begin{array}{ccc}
 \frac{\cos \theta+1}{\cos \theta+3} & -\frac{2 \sin \theta}{3 \cos \theta+9} & -\frac{2 \sin \theta}{3 \cos \theta+9} \\
 -\frac{2 \sin \theta}{3 \cos \theta+9} & \frac{4-4 \cos \theta}{3 \cos \theta+9} & 0 \\
 -\frac{2 \sin \theta}{3 \cos \theta+9} & 0 & \frac{4-4 \cos \theta}{3 \cos \theta+9} \\
\end{array}
\right)^{-1}=
\left(
\begin{array}{ccc}
 3+\frac{6}{\cos \theta+1} & \frac{3}{2} (\cos \theta+3) \csc \theta & \frac{3}{2} (\cos \theta+3) \csc \theta \\
 \frac{3}{2} (\cos \theta+3) \csc \theta & 3 \csc ^2\frac{\theta}{2}-\frac{3}{2} & \frac{3}{8} (\cos \theta+3) \csc ^2\frac{\theta}{2} \\
 \frac{3}{2} (\cos \theta+3) \csc \theta & \frac{3}{8} (\cos \theta+3) \csc ^2\frac{\theta}{2} & 3 \csc ^2\frac{\theta}{2}-\frac{3}{2} \\
\end{array}
\right).
\end{equation}
\end{widetext}
We place the full expression for the FI matrix in the Supplemental Material, with a determinant that only vanishes when $(1-p_A) (1-p_B) \sin (\theta/2) \sin \theta=0$. As with depolarization, the bound of Eq.~\eqref{eq:CRB classical} can be saturated without resorting to considerations of QFI from Eq.~\eqref{eq:how tight is multiparam QFI} because we have chosen a fixed, accessible measurement scheme. We can again conclude that ICO may grant a formally infinite advantage over DCO schemes in this multiparameter estimation context.

\subsection{Estimating amplitude damping noise and unitary phase}
Next let us show how to simultaneously estimate both amplitude damping channels' noise parameters and the unitary's phase by controlling the order of operations with a higher-dimensional quantum switch. Keeping the same nominal three orders as in the previous sections, we use the maximally mixed qubit probe state to calculate
\begin{align}
R_{01}&=\frac{1}{2} [ \eu^{-\iu\theta} (1-p_A p_B)+p_A p_B+1 ] , \nonumber \\
R_{02}&=\frac{1}{2} [-(p_A-1) \sqrt{p_B} \eu^{-\iu\theta}+p_A p_B-\sqrt{p_A} (p_B-1)+1 ] , \nonumber \\
R_{12}&=\frac{1}{2} [ -\sqrt{p_A} (p_B-1) \eu^{\iu\theta}+p_A p_B-(p_A-1) \sqrt{p_B}+1 ] \, .
\end{align}

\begin{figure}[b]
    \centering
    \includegraphics[width=\columnwidth]{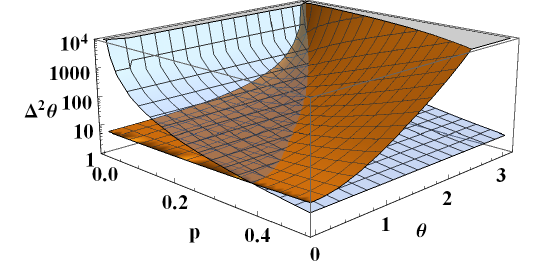}
    \caption{Minimum variance that saturates the Cram\'er-Rao bound for estimating a phase $\theta$ of a qubit rotation subject to amplitude damping noise for our ICO (orange, solid) and the optimal DCO (blue, translucent) multiparameter estimation schemes. The multiparameter schemes requires inverting the full FI matrix and inspecting the $\theta\theta$ component thereof, which is plotted as $\Delta^2\theta$ here. Our scheme with ICO is significantly better than strategies with DCO when $p$ and $\theta$ are small and vice-versa; ICO can be used to great effect in the former regime and should be avoided in the latter. These uncertainties are plotted for a single trial and will be normalized by the total number of independent trials, so one need only worry about the ratio between the variances for ICO and DCO, with less concern paid to the absolute magnitudes.}
    \label{fig:damping multiparameter phase}
\end{figure}

We can again measure the evolved control state using the POVM comprised of projectors onto states $(\ket{i}\pm\ket
j)/\sqrt{2}$ with $i<j$. This will be sensitive to the real parts of $R_{ij}$, which are all that we require for estimating $\theta$, $p_A$, and $p_B$. Computing the FI matrix with this method yields complicated expressions, so we plot the appropriate component of the inverse $\left(\mathsf{\mathbf{F}}^{-1}\right)_{\theta\theta}$ in Fig.~\ref{fig:damping multiparameter phase} to display the phase sensitivity of the multiparameter scheme as a function of $\theta$ and $p_A=p_B\equiv p$.
The minimum variance $\Delta^2\hat{\theta}$ is seen to be bounded, showing that this multiparameter estimation scheme is successful, and significantly outperforms the limit one could achieve with single-parameter DCO schemes [$1/p^2$; c.f. Eq.~\eqref{eq:QFI damp DCO}] when $p$ and $\theta$ are small.
We have license to use the components of $\mathsf{\mathbf{F}}^{-1}$ as the covariances on the estimated parameters due to having supplied a fixed POVM that can saturate the bound of Eq.~\eqref{eq:CRB classical}.
This comparison is the same whether we use a multiparameter or a single-parameter estimation scheme for the qubit probe with DCO, because the off-diagonal components $\mathsf{Q}_{\theta p_A}$ and $\mathsf{Q}_{\theta p_B}$ vanish in that case and, so, multiparameter estimation does not worsen the estimation.

\subsection{Estimation of a unitary's phase and axis}
What if one desires to simultaneously estimate more than one parameter from $U$ using ICO? Increasing the dimension of the control system will again be necessary, but we saw above that depolarization channels with maximally mixed probes only give access to one parameter from $U$ (the Supplemental Material  shows all of the dependencies to be $u(\theta)=|\Tr(U)|^2$ and $\tilde{u}(\theta,\mathbf{n};\rho)=\expct{U}\Tr(U^\dagger)$, so one could consider engineering more complicated probe states to glean information about an additional function of $(\theta,\mathbf{n})$ through $\tilde{u}$; the experimental challenge of creating other probe states must be balanced with their effectiveness at identifying requisite parameters and we leave such study to further work). Here we explore whether ICO in the presence of dephasing or amplitude damping channels, which do not act isotropically on a state, gives access to more parameters of $U$ for simultaneous estimation. It turns out that the former is insufficient while the latter can be used for such simultaneous estimation with maximally mixed probe states.

\subsubsection{Dephasing channels cannot be used to estimate phase and axis}
Consider the case of qubit rotations, in which we parametrize the unitary's axis as $\mathbf{n}=\left(\sin\Theta\cos\Phi,\sin\Theta\sin\Phi,\cos\Theta\right)$. Subjecting a maximally mixed probe state to dephasing noise as above with a coherent control of the three orders of the channels yields the following matrix elements when we consider dephasing along the $z$ axis [i.e., $\mathcal{E}(\rho)=p\rho+(1-p)\sigma_z\rho\sigma_z$]:
\begin{widetext}
\begin{align}
R_{01}&=\frac{1}{2} [(\cos 2 \Theta  + 2 \sin ^2\Theta  \cos \theta) (-2 p_A p_B+p_A+p_B)+2 p_A p_B-p_A-p_B+2 ] , \nonumber \\
R_{02}&=\frac{1}{2} [-(p_A-1) (\cos 2 \Theta + 2 \sin ^2\Theta  \cos \theta )+p_A+1 ] , \nonumber\\
R_{12}&=\frac{1}{2} [-(p_B-1) (\cos 2 \Theta + 2 \sin ^2\Theta  \cos \theta )+ p_B+1 ] , \nonumber \\
R_{24}&=\frac{1}{2} \{  (\cos 2 \Theta +2 \sin ^2\Theta  \cos \theta ) [p_A (p_B-1)+\sqrt{(p_A-1) p_A (p_B-1) p_B}-p_B+1 ] \nonumber \\
& + 3 p_A p_B+3 \sqrt{(p_A-1) p_A (p_B-1) p_B}-p_A-p_B+1 \} .
\end{align}
\end{widetext}
We have included an additional ordering by adding a control state $\ket{4}$ that sends the probe through the channels as $\mathcal{E}^{(B)}\circ\mathcal{E}^{(A)}\circ U$ to showcase a general trend (keeping the same orders as in the Supplemental Material). Again, these are independent from $\Phi$ due to the particular dephasing axis; another dephasing axis allows one to inspect other projections of $\mathbf{n}$ onto that axis. The only angular information, however, arises in the form of the single function $\sin^2\Theta\cos\theta+\cos2\Theta$. This function is indeed sensitive to the rotation angle (phase) and the projection of the dephasing axis onto the unitary's rotation axis, with this projection explaining why ICO could be used above for $x$- and $y$-axis dephasings but not $z$-axis dephasing for unitaries about the $z$ axis. Since there is only one function present, only one variable can be estimated. If the unitary's rotation angle is known, this can be used to estimate the rotation axis and vice versa, but under no circumstances can this be used to estimate two unitary parameters simultaneously. Probe states that are not maximally mixed would be necessary to perform such a simultaneous estimation with ICO.

\subsubsection{Amplitude damping channel can be used to estimate phase and axis}
Consider again the case of qubit rotations with a general axis $\mathbf{n}$. Subjecting a maximally mixed probe state to amplitude damping noise as above with a coherent control of the three orders of the channels yields the matrix elements
\begin{widetext}
\begin{align}
R_{01}&=\frac{1}{2} [ \sin ^2\Theta  \sqrt{p_A p_B} + p_A p_B \cos ^2\Theta - \cos \theta (\sin ^2 \Theta  \sqrt{p_A p_B}+p_A p_B \cos ^2 \Theta - 1 ) + \iu \cos \Theta (p_A p_B-1) \sin \theta+1 ] , \nonumber \\
R_{02}&=\frac{1}{8} \{ 2 \sqrt{p_B}  [ \cos 2 \Theta  ( p_A \sqrt{p_B}-2 \sqrt{p_A p_B}+p_A+\sqrt{p_B}-1 ) \sin^2 (\theta/2) + 2 \iu (p_A-1) \cos \Theta  \sin \theta ] \nonumber  \\
&+ [(\sqrt{p_A}-1)^2 p_B - 3 (p_A-1) \sqrt{p_B}] \cos \theta - 2 \sqrt{p_A} (p_B-2)+p_A \left(3 p_B-\sqrt{p_B}\right)-p_B+\sqrt{p_B}+4 \} , \nonumber \\
R_{12}&=\frac{1}{8} \{ 2 \cos 2 \Theta  [2 p_A (p_B- \sqrt{p_B})+\sqrt{p_A} (p_B-1)-p_B+1 ] \sin^2 (\theta/2 - 4 \iu \sqrt{p_A} (p_B-1) \cos \Theta  \sin \theta \nonumber \\
&+ [-3 \sqrt{p_A} (p_B-1)+2 p_A (p_B - \sqrt{p_B} )-p_B + 1 ] \cos \theta+ (2 p_A-\sqrt{p_A}+1 ) p_B-2 (p_A-2) \sqrt{p_B}+\sqrt{p_A}+3 \} \, .
\label{eq:Rij multiparameter damping}
\end{align}
\end{widetext} 
From these expressions, we see the importance of the interplay between the particular amplitude damping channel and $\mathbf{n}$: $\Phi$ is absent from $R_{ij}$. ICO with this particular amplitude damping channel can be used to simultaneously estimate the unitary's phase and the polar angle of its rotation axis, while another amplitude damping channel that singles out a different preferred axis could be used to learn about another projection of $\mathbf{n}$.

Suppose one wishes to simultaneously estimate both noise parameters $p_A$ and $p_B$ in addition to the two unitary parameters $\theta$ and $\Theta$ using ICO and this pair of amplitude damping channels. One must immediately be wary, as we have only computed three quantities $R_{ij}$ and seek four parameters. There are a few paths forward: a) one can perform a measurement with different POVM elements sensitive to the real and imaginary parts of $R_{ij}$, using projections onto the states $(\ket{i}\pm\iu\ket{j})/\sqrt{2}$ in addition to $(\ket{i}\pm\ket{j})/\sqrt{2}$; b) one can consider situations in which the two noise levels are known to be equal, $p_A=p_B\equiv p$, such that the total number of parameters to be estimated is three; c) one may seek to only estimate a subset of the parameters, implicitly assuming the rest to be known; or, d) one can consider expanding the dimension of the control system, such as by adding a control state $\ket{4}$ that sends the probe through the channels as $\mathcal{E}^{(B)}\circ\mathcal{E}^{(A)}\circ U$ (keeping the same orders as in the Supplemental Material), which provides new functions of the four parameters such as
\eq{
R_{04}=\frac{1}{2} \left(p_A \left(p_B-\sqrt{p_B}\right)-\sqrt{p_A} (p_B-1)+\sqrt{p_B}+1\right).
}

\begin{figure}[t]
    \includegraphics[width=\columnwidth]{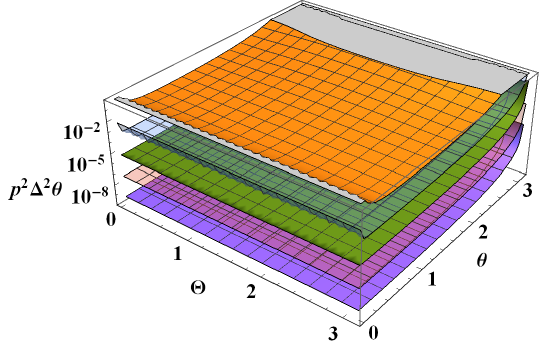}
    \caption{Decrease in uncertainty for estimating the rotation angle $\theta$ of a qubit rotation when simultaneously estimating the unitary's rotation angle, its axis's polar coordinate, and an amplitude damping noise level $p$ using ICO, relative to schemes with DCO that achieve a minimum $p^2$. This is equal to the ratio of the smallest possible inverse of the FI matrix without ICO, $p^2$, to the $\theta,\theta$ element of the inverse of the measured FI matrix with ICO. The probe, which for ICO is a maximally mixed state, goes through an amplitude damping noise channel with strength $p$ both before and after the unitary. The different sheets plotted correspond to $p$ ranging from $10^{-1}$ to $10^{-5}$ by factors of 10, with the $p$ increasing from the lowest to the highest sheet; the advantage is approximately $\mathcal{O}(p^2)$ smaller variances. The upper cutoff is set to $1$ to single out the regime of ICO-driven advantages. The advantage is most prominent when $\theta$ is further from $0$ and $\pi$.}
    \label{fig:damp multiparam angle}
\end{figure}
\begin{figure}[t]
    \includegraphics[width=\columnwidth]{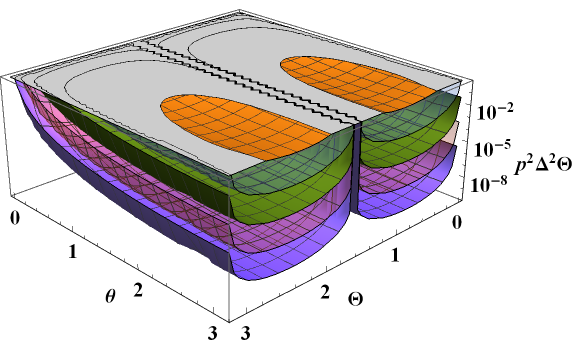}
    \caption{Same as Fig.~\ref{fig:damp multiparam angle}, but the uncertainty is plotted for estimating the polar angle $\Theta$ of a qubit rotation's rotation axis. Again, $p$ increases from the lowest to the highest sheet with approximate advantages for ICO of the order $\mathcal{O}(p^2)$. Now the advantage is most prominent when $\theta$ is further from $0$ when $\Theta$ is furthest from $0$, $\pi/2$, and $\pi$.}
    \label{fig:damp multiparam pole}
\end{figure}
\begin{figure}[t]
    \includegraphics[width=\columnwidth]{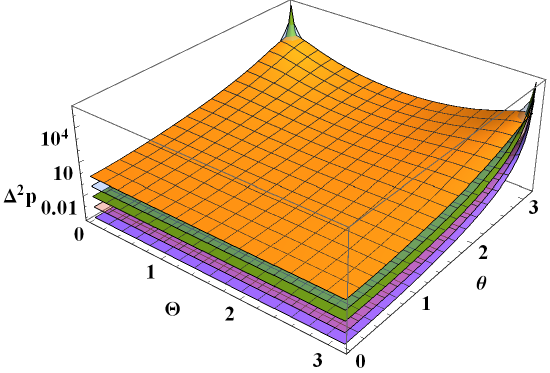}
    \caption{Same as Figs. \ref{fig:damp multiparam angle} and \ref{fig:damp multiparam pole} but the uncertainty is plotted for estimating the noise level of the amplitude damping channel and is not normalized. Again, $p$ increases from the lowest to the highest sheet. The uncertainty is lowest when $\theta$ is smallest.}
    \label{fig:damp multiparam noise}
\end{figure}

We now inspect the performance of measuring the control state in the $(\ket{i}\pm\ket{j})/\sqrt{2}$ basis as before. We consider the case where $p_A=p_B\equiv p$ to streamline the assessment, using only the coefficients from Eq. \eqref{eq:Rij multiparameter damping}. Normalizing the minimum values of $\Delta^2\theta$ and $\Delta^2\Theta$ by the inverse of Eq. \eqref{eq:QFI damp DCO}, which is the increase in uncertainty one would expect for strategies with DCO, we plot the minimum uncertainties for $\theta$, $\Theta$, and $p$ in Figs. \ref{fig:damp multiparam angle}, \ref{fig:damp multiparam pole}, \ref{fig:damp multiparam noise}, respectively for various small values of $p$. As above, we consider the components of the inverse of the FI matrices to represent the minimum uncertainties, which is justified in the asymptotic limit of saturating Eq.~\eqref{eq:CRB classical} with a fixed POVM. As discussed in the figure captions, one can observe a significant advantage relative to DCO schemes for estimating $\theta$ and $\Theta$ when $p$ is small and the former two parameters are in the proper regimes, with the advantage qualitatively corresponding to $\mathcal{O}(p^{2})$ smaller variances, while one can sensibly estimate $p$ at the same time if $\theta$ is small.

\section{Multiparameter estimation with multiple copies of identical noisy unitaries}
\label{sec:multiparam copied channels}
\begin{figure}
    \centering
    \includegraphics[width=\columnwidth]{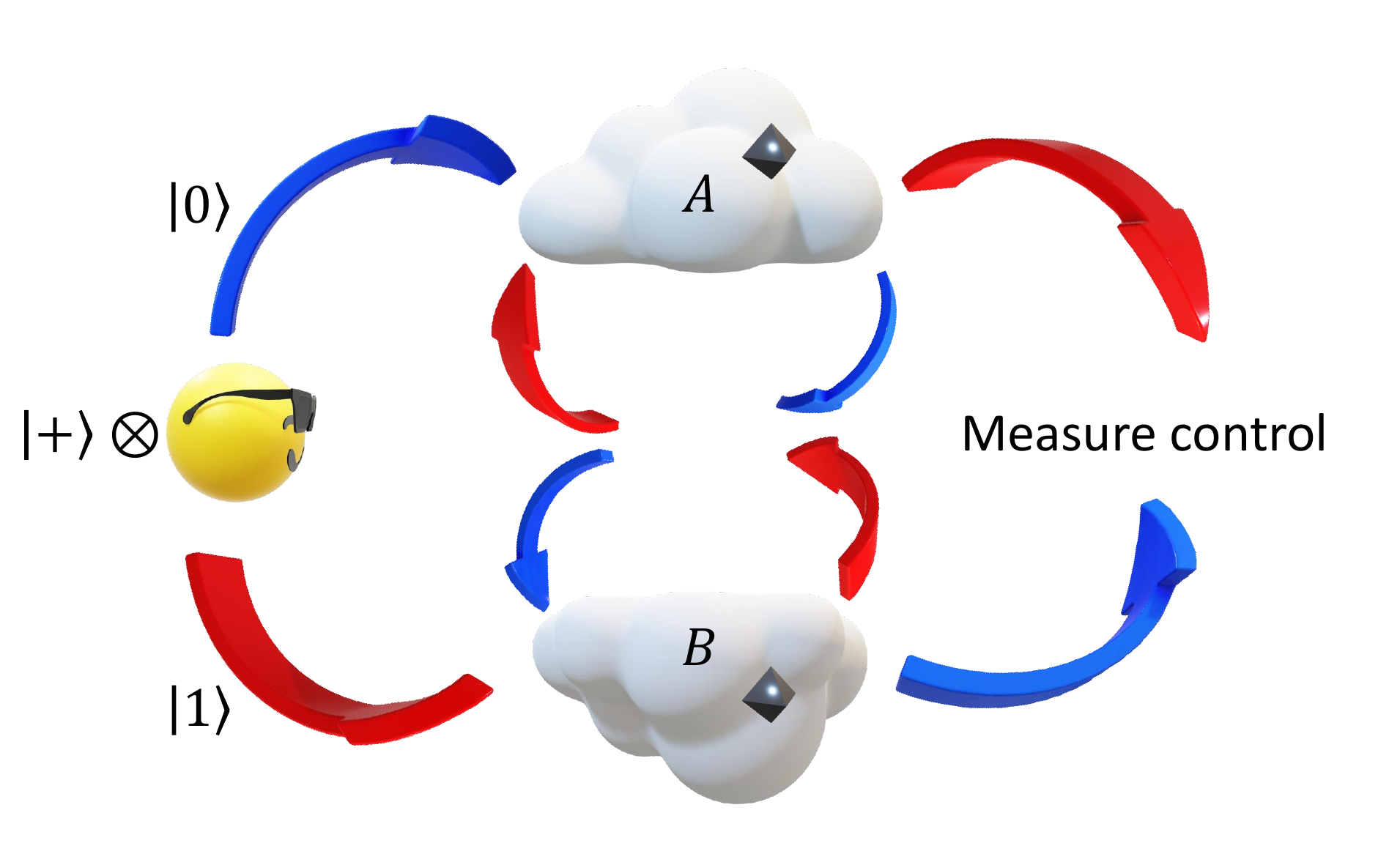}
    \caption{
    Schematic for metrology with ICO, given two copies $A$ and $B$ of a noisy unitary channel. As in Fig. \ref{fig:general one unitary two noise}, a control system dictates the order in which the channels are traversed by the probe, while a measurement on the control alone that has not interacted with the noisy channels is sufficient to infer properties of the channels.
    }
    \label{fig:general two unitary two noise}
\end{figure}

Our above analyses used ICO to crucially control the order in which a unitary and noise channels were applied, schematized in Fig. \ref{fig:general one unitary two noise}. Other studies of ICO for noisy metrology, in contrast, assumed multiple identical copies of the same noisy channel, without the possibility of controlling the order of the noise and unitary within one joint channel. An example scheme can be seen in Fig. \ref{fig:general two unitary two noise}, where now each one unitary is embedded in noise channel $A$ and another identical unitary in noise channel $B$, with ICO merely controlling the order of overall channels $A$ and $B$. For such schemes, no information about the unitary can be found if the noise channels are completely depolarizing or completely amplitude damping, in contrast to our earlier schemes, even in the limit of large numbers of copies of the channels \cite{Kurdzialeketal2022arxiv,Liuetal2023}. In this section, we show how identical-channel schemes with fixed causal orders \textit{within} each channel can be extended to multiparameter estimation in arbitrary dimensions using ICO. We also show how such strategies can retain FI of order $\mathcal{O}(p)$ for any number $D$ depolarization channels, even though naive schemes with DCO  would have FI dramatically lower at order $\mathcal{O}(p^D)$; even though such an advantage should also be attainable in in the limit of arbitrary copies of the channels by using adaptive techniques or ancilla-entangled strategies \cite{Kurdzialeketal2022arxiv}, we provide an explicit procedure to attain such an advantage here.

Consider the joint unitary-depolarization channel
\begin{equation}
\mathcal{E}_{\mathrm{U-depol}}(\rho)=pU\rho U^\dagger+(1-p)\frac{\openone}{d} \, .
\end{equation} 
This can be achieved by concatenating the unitary and depolarization channels above in either fixed order (unitary then depolarization or depolarization then unitary), so its Kraus operators can be chosen to be
\begin{equation}
K_{kl}(p)=\sqrt{\frac{1-p}{d}}\ket{k}\bra{l}U
\end{equation} 
and $K_{\openone}=\sqrt{p}U$.

What happens when a control system controls the order in which two copies of $\mathcal{E}_{\mathrm{U-depol}}$ are applied to a probe? For $d=2$ this has been studied in Ref. \cite{ChapeauBlondeau2021ICO}. Rather than simply extend this result to arbitrary $d$, we also allow for three copies of the same channel, increasing the dimension of the control state to allow for multiple parameters to simultaneously be estimated. With control states $\ket{0}$, $\ket{1}$, and $\ket{2}$ dictating that the probe experiences the noisy unitary channels in orders $\mathcal{E}^{(A)}\circ \mathcal{E}^{(B)}\circ \mathcal{E}^{(C)}$, $\mathcal{E}^{(C)}\circ \mathcal{E}^{(B)}\circ \mathcal{E}^{(A)}$, and $\mathcal{E}^{(B)}\circ \mathcal{E}^{(C)}\circ \mathcal{E}^{(A)}$, respectively, the three off-diagonal matrix elements of the evolved control state $\rhoc^\prime$ require the three functions
\begin{widetext}
\begin{align}
R_{01}&=p_Ap_Bp_C+\frac{p_Ap_B+p_Ap_C+p_Bp_C-3p_Ap_Bp_C}{d}\Tr(U^2)\expct{U^{\dagger 2}}+\frac{1-p_Ap_B-p_Ap_C-p_Bp_C+2p_Ap_Bp_C}{d^2} \nonumber \\
&=p^3+3p^2\frac{1-p}{d}\Tr(U^2)\expct{U^{\dagger 2}}+\frac{1-3p^2+2p^3}{d^2} , \nonumber \\
R_{02}&=p_Ap_Bp_C+\frac{p_A(1-p_Bp_C)}{d}\Tr(U)\expct{U^{\dagger }}+\frac{(1-p_A)p_Bp_C}{d}\Tr(U^2)\expct{U^{\dagger 2}}+\frac{(1-p_A)(1-p_B p_C)}{d^2} \nonumber \\
&=p^3+\frac{p(1-p^2)}{d}\Tr(U)\expct{U^{\dagger }}+p^2\frac{(1-p)}{d}\Tr(U^2)\expct{U^{\dagger 2}}+\frac{(1-p)(1-p^2)}{d^2} ,\nonumber \\
R_{12}&=p_C+\frac{p_Ap_B(1-p_C)}{d}\Tr(U^\dagger)\expct{U}+\frac{(1-p_A)p_B(1-p_C)}{d^2}\left|\Tr(U)\right|^2+\frac{(1-p_B)(1-p_C)}{d^2}\nonumber\\
&=p+\frac{p^2(1-p)}{d}\Tr(U^\dagger)\expct{U}+\frac{(1-p)^2p}{d^2}\left|\Tr(U)\right|^2+\frac{(1-p)^2}{d^2},
\end{align}
\end{widetext} 
where we have kept distinct values of $p_O$ on the first lines of the equations to show where the different terms originate; when the three channels are truly identical, we can set them each to be the same variable $p$. All three functions are linearly independent, even with maximally mixed probe states that will makes these three into real functions that depend on $u(\theta)$, $u(2\theta)$, $p$, and $d$.

A measurement of the control system in the $P_{ij\pm}$ basis will thus yield information from which the unitary's phase and the depolarization noise's strength can simultaneously be estimated. One only needs to measure two of the off-diagonal components to find these two parameters, such as $(\ket{0}\pm\ket{1})/\sqrt{2}$ and $(\ket{0}\pm\ket{2})/\sqrt{2}$, but redundant information can be obtained by measuring the other components and one can also use these to simultaneously estimate the dimension parameter $d$ if it is unknown. 

These demonstrate how increasing the dimension of the control system gives access to multiparameter estimation techniques for ICO strategies with multiple copies of the same noisy unitaries. Similar results can straightforwardly be obtained for other noise channels, where one can also revisit the question of measuring more than one parameter of the unitary simultaneously with only maximally mixed probe states. Here, one can also learn about multiple properties of the unitary simultaneously by using a probe state other than the maximally mixed one, as there are three different complex functional dependencies of the $R_{ij}$ on $U$ when the control system dimension was simply increased from 2 to 3. Higher dimensional controls lead to more simultaneously estimable parameters.

Next, we consider the extension to $D$ copies of the depolarization channel by allowing the control state $\ket{0}$ to dictate the order $\mathcal{E}^{(A_0)}\circ \mathcal{E}^{(A_1)}\circ\mathcal{E}^{(A_2)}\circ\cdots\circ\mathcal{E}^{(A_{D-1})}$ and $\ket{1}$ to dictate $\mathcal{E}^{(A_1)}\circ\mathcal{E}^{(A_2)}\circ\cdots\circ\mathcal{E}^{(A_{D-1})}\circ \mathcal{E}^{(A_0)}$. Keeping terms to lowest order in $p$ means that we need only consider at most one of the channels to contribute the Kraus operator $K_{\openone}$. These terms are identical when that single identity Kraus operator comes from any channel other than $A_0$, so we can readily compute
\begin{equation}
R_{01}\approx\frac{1-p}{d^{2}}+\frac{p}{d}\Tr\left(U \right)\expct{U^\dagger}.
\end{equation}
The dependence on $U$ is only diminished by $\mathcal{O}(p)$, instead of by $\mathcal{O}(p^D)$ for $D$ passes through a depolarization channel, which constitutes another large advantage over naive DCO  schemes, with the caveat that adaptive or ancilla-assisted schemes should be able to attain our scaling in the limit of unlimited copies of the noisy channel $U$. 

\section{Results independent from probe state: When the probe is a qubit}
\label{sec:independent from probe}

In most of the examples above, the probe state was chosen to be the maximally mixed state to showcase the  capabilities of ICO: ICO allows a maximally insensitive state to become sensitive. In a related but different context of ICO metrology with qubit probe states, it was found that the results were independent from the probe state~\cite{ChapeauBlondeau2021ICO,ChapeauBlondeau2022}, implying that maximally mixed states would achieve the same results as any other probe state. We show here how this trend can be generalized to ICO schemes with arbitrary numbers of channels, including having multiple copies of $U$, multiple copies of the noise channels, and more than two different noise channels.

Arbitrary qubit states can be decomposed as
\begin{equation}
\rhop=\frac{1}{2}\left(\openone +\mathbf{r}\cdot\boldsymbol{\sigma}\right),
\end{equation} 
so we desire to show that all of the parameter dependence can be imprinted onto the control system $\rhoc^\prime$ in a manner independent of $\mathbf{r}$. This is equivalent to showing that $R_{j_1 j_2}(\rhop)=R_{j_1 j_2}(\openone/2)$ or that $R_{j_1 j_2}(\sigma_i)=0\, \forall i\in(1,2,3)$. Actually, this property only holds true for some specific sequences of causal orders and some particular noisy channels. What we can instead prove is that
\begin{equation}
\RE [ R_{j_1 j_2}(\rhop) ] = \RE [ R_{j_1 j_2}(\openone/2 ) ] \, \Leftrightarrow \, \RE [
R_{j_1 j_2}(\sigma_i) ]=0 
\end{equation}
for all channels with Kraus operators 
\begin{equation}
K_{i}^{(A_j)} K_{i}^{(A_j)\, \dagger}= K_{i}^{(A_j)\, \dagger}K_{i}^{(A_j)}\propto\openone \, .
\end{equation} 
That is,  we show that measuring the real parts of the off-diagonal elements of $\rhoc^\prime$, equivalent to measuring $\rhoc^\prime$ in the $(\ket{i}\pm\ket {j})/\sqrt{2}$ basis when $\rhoc$ is initialized with coefficients of equal phase, gives information about the channels that is independent of the probe state whenever each Kraus operator from the channels can be written as some constant 
multiplied some unitary. No relationships between any two Kraus operators $K_i^{(A_j)}$ and $K_{i^\prime}^{(A_{j^\prime})}$ are necessary to enable our broad result.
 
In consequence, we seek a proof that
\begin{widetext}
\begin{equation}
\RE \left[\mathrm{Tr}\left(
\sum_{i_1,\cdots,i_3}
K_{i_{\pi_{j_1}(0)}}^{(A_{\pi_{j_1}(0)})}\cdots K_{i_{\pi_{j_1}(D-1)}}^{(A_{\pi_{j_1}(D-1)})}
\boldsymbol{\sigma}
\left(K_{i_{\pi_{j_2}(0)}}^{(A_{\pi_{j_2}(0)})}\cdots K_{i_{\pi_{j_2}(D-1)}}^{(A_{\pi_{j_2}(D-1)})}\right)^\dagger\right)\right]=\mathbf{0},\quad \forall K_{i}^{(A_j)} 
=\alpha(i,j) U(i,j),
\label{eq:real trace Kraus sigma vanishes}
\end{equation}
\end{widetext} 
where each value of $\alpha$ 
and $U$ can vary with $i$ and $j$
. 
Although this expression looks formidable, it can be proven using routine properties of Pauli matrices:
\begin{equation}
\sigma_\mu\sigma_\nu=\delta_{\mu \nu}\sigma_0+\iu\sum_{\lambda=1}^3\epsilon_{\mu \nu \lambda}\sigma_\lambda \, ,
\label{eq:Pauli matrix product property}
\end{equation}
where we have used $\sigma_0=\openone$, the Kronecker delta $\delta_{\mu \nu}$, and the fully antisymmetric Levi-Civita tensor $\epsilon_{\mu \nu \lambda}$. The constants $\alpha$ are immaterial to the proof of Eq.~\eqref{eq:real trace Kraus sigma vanishes} so we need not keep track of them. In fact, because each Kraus operator and its Hermitian conjugate appears in Eq. \eqref{eq:real trace Kraus sigma vanishes}, the global phase of each Kraus operator is irrelevant, so we can always consider $\alpha$ to be real.

A single Kraus operator takes the form
\begin{equation}
K= \alpha  \sigma_0 + \iu \boldsymbol{v}\cdot \boldsymbol{\sigma} 
\label{eq:basic Kraus propto unitary}
\end{equation}
for some real vector $\boldsymbol{v}=(v_1,v_2,v_3)$ and real constant $\alpha$. 
%
The product of two Kraus operators of the form of Eq.~\eqref{eq:basic Kraus propto unitary} is another Kraus operator of the same form:
\begin{align}
(\alpha  \sigma_0   & + \iu \boldsymbol{v} \cdot \boldsymbol{\sigma} ) (\alpha^\prime  \sigma_0 + \iu \boldsymbol{v}^\prime\cdot \boldsymbol{\sigma} )  = 
(\alpha \alpha^\prime - \boldsymbol{v}\cdot\boldsymbol{v}^\prime )\sigma_0 \nonumber \\
& +\iu (\alpha\boldsymbol{v}^\prime+\alpha^\prime\boldsymbol{v}-\boldsymbol{v}\times\boldsymbol{v}^\prime )\cdot\boldsymbol{\sigma}  \equiv \alpha^{\prime\prime}\sigma_0+\iu\boldsymbol{v}^{\prime\prime}\cdot\boldsymbol{\sigma} \, .
\end{align}
The important property is that $\alpha^{\prime\prime}$ is still real, which is not true for a generic multiplication of two unitary operators.
We therefore infer that
\begin{equation}
K_{i_{\pi_{j_1}(0)}}^{(A_{\pi_{j_1}(0)})}\cdots K_{i_{\pi_{j_1}(D-1)}}^{(A_{\pi_{j_1}(D-1)})}=\beta\sigma_0+\iu \mathbf{u}\cdot\boldsymbol{\sigma} \, ,
\end{equation} 
for some real $\beta$ and vector $\mathbf{u}$, and similarly for the Hermitian conjugates of these Kraus operators with another real $\beta^\prime$ and vector $\mathbf{u}^\prime$. 

We are now equipped to tackle the expression in Eq.~\eqref{eq:real trace Kraus sigma vanishes}. By the cyclic nature of the trace, we simply need to show that
\begin{equation}
\RE \left\{\mathrm{Tr}\left[
\left(\beta^{\prime\prime}\sigma_0+\iu \mathbf{u}^{\prime\prime}\cdot\boldsymbol{\sigma}\right) \boldsymbol{\sigma} \right]\right\}=\boldsymbol{0} \, .
\end{equation} 
Each of the components of $\boldsymbol{\sigma}$ is traceless, so $\Tr (\sigma_0\boldsymbol{\sigma})=\boldsymbol{0}$. Similarly, using Eq.~\eqref{eq:Pauli matrix product property}, we find that
\begin{equation}
\Tr [ (\mathbf{u}^{\prime\prime}\cdot\boldsymbol{\sigma} ) \boldsymbol{\sigma} ]=2\mathbf{u}^{\prime\prime} \, ,
\end{equation}
where the factor of 2 comes from $\Tr(\sigma_0)=2$. The vector $\mathbf{u}^{\prime\prime}$ is always real, as explained above, which immediately proves Eq. \eqref{eq:real trace Kraus sigma vanishes}.


How common are such Kraus operators that satisfy $K_i^\dagger K_i\propto\mathds{1}$? This is manifestly satisfied by the dephasing channel, with Kraus operators proportional to $\mathds{I}$ and $\sigma_{\mathbf{u}}$. For the depolarizing channel, we need a Kraus-operator decomposition other than the one used above to show that it also satisfies this condition, remembering that different sets of Kraus operators can lead to identical dynamics if they are related by unitary transformations as $K_i^\prime=\sum_j \mathsf{U}_{ij}K_j$ with unitary matrices $\mathsf{\mathbf{U}}$. A possible decomposition of the depolarizing channel is into the four Kraus operators $\frac{\sqrt{1+3p}}{2}\mathds{1}$, $\frac{\sqrt{1-p}}{2}\sigma_1$, $\frac{\sqrt{1-p}}{2}\sigma_2$, and $\frac{\sqrt{1-p}}{2}\sigma_3$, which are manifestly proportional to unitary matrices. As for the amplitude damping channel, one can verify that matrix elements such as $R_{01}$ do depend on the initial probe state, changing from its expression in Eq. \eqref{eq:R01 amp damp qubit} for maximally mixed probes to $R_{01}=p_Ap_B$ for probe state $\rhop=|1\rangle\langle 1|$, from which one can conclude that there is no Kraus-operator decomposition for amplitude damping channels in which the Kraus operators are proportional to unitary matrices. Any number of of depolarization channels and dephasing channels, as well as their generalizations into Pauli channels \cite{Cubittetal2008,LeungWatrous2017}, supplied in any number of coherently controlled orders, will lead to control states the real part of whose elements will be independent from the probe state that traversed the channels.


\section{Concluding remarks}

Indefinite causal order opens many doors to quantum-enhanced metrology. We showed how a variety of noise channels that would otherwise eradicate all hopes of measuring parameters could be circumvented by ICO to allow those parameters to be estimated, with dramatic scaling advantages over any causally ordered scheme. Our protocols only require measurement of a control system that did not probe the unitary in question and allow one to simultaneously estimate multiple unitary and noise parameters. All of the protocols detailed here are readily accessible to experiments that have already investigated ICO using a quantum switch, especially those that studied communication through noisy channels with ICO. They are especially experimentally friendly due to the probe states being maximally mixed and the measurements being projections on superposition states standard to interferometry. We hope this incorporation of multiparameter estimation to ICO continues to be a fruitful breeding ground for many more quantum advantages.

The important distinction between this and previous works that studied metrology augmented by ICO are our allowance of the order between the noise and unitary channels to also be controlled; whereas, previous works only had access to controlling the order of multiple copies of identical noisy operations, so they could never achieve the results of this paper in the limit of completely noisy channels. In the case of identical noisy operations and single-parameter estimation, landmark studies showed that ICO should always provide at least a small advantage~\cite{Liuetal2023}, but that such an advantage disappears in the asymptotic limit of infinite copies of the noisy operations, where adaptive and entangled-ancilla protocols are equally as effective as ICO \cite{Kurdzialeketal2022arxiv}. These leave open intriguing questions. In terms of having multiple copies of identical channels, as in our Sec.~\ref{sec:multiparam copied channels}, is there a hierarchy of estimation strategies for multiparameter estimation? Does ICO retain any advantage in the asymptotic limit for multiparameter estimation? And, for all of our findings in Secs.~\ref{sec:single param} and \ref{sec:multiparam}, where we allow for the order in which the noise channels and the unitary are applied to be controlled, how do the advantages of ICO evolve when multiple copies of the noise and unitary channels are allowed? With two identical unitaries and four noise channels, does the ICO advantage increase or decrease? In the asymptotic limit of a large number of identical unitaries and noise channels, can rigorous inequalities or equalities be proven between different classes of estimation strategies? We know that controlling the order of a single unitary and a single completely depolarizing channel, as described here, will outperform even an infinite number of applications of identical unitary channels that are each always subject to complete depolarization noise; we thus expect ICO to have even greater advantages as the number of copies of the channels is increased.

\begin{acknowledgments}
The authors are grateful for discussions with Kent Bonsma-Fisher, Fr\'ed\'eric Bouchard,  Duncan England, Kate Fenwick, and Brayden Freitas, and Benjamin Sussman. They also thank the International Network of Acausal Quantum Technology, funded by the Engineering and Physical Sciences Research Council (EPSRC),  for support. AZG and KH  acknowledge that the NRC headquarters is located on the traditional unceded territory of the Algonquin Anishinaabe and Mohawk people, as well as support from NRC's Quantum Sensors Challenge Program. AZG acknowledges funding from the NSERC PDF program. LLSS acknowledges support from Ministerio de Ciencia e Innovaci\'on (Grant  PID2021-127781NB-I00).
\end{acknowledgments}


\begin{thebibliography}{120}%
\makeatletter
\providecommand \@ifxundefined [1]{%
 \@ifx{#1\undefined}
}%
\providecommand \@ifnum [1]{%
 \ifnum #1\expandafter \@firstoftwo
 \else \expandafter \@secondoftwo
 \fi
}%
\providecommand \@ifx [1]{%
 \ifx #1\expandafter \@firstoftwo
 \else \expandafter \@secondoftwo
 \fi
}%
\providecommand \natexlab [1]{#1}%
\providecommand \enquote  [1]{``#1''}%
\providecommand \bibnamefont  [1]{#1}%
\providecommand \bibfnamefont [1]{#1}%
\providecommand \citenamefont [1]{#1}%
\providecommand \href@noop [0]{\@secondoftwo}%
\providecommand \href [0]{\begingroup \@sanitize@url \@href}%
\providecommand \@href[1]{\@@startlink{#1}\@@href}%
\providecommand \@@href[1]{\endgroup#1\@@endlink}%
\providecommand \@sanitize@url [0]{\catcode `\\12\catcode `\$12\catcode
  `\&12\catcode `\#12\catcode `\^12\catcode `\_12\catcode `\%12\relax}%
\providecommand \@@startlink[1]{}%
\providecommand \@@endlink[0]{}%
\providecommand \url  [0]{\begingroup\@sanitize@url \@url }%
\providecommand \@url [1]{\endgroup\@href {#1}{\urlprefix }}%
\providecommand \urlprefix  [0]{URL }%
\providecommand \Eprint [0]{\href }%
\providecommand \doibase [0]{https://doi.org/}%
\providecommand \selectlanguage [0]{\@gobble}%
\providecommand \bibinfo  [0]{\@secondoftwo}%
\providecommand \bibfield  [0]{\@secondoftwo}%
\providecommand \translation [1]{[#1]}%
\providecommand \BibitemOpen [0]{}%
\providecommand \bibitemStop [0]{}%
\providecommand \bibitemNoStop [0]{.\EOS\space}%
\providecommand \EOS [0]{\spacefactor3000\relax}%
\providecommand \BibitemShut  [1]{\csname bibitem#1\endcsname}%
\let\auto@bib@innerbib\@empty
\bibitem [{\citenamefont {Escher}\ \emph {et~al.}(2011)\citenamefont {Escher},
  \citenamefont {de~Matos~Filho},\ and\ \citenamefont
  {Davidovich}}]{Escheretal2011}%
  \BibitemOpen
  \bibfield  {author} {\bibinfo {author} {\bibfnamefont {B.~M.}\ \bibnamefont
  {Escher}}, \bibinfo {author} {\bibfnamefont {R.~L.}\ \bibnamefont
  {de~Matos~Filho}},\ and\ \bibinfo {author} {\bibfnamefont {L.}~\bibnamefont
  {Davidovich}},\ }\bibfield  {title} {\bibinfo {title} {General framework for
  estimating the ultimate precision limit in noisy quantum-enhanced
  metrology},\ }\href {https://doi.org/10.1038/nphys1958} {\bibfield  {journal}
  {\bibinfo  {journal} {Nat. Phys.}\ }\textbf {\bibinfo {volume} {7}},\
  \bibinfo {pages} {406} (\bibinfo {year} {2011})}\BibitemShut {NoStop}%
\bibitem [{\citenamefont {Caves}(1981)}]{Caves1981}%
  \BibitemOpen
  \bibfield  {author} {\bibinfo {author} {\bibfnamefont {C.~M.}\ \bibnamefont
  {Caves}},\ }\bibfield  {title} {\bibinfo {title} {Quantum-mechanical noise in
  an interferometer},\ }\href {https://doi.org/10.1103/PhysRevD.23.1693}
  {\bibfield  {journal} {\bibinfo  {journal} {Phys. Rev. D}\ }\textbf {\bibinfo
  {volume} {23}},\ \bibinfo {pages} {1693} (\bibinfo {year}
  {1981})}\BibitemShut {NoStop}%
\bibitem [{\citenamefont {Dowling}(1998)}]{Dowling1998}%
  \BibitemOpen
  \bibfield  {author} {\bibinfo {author} {\bibfnamefont {J.~P.}\ \bibnamefont
  {Dowling}},\ }\bibfield  {title} {\bibinfo {title} {Correlated input-port,
  matter-wave interferometer: Quantum-noise limits to the atom-laser
  gyroscope},\ }\href {https://doi.org/10.1103/PhysRevA.57.4736} {\bibfield
  {journal} {\bibinfo  {journal} {Phys. Rev. A}\ }\textbf {\bibinfo {volume}
  {57}},\ \bibinfo {pages} {4736} (\bibinfo {year} {1998})}\BibitemShut
  {NoStop}%
\bibitem [{\citenamefont {Giovannetti}\ \emph {et~al.}(2004)\citenamefont
  {Giovannetti}, \citenamefont {Lloyd},\ and\ \citenamefont
  {Maccone}}]{Giovannettietal2004}%
  \BibitemOpen
  \bibfield  {author} {\bibinfo {author} {\bibfnamefont {V.}~\bibnamefont
  {Giovannetti}}, \bibinfo {author} {\bibfnamefont {S.}~\bibnamefont {Lloyd}},\
  and\ \bibinfo {author} {\bibfnamefont {L.}~\bibnamefont {Maccone}},\
  }\bibfield  {title} {\bibinfo {title} {Quantum-enhanced measurements: Beating
  the standard quantum limit},\ }\href
  {https://doi.org/10.1126/science.1104149} {\bibfield  {journal} {\bibinfo
  {journal} {Science}\ }\textbf {\bibinfo {volume} {306}},\ \bibinfo {pages}
  {1330} (\bibinfo {year} {2004})}\BibitemShut {NoStop}%
\bibitem [{\citenamefont {Berry}\ \emph {et~al.}(2009)\citenamefont {Berry},
  \citenamefont {Higgins}, \citenamefont {Bartlett}, \citenamefont {Mitchell},
  \citenamefont {Pryde},\ and\ \citenamefont {Wiseman}}]{Berryetal2009}%
  \BibitemOpen
  \bibfield  {author} {\bibinfo {author} {\bibfnamefont {D.~W.}\ \bibnamefont
  {Berry}}, \bibinfo {author} {\bibfnamefont {B.~L.}\ \bibnamefont {Higgins}},
  \bibinfo {author} {\bibfnamefont {S.~D.}\ \bibnamefont {Bartlett}}, \bibinfo
  {author} {\bibfnamefont {M.~W.}\ \bibnamefont {Mitchell}}, \bibinfo {author}
  {\bibfnamefont {G.~J.}\ \bibnamefont {Pryde}},\ and\ \bibinfo {author}
  {\bibfnamefont {H.~M.}\ \bibnamefont {Wiseman}},\ }\bibfield  {title}
  {\bibinfo {title} {How to perform the most accurate possible phase
  measurements},\ }\href {https://doi.org/10.1103/PhysRevA.80.052114}
  {\bibfield  {journal} {\bibinfo  {journal} {Phys. Rev. A}\ }\textbf {\bibinfo
  {volume} {80}},\ \bibinfo {pages} {052114} (\bibinfo {year}
  {2009})}\BibitemShut {NoStop}%
\bibitem [{\citenamefont {Taylor}\ \emph {et~al.}(2013)\citenamefont {Taylor},
  \citenamefont {Janousek}, \citenamefont {Daria}, \citenamefont {Knittel},
  \citenamefont {Hage}, \citenamefont {Bachor},\ and\ \citenamefont
  {Bowen}}]{Tayloretal2013}%
  \BibitemOpen
  \bibfield  {author} {\bibinfo {author} {\bibfnamefont {M.~A.}\ \bibnamefont
  {Taylor}}, \bibinfo {author} {\bibfnamefont {J.}~\bibnamefont {Janousek}},
  \bibinfo {author} {\bibfnamefont {V.}~\bibnamefont {Daria}}, \bibinfo
  {author} {\bibfnamefont {J.}~\bibnamefont {Knittel}}, \bibinfo {author}
  {\bibfnamefont {B.}~\bibnamefont {Hage}}, \bibinfo {author} {\bibfnamefont
  {H.-A.}\ \bibnamefont {Bachor}},\ and\ \bibinfo {author} {\bibfnamefont
  {W.~P.}\ \bibnamefont {Bowen}},\ }\bibfield  {title} {\bibinfo {title}
  {Biological measurement beyond the quantum limit},\ }\href
  {http://dx.doi.org/10.1038/nphoton.2012.346} {\bibfield  {journal} {\bibinfo
  {journal} {Nat. Photon.}\ }\textbf {\bibinfo {volume} {7}},\ \bibinfo {pages}
  {229} (\bibinfo {year} {2013})}\BibitemShut {NoStop}%
\bibitem [{\citenamefont {Tsang}\ \emph {et~al.}(2016)\citenamefont {Tsang},
  \citenamefont {Nair},\ and\ \citenamefont {Lu}}]{Tsangetal2016}%
  \BibitemOpen
  \bibfield  {author} {\bibinfo {author} {\bibfnamefont {M.}~\bibnamefont
  {Tsang}}, \bibinfo {author} {\bibfnamefont {R.}~\bibnamefont {Nair}},\ and\
  \bibinfo {author} {\bibfnamefont {X.-M.}\ \bibnamefont {Lu}},\ }\bibfield
  {title} {\bibinfo {title} {Quantum theory of superresolution for two
  incoherent optical point sources},\ }\href
  {https://doi.org/10.1103/PhysRevX.6.031033} {\bibfield  {journal} {\bibinfo
  {journal} {Phys. Rev. X}\ }\textbf {\bibinfo {volume} {6}},\ \bibinfo {pages}
  {031033} (\bibinfo {year} {2016})}\BibitemShut {NoStop}%
\bibitem [{\citenamefont {Liu}\ \emph {et~al.}(2020)\citenamefont {Liu},
  \citenamefont {Zhang}, \citenamefont {Li}, \citenamefont {Zhang},
  \citenamefont {Yin}, \citenamefont {Fei}, \citenamefont {Li}, \citenamefont
  {Liu}, \citenamefont {Xu}, \citenamefont {Chen},\ and\ \citenamefont
  {Pan}}]{Liuetal2020}%
  \BibitemOpen
  \bibfield  {author} {\bibinfo {author} {\bibfnamefont {L.-Z.}\ \bibnamefont
  {Liu}}, \bibinfo {author} {\bibfnamefont {Y.-Z.}\ \bibnamefont {Zhang}},
  \bibinfo {author} {\bibfnamefont {Z.-D.}\ \bibnamefont {Li}}, \bibinfo
  {author} {\bibfnamefont {R.}~\bibnamefont {Zhang}}, \bibinfo {author}
  {\bibfnamefont {X.-F.}\ \bibnamefont {Yin}}, \bibinfo {author} {\bibfnamefont
  {Y.-Y.}\ \bibnamefont {Fei}}, \bibinfo {author} {\bibfnamefont
  {L.}~\bibnamefont {Li}}, \bibinfo {author} {\bibfnamefont {N.-L.}\
  \bibnamefont {Liu}}, \bibinfo {author} {\bibfnamefont {F.}~\bibnamefont
  {Xu}}, \bibinfo {author} {\bibfnamefont {Y.-A.}\ \bibnamefont {Chen}},\ and\
  \bibinfo {author} {\bibfnamefont {J.-W.}\ \bibnamefont {Pan}},\ }\bibfield
  {title} {\bibinfo {title} {Distributed quantum phase estimation with
  entangled photons},\ }\href {https://doi.org/10.1038/s41566-020-00718-2}
  {\bibfield  {journal} {\bibinfo  {journal} {Nat. Photon.}\ }\textbf {\bibinfo
  {volume} {15}},\ \bibinfo {pages} {137} (\bibinfo {year} {2020})}\BibitemShut
  {NoStop}%
\bibitem [{\citenamefont {Mitchell}\ \emph {et~al.}(2004)\citenamefont
  {Mitchell}, \citenamefont {Lundeen},\ and\ \citenamefont
  {Steinberg}}]{Mitchelletal2004}%
  \BibitemOpen
  \bibfield  {author} {\bibinfo {author} {\bibfnamefont {M.~W.}\ \bibnamefont
  {Mitchell}}, \bibinfo {author} {\bibfnamefont {J.~S.}\ \bibnamefont
  {Lundeen}},\ and\ \bibinfo {author} {\bibfnamefont {A.~M.}\ \bibnamefont
  {Steinberg}},\ }\bibfield  {title} {\bibinfo {title} {Super-resolving phase
  measurements with a multiphoton entangled state},\ }\href
  {https://doi.org/10.1038/nature02493} {\bibfield  {journal} {\bibinfo
  {journal} {Nature}\ }\textbf {\bibinfo {volume} {429}},\ \bibinfo {pages}
  {161} (\bibinfo {year} {2004})}\BibitemShut {NoStop}%
\bibitem [{\citenamefont {{The L I G O Scientific
  Collaboration}}(2011)}]{LIGO2011}%
  \BibitemOpen
  \bibfield  {author} {\bibinfo {author} {\bibnamefont {{The L I G O Scientific
  Collaboration}}},\ }\bibfield  {title} {\bibinfo {title} {A gravitational
  wave observatory operating beyond the quantum shot-noise limit},\ }\href
  {http://dx.doi.org/10.1038/nphys2083} {\bibfield  {journal} {\bibinfo
  {journal} {Nat. Phys.}\ }\textbf {\bibinfo {volume} {7}},\ \bibinfo {pages}
  {962} (\bibinfo {year} {2011})}\BibitemShut {NoStop}%
\bibitem [{\citenamefont {Whittaker}\ \emph {et~al.}(2017)\citenamefont
  {Whittaker}, \citenamefont {Erven}, \citenamefont {Neville}, \citenamefont
  {Berry}, \citenamefont {O'Brien}, \citenamefont {Cable},\ and\ \citenamefont
  {Matthews}}]{Whittakeretal2017}%
  \BibitemOpen
  \bibfield  {author} {\bibinfo {author} {\bibfnamefont {R.}~\bibnamefont
  {Whittaker}}, \bibinfo {author} {\bibfnamefont {C.}~\bibnamefont {Erven}},
  \bibinfo {author} {\bibfnamefont {A.}~\bibnamefont {Neville}}, \bibinfo
  {author} {\bibfnamefont {M.}~\bibnamefont {Berry}}, \bibinfo {author}
  {\bibfnamefont {J.~L.}\ \bibnamefont {O'Brien}}, \bibinfo {author}
  {\bibfnamefont {H.}~\bibnamefont {Cable}},\ and\ \bibinfo {author}
  {\bibfnamefont {J.~C.~F.}\ \bibnamefont {Matthews}},\ }\bibfield  {title}
  {\bibinfo {title} {Absorption spectroscopy at the ultimate quantum limit from
  single-photon states},\ }\href {https://doi.org/10.1088/1367-2630/aa5512}
  {\bibfield  {journal} {\bibinfo  {journal} {New J. Phys.}\ }\textbf {\bibinfo
  {volume} {19}},\ \bibinfo {pages} {023013} (\bibinfo {year}
  {2017})}\BibitemShut {NoStop}%
\bibitem [{\citenamefont {You}\ \emph {et~al.}(2021)\citenamefont {You},
  \citenamefont {Hong}, \citenamefont {Bierhorst}, \citenamefont {Lita},
  \citenamefont {Glancy}, \citenamefont {Kolthammer}, \citenamefont {Knill},
  \citenamefont {Nam}, \citenamefont {Mirin}, \citenamefont
  {Maga{\~n}a-Loaiza},\ and\ \citenamefont {Gerrits}}]{Youetal2021}%
  \BibitemOpen
  \bibfield  {author} {\bibinfo {author} {\bibfnamefont {C.}~\bibnamefont
  {You}}, \bibinfo {author} {\bibfnamefont {M.}~\bibnamefont {Hong}}, \bibinfo
  {author} {\bibfnamefont {P.}~\bibnamefont {Bierhorst}}, \bibinfo {author}
  {\bibfnamefont {A.~E.}\ \bibnamefont {Lita}}, \bibinfo {author}
  {\bibfnamefont {S.}~\bibnamefont {Glancy}}, \bibinfo {author} {\bibfnamefont
  {S.}~\bibnamefont {Kolthammer}}, \bibinfo {author} {\bibfnamefont
  {E.}~\bibnamefont {Knill}}, \bibinfo {author} {\bibfnamefont {S.~W.}\
  \bibnamefont {Nam}}, \bibinfo {author} {\bibfnamefont {R.~P.}\ \bibnamefont
  {Mirin}}, \bibinfo {author} {\bibfnamefont {O.~S.}\ \bibnamefont
  {Maga{\~n}a-Loaiza}},\ and\ \bibinfo {author} {\bibfnamefont
  {T.}~\bibnamefont {Gerrits}},\ }\bibfield  {title} {\bibinfo {title}
  {Scalable multiphoton quantum metrology with neither pre- nor post-selected
  measurements},\ }\href {https://doi.org/10.1063/5.0063294} {\bibfield
  {journal} {\bibinfo  {journal} {Appl. Phys. Rev.}\ }\textbf {\bibinfo
  {volume} {8}},\ \bibinfo {pages} {041406} (\bibinfo {year}
  {2021})}\BibitemShut {NoStop}%
\bibitem [{\citenamefont {Qin}\ \emph {et~al.}(2023)\citenamefont {Qin},
  \citenamefont {Deng}, \citenamefont {Zhong}, \citenamefont {Peng},
  \citenamefont {Su}, \citenamefont {Luo}, \citenamefont {Xu}, \citenamefont
  {Wu}, \citenamefont {Gong}, \citenamefont {Liu}, \citenamefont {Wang},
  \citenamefont {Chen}, \citenamefont {Li}, \citenamefont {Liu}, \citenamefont
  {Lu},\ and\ \citenamefont {Pan}}]{Qinetal2023}%
  \BibitemOpen
  \bibfield  {author} {\bibinfo {author} {\bibfnamefont {J.}~\bibnamefont
  {Qin}}, \bibinfo {author} {\bibfnamefont {Y.-H.}\ \bibnamefont {Deng}},
  \bibinfo {author} {\bibfnamefont {H.-S.}\ \bibnamefont {Zhong}}, \bibinfo
  {author} {\bibfnamefont {L.-C.}\ \bibnamefont {Peng}}, \bibinfo {author}
  {\bibfnamefont {H.}~\bibnamefont {Su}}, \bibinfo {author} {\bibfnamefont
  {Y.-H.}\ \bibnamefont {Luo}}, \bibinfo {author} {\bibfnamefont {J.-M.}\
  \bibnamefont {Xu}}, \bibinfo {author} {\bibfnamefont {D.}~\bibnamefont {Wu}},
  \bibinfo {author} {\bibfnamefont {S.-Q.}\ \bibnamefont {Gong}}, \bibinfo
  {author} {\bibfnamefont {H.-L.}\ \bibnamefont {Liu}}, \bibinfo {author}
  {\bibfnamefont {H.}~\bibnamefont {Wang}}, \bibinfo {author} {\bibfnamefont
  {M.-C.}\ \bibnamefont {Chen}}, \bibinfo {author} {\bibfnamefont
  {L.}~\bibnamefont {Li}}, \bibinfo {author} {\bibfnamefont {N.-L.}\
  \bibnamefont {Liu}}, \bibinfo {author} {\bibfnamefont {C.-Y.}\ \bibnamefont
  {Lu}},\ and\ \bibinfo {author} {\bibfnamefont {J.-W.}\ \bibnamefont {Pan}},\
  }\bibfield  {title} {\bibinfo {title} {Unconditional and robust quantum
  metrological advantage beyond {N00N} states},\ }\href
  {https://doi.org/10.1103/PhysRevLett.130.070801} {\bibfield  {journal}
  {\bibinfo  {journal} {Phys. Rev. Lett.}\ }\textbf {\bibinfo {volume} {130}},\
  \bibinfo {pages} {070801} (\bibinfo {year} {2023})}\BibitemShut {NoStop}%
\bibitem [{\citenamefont {Raymer}\ and\ \citenamefont
  {Monroe}(2019)}]{RaymerMonroe2019}%
  \BibitemOpen
  \bibfield  {author} {\bibinfo {author} {\bibfnamefont {M.~G.}\ \bibnamefont
  {Raymer}}\ and\ \bibinfo {author} {\bibfnamefont {C.}~\bibnamefont
  {Monroe}},\ }\bibfield  {title} {\bibinfo {title} {The {US} national quantum
  initiative},\ }\href {https://doi.org/10.1088/2058-9565/ab0441} {\bibfield
  {journal} {\bibinfo  {journal} {Quantum Sci. Technol.}\ }\textbf {\bibinfo
  {volume} {4}},\ \bibinfo {pages} {020504} (\bibinfo {year}
  {2019})}\BibitemShut {NoStop}%
\bibitem [{\citenamefont {Sussman}\ \emph {et~al.}(2019)\citenamefont
  {Sussman}, \citenamefont {Corkum}, \citenamefont {Blais}, \citenamefont
  {Cory},\ and\ \citenamefont {Damascelli}}]{Sussmanetal2019}%
  \BibitemOpen
  \bibfield  {author} {\bibinfo {author} {\bibfnamefont {B.}~\bibnamefont
  {Sussman}}, \bibinfo {author} {\bibfnamefont {P.}~\bibnamefont {Corkum}},
  \bibinfo {author} {\bibfnamefont {A.}~\bibnamefont {Blais}}, \bibinfo
  {author} {\bibfnamefont {D.}~\bibnamefont {Cory}},\ and\ \bibinfo {author}
  {\bibfnamefont {A.}~\bibnamefont {Damascelli}},\ }\bibfield  {title}
  {\bibinfo {title} {Quantum Canada},\ }\href
  {https://doi.org/10.1088/2058-9565/ab029d} {\bibfield  {journal} {\bibinfo
  {journal} {Quantum Sci. Technol.}\ }\textbf {\bibinfo {volume} {4}},\
  \bibinfo {pages} {020503} (\bibinfo {year} {2019})}\BibitemShut {NoStop}%
\bibitem [{\citenamefont {Yamamoto}\ \emph {et~al.}(2019)\citenamefont
  {Yamamoto}, \citenamefont {Sasaki},\ and\ \citenamefont
  {Takesue}}]{Yamamotoetal2019}%
  \BibitemOpen
  \bibfield  {author} {\bibinfo {author} {\bibfnamefont {Y.}~\bibnamefont
  {Yamamoto}}, \bibinfo {author} {\bibfnamefont {M.}~\bibnamefont {Sasaki}},\
  and\ \bibinfo {author} {\bibfnamefont {H.}~\bibnamefont {Takesue}},\
  }\bibfield  {title} {\bibinfo {title} {Quantum information science and
  technology in {J}apan},\ }\href {https://doi.org/10.1088/2058-9565/ab0077}
  {\bibfield  {journal} {\bibinfo  {journal} {Quantum Sci. Technol.}\ }\textbf
  {\bibinfo {volume} {4}},\ \bibinfo {pages} {020502} (\bibinfo {year}
  {2019})}\BibitemShut {NoStop}%
\bibitem [{\citenamefont {OIDA}(2020)}]{OIDA2020}%
  \BibitemOpen
  \bibfield  {author} {\bibinfo {author} {\bibnamefont {OIDA}},\ }\bibfield
  {title} {\bibinfo {title} {Oida quantum photonics roadmap: Every photon
  counts},\ }\href {http://www.osapublishing.org/abstract.cfm?URI=OIDA-2020-3}
  {\bibfield  {journal} {\bibinfo  {journal} {OIDA}\ }\bibinfo {series} {Optica
  Industry Report},\ \bibinfo {pages} {3} (\bibinfo {year} {2020})}\BibitemShut
  {NoStop}%
\bibitem [{\citenamefont {Knight}\ and\ \citenamefont
  {Walmsley}(2019)}]{KnightWalmsley2019}%
  \BibitemOpen
  \bibfield  {author} {\bibinfo {author} {\bibfnamefont {P.}~\bibnamefont
  {Knight}}\ and\ \bibinfo {author} {\bibfnamefont {I.}~\bibnamefont
  {Walmsley}},\ }\bibfield  {title} {\bibinfo {title} {{UK} national quantum
  technology programme},\ }\href {https://doi.org/10.1088/2058-9565/ab4346}
  {\bibfield  {journal} {\bibinfo  {journal} {Quantum Sci. Technol.}\ }\textbf
  {\bibinfo {volume} {4}},\ \bibinfo {pages} {040502} (\bibinfo {year}
  {2019})}\BibitemShut {NoStop}%
\bibitem [{\citenamefont {Matsumoto}(2002)}]{Matsumoto2002}%
  \BibitemOpen
  \bibfield  {author} {\bibinfo {author} {\bibfnamefont {K.}~\bibnamefont
  {Matsumoto}},\ }\bibfield  {title} {\bibinfo {title} {{A new approach to the
  Cram{\'{e}}r-Rao-type bound of the pure-state model}},\ }\href
  {https://doi.org/10.1088/0305-4470/35/13/307} {\bibfield  {journal} {\bibinfo
   {journal} {J. Phys. A: Math. Theor.}\ }\textbf {\bibinfo {volume} {35}},\
  \bibinfo {pages} {3111} (\bibinfo {year} {2002})}\BibitemShut {NoStop}%
\bibitem [{\citenamefont {Paris}(2009)}]{Paris2009}%
  \BibitemOpen
  \bibfield  {author} {\bibinfo {author} {\bibfnamefont {M.~G.~A.}\
  \bibnamefont {Paris}},\ }\bibfield  {title} {\bibinfo {title} {Quantum
  estimation for quantum technology},\ }\href
  {https://doi.org/10.1142/S0219749909004839} {\bibfield  {journal} {\bibinfo
  {journal} {Int. J. Quantum Inform.}\ }\textbf {\bibinfo {volume} {07}},\
  \bibinfo {pages} {125} (\bibinfo {year} {2009})}\BibitemShut {NoStop}%
\bibitem [{\citenamefont {T{\'{o}}th}\ and\ \citenamefont
  {Apellaniz}(2014)}]{TothApellaniz2014}%
  \BibitemOpen
  \bibfield  {author} {\bibinfo {author} {\bibfnamefont {G.}~\bibnamefont
  {T{\'{o}}th}}\ and\ \bibinfo {author} {\bibfnamefont {I.}~\bibnamefont
  {Apellaniz}},\ }\bibfield  {title} {\bibinfo {title} {Quantum metrology from
  a quantum information science perspective},\ }\href
  {https://doi.org/10.1088/1751-8113/47/42/424006} {\bibfield  {journal}
  {\bibinfo  {journal} {J. Phys. A: Math. Theor.}\ }\textbf {\bibinfo {volume}
  {47}},\ \bibinfo {pages} {424006} (\bibinfo {year} {2014})}\BibitemShut
  {NoStop}%
\bibitem [{\citenamefont {Szczykulska}\ \emph {et~al.}(2016)\citenamefont
  {Szczykulska}, \citenamefont {Baumgratz},\ and\ \citenamefont
  {Datta}}]{Szczykulskaetal2016}%
  \BibitemOpen
  \bibfield  {author} {\bibinfo {author} {\bibfnamefont {M.}~\bibnamefont
  {Szczykulska}}, \bibinfo {author} {\bibfnamefont {T.}~\bibnamefont
  {Baumgratz}},\ and\ \bibinfo {author} {\bibfnamefont {A.}~\bibnamefont
  {Datta}},\ }\bibfield  {title} {\bibinfo {title} {Multi-parameter quantum
  metrology},\ }\href {https://doi.org/10.1080/23746149.2016.1230476}
  {\bibfield  {journal} {\bibinfo  {journal} {Adv. Phys. X}\ }\textbf {\bibinfo
  {volume} {1}},\ \bibinfo {pages} {621} (\bibinfo {year} {2016})}\BibitemShut
  {NoStop}%
\bibitem [{\citenamefont {Braun}\ \emph {et~al.}(2018)\citenamefont {Braun},
  \citenamefont {Adesso}, \citenamefont {Benatti}, \citenamefont {Floreanini},
  \citenamefont {Marzolino}, \citenamefont {Mitchell},\ and\ \citenamefont
  {Pirandola}}]{Braunetal2018}%
  \BibitemOpen
  \bibfield  {author} {\bibinfo {author} {\bibfnamefont {D.}~\bibnamefont
  {Braun}}, \bibinfo {author} {\bibfnamefont {G.}~\bibnamefont {Adesso}},
  \bibinfo {author} {\bibfnamefont {F.}~\bibnamefont {Benatti}}, \bibinfo
  {author} {\bibfnamefont {R.}~\bibnamefont {Floreanini}}, \bibinfo {author}
  {\bibfnamefont {U.}~\bibnamefont {Marzolino}}, \bibinfo {author}
  {\bibfnamefont {M.~W.}\ \bibnamefont {Mitchell}},\ and\ \bibinfo {author}
  {\bibfnamefont {S.}~\bibnamefont {Pirandola}},\ }\bibfield  {title} {\bibinfo
  {title} {Quantum-enhanced measurements without entanglement},\ }\href
  {https://doi.org/10.1103/RevModPhys.90.035006} {\bibfield  {journal}
  {\bibinfo  {journal} {Rev. Mod. Phys.}\ }\textbf {\bibinfo {volume} {90}},\
  \bibinfo {pages} {035006} (\bibinfo {year} {2018})}\BibitemShut {NoStop}%
\bibitem [{\citenamefont {Liu}\ \emph {et~al.}(2019)\citenamefont {Liu},
  \citenamefont {Yuan}, \citenamefont {Lu},\ and\ \citenamefont
  {Wang}}]{Liuetal2019}%
  \BibitemOpen
  \bibfield  {author} {\bibinfo {author} {\bibfnamefont {J.}~\bibnamefont
  {Liu}}, \bibinfo {author} {\bibfnamefont {H.}~\bibnamefont {Yuan}}, \bibinfo
  {author} {\bibfnamefont {X.-M.}\ \bibnamefont {Lu}},\ and\ \bibinfo {author}
  {\bibfnamefont {X.}~\bibnamefont {Wang}},\ }\bibfield  {title} {\bibinfo
  {title} {Quantum {Fisher} information matrix and multiparameter estimation},\
  }\href {https://doi.org/10.1088/1751-8121/ab5d4d} {\bibfield  {journal}
  {\bibinfo  {journal} {J. Phys. A: Math. Theor.}\ }\textbf {\bibinfo {volume}
  {53}},\ \bibinfo {pages} {023001} (\bibinfo {year} {2019})}\BibitemShut
  {NoStop}%
\bibitem [{\citenamefont {Sidhu}\ and\ \citenamefont
  {Kok}(2020)}]{SidhuKok2020}%
  \BibitemOpen
  \bibfield  {author} {\bibinfo {author} {\bibfnamefont {J.~S.}\ \bibnamefont
  {Sidhu}}\ and\ \bibinfo {author} {\bibfnamefont {P.}~\bibnamefont {Kok}},\
  }\bibfield  {title} {\bibinfo {title} {Geometric perspective on quantum
  parameter estimation},\ }\href {https://doi.org/10.1116/1.5119961} {\bibfield
   {journal} {\bibinfo  {journal} {AVS Quantum Sci.}\ }\textbf {\bibinfo
  {volume} {2}},\ \bibinfo {pages} {014701} (\bibinfo {year}
  {2020})}\BibitemShut {NoStop}%
\bibitem [{\citenamefont {Albarelli}\ \emph {et~al.}(2020)\citenamefont
  {Albarelli}, \citenamefont {Barbieri}, \citenamefont {Genoni},\ and\
  \citenamefont {Gianani}}]{Albarellietal2020}%
  \BibitemOpen
  \bibfield  {author} {\bibinfo {author} {\bibfnamefont {F.}~\bibnamefont
  {Albarelli}}, \bibinfo {author} {\bibfnamefont {M.}~\bibnamefont {Barbieri}},
  \bibinfo {author} {\bibfnamefont {M.}~\bibnamefont {Genoni}},\ and\ \bibinfo
  {author} {\bibfnamefont {I.}~\bibnamefont {Gianani}},\ }\bibfield  {title}
  {\bibinfo {title} {A perspective on multiparameter quantum metrology: From
  theoretical tools to applications in quantum imaging},\ }\href
  {https://doi.org/https://doi.org/10.1016/j.physleta.2020.126311} {\bibfield
  {journal} {\bibinfo  {journal} {Phys. Lett. A}\ }\textbf {\bibinfo {volume}
  {384}},\ \bibinfo {pages} {126311} (\bibinfo {year} {2020})}\BibitemShut
  {NoStop}%
\bibitem [{\citenamefont {Polino}\ \emph {et~al.}(2020)\citenamefont {Polino},
  \citenamefont {Valeri}, \citenamefont {Spagnolo},\ and\ \citenamefont
  {Sciarrino}}]{Polinoetal2020}%
  \BibitemOpen
  \bibfield  {author} {\bibinfo {author} {\bibfnamefont {E.}~\bibnamefont
  {Polino}}, \bibinfo {author} {\bibfnamefont {M.}~\bibnamefont {Valeri}},
  \bibinfo {author} {\bibfnamefont {N.}~\bibnamefont {Spagnolo}},\ and\
  \bibinfo {author} {\bibfnamefont {F.}~\bibnamefont {Sciarrino}},\ }\bibfield
  {title} {\bibinfo {title} {Photonic quantum metrology},\ }\href
  {https://doi.org/10.1116/5.0007577} {\bibfield  {journal} {\bibinfo
  {journal} {AVS Quantum Sci.}\ }\textbf {\bibinfo {volume} {2}},\ \bibinfo
  {pages} {024703} (\bibinfo {year} {2020})}\BibitemShut {NoStop}%
\bibitem [{\citenamefont {Demkowicz-Dobrza{\'{n}}ski}\ \emph
  {et~al.}(2020)\citenamefont {Demkowicz-Dobrza{\'{n}}ski}, \citenamefont
  {G{\'{o}}recki},\ and\ \citenamefont
  {Gu{\c{t}}{\u{a}}}}]{DemkowiczDobrzanskietal2020}%
  \BibitemOpen
  \bibfield  {author} {\bibinfo {author} {\bibfnamefont {R.}~\bibnamefont
  {Demkowicz-Dobrza{\'{n}}ski}}, \bibinfo {author} {\bibfnamefont
  {W.}~\bibnamefont {G{\'{o}}recki}},\ and\ \bibinfo {author} {\bibfnamefont
  {M.}~\bibnamefont {Gu{\c{t}}{\u{a}}}},\ }\bibfield  {title} {\bibinfo {title}
  {Multi-parameter estimation beyond quantum {F}isher information},\ }\href
  {https://doi.org/10.1088/1751-8121/ab8ef3} {\bibfield  {journal} {\bibinfo
  {journal} {J. Phys. A: Math. Theor.}\ }\textbf {\bibinfo {volume} {53}},\
  \bibinfo {pages} {363001} (\bibinfo {year} {2020})}\BibitemShut {NoStop}%
\bibitem [{\citenamefont {Liu}\ \emph {et~al.}(2022)\citenamefont {Liu},
  \citenamefont {Zhang}, \citenamefont {Chen}, \citenamefont {Wang},\ and\
  \citenamefont {Yuan}}]{Liuetal2022}%
  \BibitemOpen
  \bibfield  {author} {\bibinfo {author} {\bibfnamefont {J.}~\bibnamefont
  {Liu}}, \bibinfo {author} {\bibfnamefont {M.}~\bibnamefont {Zhang}}, \bibinfo
  {author} {\bibfnamefont {H.}~\bibnamefont {Chen}}, \bibinfo {author}
  {\bibfnamefont {L.}~\bibnamefont {Wang}},\ and\ \bibinfo {author}
  {\bibfnamefont {H.}~\bibnamefont {Yuan}},\ }\bibfield  {title} {\bibinfo
  {title} {Optimal scheme for quantum metrology},\ }\href
  {https://doi.org/https://doi.org/10.1002/qute.202100080} {\bibfield
  {journal} {\bibinfo  {journal} {Adv. Quantum Technol.}\ }\textbf {\bibinfo
  {volume} {5}},\ \bibinfo {pages} {2100080} (\bibinfo {year}
  {2022})}\BibitemShut {NoStop}%
\bibitem [{\citenamefont {Goldberg}\ \emph
  {et~al.}(2021{\natexlab{a}})\citenamefont {Goldberg}, \citenamefont {Romero},
  \citenamefont {Sanz},\ and\ \citenamefont
  {S\'{a}nchez-Soto}}]{Goldbergetal2021singular}%
  \BibitemOpen
  \bibfield  {author} {\bibinfo {author} {\bibfnamefont {A.~Z.}\ \bibnamefont
  {Goldberg}}, \bibinfo {author} {\bibfnamefont {J.~L.}\ \bibnamefont
  {Romero}}, \bibinfo {author} {\bibfnamefont {A.~S.}\ \bibnamefont {Sanz}},\
  and\ \bibinfo {author} {\bibfnamefont {L.~L.}\ \bibnamefont
  {S\'{a}nchez-Soto}},\ }\bibfield  {title} {\bibinfo {title} {Taming
  singularities of the quantum {F}isher information},\ }\href
  {https://doi.org/10.1142/S0219749921400049} {\bibfield  {journal} {\bibinfo
  {journal} {Int. J. Quantum Inform.}\ }\textbf {\bibinfo {volume} {19}},\
  \bibinfo {pages} {2140004} (\bibinfo {year}
  {2021}{\natexlab{a}})}\BibitemShut {NoStop}%
\bibitem [{\citenamefont {Zhu}(2015)}]{Zhu2015}%
  \BibitemOpen
  \bibfield  {author} {\bibinfo {author} {\bibfnamefont {H.}~\bibnamefont
  {Zhu}},\ }\bibfield  {title} {\bibinfo {title} {Information complementarity:
  A new paradigm for decoding quantum incompatibility},\ }\href
  {https://doi.org/10.1038/srep14317} {\bibfield  {journal} {\bibinfo
  {journal} {Sci. Rep.}\ }\textbf {\bibinfo {volume} {5}},\ \bibinfo {pages}
  {14317} (\bibinfo {year} {2015})}\BibitemShut {NoStop}%
\bibitem [{\citenamefont {Ragy}\ \emph {et~al.}(2016)\citenamefont {Ragy},
  \citenamefont {Jarzyna},\ and\ \citenamefont
  {Demkowicz-Dobrza\ifmmode~\acute{n}\else \'{n}\fi{}ski}}]{Ragyetal2016}%
  \BibitemOpen
  \bibfield  {author} {\bibinfo {author} {\bibfnamefont {S.}~\bibnamefont
  {Ragy}}, \bibinfo {author} {\bibfnamefont {M.}~\bibnamefont {Jarzyna}},\ and\
  \bibinfo {author} {\bibfnamefont {R.}~\bibnamefont
  {Demkowicz-Dobrza\ifmmode~\acute{n}\else \'{n}\fi{}ski}},\ }\bibfield
  {title} {\bibinfo {title} {Compatibility in multiparameter quantum
  metrology},\ }\href {https://doi.org/10.1103/PhysRevA.94.052108} {\bibfield
  {journal} {\bibinfo  {journal} {Phys. Rev. A}\ }\textbf {\bibinfo {volume}
  {94}},\ \bibinfo {pages} {052108} (\bibinfo {year} {2016})}\BibitemShut
  {NoStop}%
\bibitem [{\citenamefont {Heinosaari}\ \emph {et~al.}(2016)\citenamefont
  {Heinosaari}, \citenamefont {Miyadera},\ and\ \citenamefont
  {Ziman}}]{Heinosaarietal2016}%
  \BibitemOpen
  \bibfield  {author} {\bibinfo {author} {\bibfnamefont {T.}~\bibnamefont
  {Heinosaari}}, \bibinfo {author} {\bibfnamefont {T.}~\bibnamefont
  {Miyadera}},\ and\ \bibinfo {author} {\bibfnamefont {M.}~\bibnamefont
  {Ziman}},\ }\bibfield  {title} {\bibinfo {title} {An invitation to quantum
  incompatibility},\ }\href {https://doi.org/10.1088/1751-8113/49/12/123001}
  {\bibfield  {journal} {\bibinfo  {journal} {J. Phys. A: Math. Theor.}\
  }\textbf {\bibinfo {volume} {49}},\ \bibinfo {pages} {123001} (\bibinfo
  {year} {2016})}\BibitemShut {NoStop}%
\bibitem [{\citenamefont {Albarelli}\ \emph {et~al.}(2019)\citenamefont
  {Albarelli}, \citenamefont {Friel},\ and\ \citenamefont
  {Datta}}]{Albarellietal2019}%
  \BibitemOpen
  \bibfield  {author} {\bibinfo {author} {\bibfnamefont {F.}~\bibnamefont
  {Albarelli}}, \bibinfo {author} {\bibfnamefont {J.~F.}\ \bibnamefont
  {Friel}},\ and\ \bibinfo {author} {\bibfnamefont {A.}~\bibnamefont {Datta}},\
  }\bibfield  {title} {\bibinfo {title} {Evaluating the {H}olevo
  {C}ram\'er-{R}ao bound for multiparameter quantum metrology},\ }\href
  {https://doi.org/10.1103/PhysRevLett.123.200503} {\bibfield  {journal}
  {\bibinfo  {journal} {Phys. Rev. Lett.}\ }\textbf {\bibinfo {volume} {123}},\
  \bibinfo {pages} {200503} (\bibinfo {year} {2019})}\BibitemShut {NoStop}%
\bibitem [{\citenamefont {Belliardo}\ and\ \citenamefont
  {Giovannetti}(2021)}]{BelliardoGiovannetti2021}%
  \BibitemOpen
  \bibfield  {author} {\bibinfo {author} {\bibfnamefont {F.}~\bibnamefont
  {Belliardo}}\ and\ \bibinfo {author} {\bibfnamefont {V.}~\bibnamefont
  {Giovannetti}},\ }\bibfield  {title} {\bibinfo {title} {Incompatibility in
  quantum parameter estimation},\ }\href
  {https://doi.org/10.1088/1367-2630/ac04ca} {\bibfield  {journal} {\bibinfo
  {journal} {New J. Phys.}\ }\textbf {\bibinfo {volume} {23}},\ \bibinfo
  {pages} {063055} (\bibinfo {year} {2021})}\BibitemShut {NoStop}%
\bibitem [{\citenamefont {Suzuki}(2020)}]{Suzuki2020}%
  \BibitemOpen
  \bibfield  {author} {\bibinfo {author} {\bibfnamefont {J.}~\bibnamefont
  {Suzuki}},\ }\bibfield  {title} {\bibinfo {title} {Nuisance parameter problem
  in quantum estimation theory: tradeoff relation and qubit examples},\ }\href
  {https://doi.org/10.1088/1751-8121/ab8672} {\bibfield  {journal} {\bibinfo
  {journal} {J. Phys. A: Math. Theor.}\ }\textbf {\bibinfo {volume} {53}},\
  \bibinfo {pages} {264001} (\bibinfo {year} {2020})}\BibitemShut {NoStop}%
\bibitem [{\citenamefont {Suzuki}\ \emph {et~al.}(2020)\citenamefont {Suzuki},
  \citenamefont {Yang},\ and\ \citenamefont {Hayashi}}]{Suzukietal2020}%
  \BibitemOpen
  \bibfield  {author} {\bibinfo {author} {\bibfnamefont {J.}~\bibnamefont
  {Suzuki}}, \bibinfo {author} {\bibfnamefont {Y.}~\bibnamefont {Yang}},\ and\
  \bibinfo {author} {\bibfnamefont {M.}~\bibnamefont {Hayashi}},\ }\bibfield
  {title} {\bibinfo {title} {Quantum state estimation with nuisance
  parameters},\ }\href {https://doi.org/10.1088/1751-8121/ab8b78} {\bibfield
  {journal} {\bibinfo  {journal} {J. Phys. A: Math. Theor.}\ }\textbf {\bibinfo
  {volume} {53}},\ \bibinfo {pages} {453001} (\bibinfo {year}
  {2020})}\BibitemShut {NoStop}%
\bibitem [{\citenamefont {Goldberg}\ \emph
  {et~al.}(2021{\natexlab{b}})\citenamefont {Goldberg}, \citenamefont
  {S\'anchez-Soto},\ and\ \citenamefont
  {Ferretti}}]{Goldbergetal2021intrinsic}%
  \BibitemOpen
  \bibfield  {author} {\bibinfo {author} {\bibfnamefont {A.~Z.}\ \bibnamefont
  {Goldberg}}, \bibinfo {author} {\bibfnamefont {L.~L.}\ \bibnamefont
  {S\'anchez-Soto}},\ and\ \bibinfo {author} {\bibfnamefont {H.}~\bibnamefont
  {Ferretti}},\ }\bibfield  {title} {\bibinfo {title} {Intrinsic sensitivity
  limits for multiparameter quantum metrology},\ }\href
  {https://doi.org/10.1103/PhysRevLett.127.110501} {\bibfield  {journal}
  {\bibinfo  {journal} {Phys. Rev. Lett.}\ }\textbf {\bibinfo {volume} {127}},\
  \bibinfo {pages} {110501} (\bibinfo {year} {2021}{\natexlab{b}})}\BibitemShut
  {NoStop}%
\bibitem [{\citenamefont {Koschorreck}\ \emph {et~al.}(2011)\citenamefont
  {Koschorreck}, \citenamefont {Napolitano}, \citenamefont {Dubost},\ and\
  \citenamefont {Mitchell}}]{Koschorrecketal2011}%
  \BibitemOpen
  \bibfield  {author} {\bibinfo {author} {\bibfnamefont {M.}~\bibnamefont
  {Koschorreck}}, \bibinfo {author} {\bibfnamefont {M.}~\bibnamefont
  {Napolitano}}, \bibinfo {author} {\bibfnamefont {B.}~\bibnamefont {Dubost}},\
  and\ \bibinfo {author} {\bibfnamefont {M.~W.}\ \bibnamefont {Mitchell}},\
  }\bibfield  {title} {\bibinfo {title} {High resolution magnetic vector-field
  imaging with cold atomic ensembles},\ }\href
  {https://doi.org/10.1063/1.3555459} {\bibfield  {journal} {\bibinfo
  {journal} {Appl. Phys. Lett.}\ }\textbf {\bibinfo {volume} {98}},\ \bibinfo
  {pages} {074101} (\bibinfo {year} {2011})}\BibitemShut {NoStop}%
\bibitem [{\citenamefont {Humphreys}\ \emph {et~al.}(2013)\citenamefont
  {Humphreys}, \citenamefont {Barbieri}, \citenamefont {Datta},\ and\
  \citenamefont {Walmsley}}]{Humphreysetal2013}%
  \BibitemOpen
  \bibfield  {author} {\bibinfo {author} {\bibfnamefont {P.~C.}\ \bibnamefont
  {Humphreys}}, \bibinfo {author} {\bibfnamefont {M.}~\bibnamefont {Barbieri}},
  \bibinfo {author} {\bibfnamefont {A.}~\bibnamefont {Datta}},\ and\ \bibinfo
  {author} {\bibfnamefont {I.~A.}\ \bibnamefont {Walmsley}},\ }\bibfield
  {title} {\bibinfo {title} {Quantum enhanced multiple phase estimation},\
  }\href {https://doi.org/10.1103/PhysRevLett.111.070403} {\bibfield  {journal}
  {\bibinfo  {journal} {Phys. Rev. Lett.}\ }\textbf {\bibinfo {volume} {111}},\
  \bibinfo {pages} {070403} (\bibinfo {year} {2013})}\BibitemShut {NoStop}%
\bibitem [{\citenamefont {Baumgratz}\ and\ \citenamefont
  {Datta}(2016)}]{BaumgratzDatta2016}%
  \BibitemOpen
  \bibfield  {author} {\bibinfo {author} {\bibfnamefont {T.}~\bibnamefont
  {Baumgratz}}\ and\ \bibinfo {author} {\bibfnamefont {A.}~\bibnamefont
  {Datta}},\ }\bibfield  {title} {\bibinfo {title} {Quantum enhanced estimation
  of a multidimensional field},\ }\href
  {https://doi.org/10.1103/PhysRevLett.116.030801} {\bibfield  {journal}
  {\bibinfo  {journal} {Phys. Rev. Lett.}\ }\textbf {\bibinfo {volume} {116}},\
  \bibinfo {pages} {030801} (\bibinfo {year} {2016})}\BibitemShut {NoStop}%
\bibitem [{\citenamefont {\ifmmode \check{R}\else
  \v{R}\fi{}eha\ifmmode~\check{c}\else \v{c}\fi{}ek}\ \emph
  {et~al.}(2017)\citenamefont {\ifmmode \check{R}\else
  \v{R}\fi{}eha\ifmmode~\check{c}\else \v{c}\fi{}ek}, \citenamefont {Hradil},
  \citenamefont {Stoklasa}, \citenamefont {Pa\'ur}, \citenamefont {Grover},
  \citenamefont {Krzic},\ and\ \citenamefont
  {S\'anchez-Soto}}]{Rehaceketal2017}%
  \BibitemOpen
  \bibfield  {author} {\bibinfo {author} {\bibfnamefont {J.}~\bibnamefont
  {\ifmmode \check{R}\else \v{R}\fi{}eha\ifmmode~\check{c}\else \v{c}\fi{}ek}},
  \bibinfo {author} {\bibfnamefont {Z.}~\bibnamefont {Hradil}}, \bibinfo
  {author} {\bibfnamefont {B.}~\bibnamefont {Stoklasa}}, \bibinfo {author}
  {\bibfnamefont {M.}~\bibnamefont {Pa\'ur}}, \bibinfo {author} {\bibfnamefont
  {J.}~\bibnamefont {Grover}}, \bibinfo {author} {\bibfnamefont
  {A.}~\bibnamefont {Krzic}},\ and\ \bibinfo {author} {\bibfnamefont {L.~L.}\
  \bibnamefont {S\'anchez-Soto}},\ }\bibfield  {title} {\bibinfo {title}
  {Multiparameter quantum metrology of incoherent point sources: Towards
  realistic superresolution},\ }\href
  {https://doi.org/10.1103/PhysRevA.96.062107} {\bibfield  {journal} {\bibinfo
  {journal} {Phys. Rev. A}\ }\textbf {\bibinfo {volume} {96}},\ \bibinfo
  {pages} {062107} (\bibinfo {year} {2017})}\BibitemShut {NoStop}%
\bibitem [{\citenamefont {Proctor}\ \emph {et~al.}(2018)\citenamefont
  {Proctor}, \citenamefont {Knott},\ and\ \citenamefont
  {Dunningham}}]{Proctoretal2018}%
  \BibitemOpen
  \bibfield  {author} {\bibinfo {author} {\bibfnamefont {T.~J.}\ \bibnamefont
  {Proctor}}, \bibinfo {author} {\bibfnamefont {P.~A.}\ \bibnamefont {Knott}},\
  and\ \bibinfo {author} {\bibfnamefont {J.~A.}\ \bibnamefont {Dunningham}},\
  }\bibfield  {title} {\bibinfo {title} {Multiparameter estimation in networked
  quantum sensors},\ }\href {https://doi.org/10.1103/PhysRevLett.120.080501}
  {\bibfield  {journal} {\bibinfo  {journal} {Phys. Rev. Lett.}\ }\textbf
  {\bibinfo {volume} {120}},\ \bibinfo {pages} {080501} (\bibinfo {year}
  {2018})}\BibitemShut {NoStop}%
\bibitem [{\citenamefont {Rubio}\ \emph {et~al.}(2020)\citenamefont {Rubio},
  \citenamefont {Knott}, \citenamefont {Proctor},\ and\ \citenamefont
  {Dunningham}}]{Rubioetal2020}%
  \BibitemOpen
  \bibfield  {author} {\bibinfo {author} {\bibfnamefont {J.}~\bibnamefont
  {Rubio}}, \bibinfo {author} {\bibfnamefont {P.~A.}\ \bibnamefont {Knott}},
  \bibinfo {author} {\bibfnamefont {T.~J.}\ \bibnamefont {Proctor}},\ and\
  \bibinfo {author} {\bibfnamefont {J.~A.}\ \bibnamefont {Dunningham}},\
  }\bibfield  {title} {\bibinfo {title} {Quantum sensing networks for the
  estimation of linear functions},\ }\href
  {https://doi.org/10.1088/1751-8121/ab9d46} {\bibfield  {journal} {\bibinfo
  {journal} {J. Phys. A: Math. Theor.}\ }\textbf {\bibinfo {volume} {53}},\
  \bibinfo {pages} {344001} (\bibinfo {year} {2020})}\BibitemShut {NoStop}%
\bibitem [{\citenamefont {Goldberg}\ \emph
  {et~al.}(2021{\natexlab{c}})\citenamefont {Goldberg}, \citenamefont {Klimov},
  \citenamefont {Leuchs},\ and\ \citenamefont
  {S{\'{a}}nchez-Soto}}]{Goldbergetal2021rotationspublished}%
  \BibitemOpen
  \bibfield  {author} {\bibinfo {author} {\bibfnamefont {A.~Z.}\ \bibnamefont
  {Goldberg}}, \bibinfo {author} {\bibfnamefont {A.~B.}\ \bibnamefont
  {Klimov}}, \bibinfo {author} {\bibfnamefont {G.}~\bibnamefont {Leuchs}},\
  and\ \bibinfo {author} {\bibfnamefont {L.~L.}\ \bibnamefont
  {S{\'{a}}nchez-Soto}},\ }\bibfield  {title} {\bibinfo {title} {Rotation
  sensing at the ultimate limit},\ }\href
  {https://doi.org/10.1088/2515-7647/abeb54} {\bibfield  {journal} {\bibinfo
  {journal} {J. Phys.: Photonics}\ }\textbf {\bibinfo {volume} {3}},\ \bibinfo
  {pages} {022008} (\bibinfo {year} {2021}{\natexlab{c}})}\BibitemShut
  {NoStop}%
\bibitem [{\citenamefont {Fiderer}\ \emph {et~al.}(2021)\citenamefont
  {Fiderer}, \citenamefont {Tufarelli}, \citenamefont {Piano},\ and\
  \citenamefont {Adesso}}]{Fidereretal2021}%
  \BibitemOpen
  \bibfield  {author} {\bibinfo {author} {\bibfnamefont {L.~J.}\ \bibnamefont
  {Fiderer}}, \bibinfo {author} {\bibfnamefont {T.}~\bibnamefont {Tufarelli}},
  \bibinfo {author} {\bibfnamefont {S.}~\bibnamefont {Piano}},\ and\ \bibinfo
  {author} {\bibfnamefont {G.}~\bibnamefont {Adesso}},\ }\bibfield  {title}
  {\bibinfo {title} {General expressions for the quantum {Fisher} information
  matrix with applications to discrete quantum imaging},\ }\href
  {https://doi.org/10.1103/PRXQuantum.2.020308} {\bibfield  {journal} {\bibinfo
   {journal} {PRX Quantum}\ }\textbf {\bibinfo {volume} {2}},\ \bibinfo {pages}
  {020308} (\bibinfo {year} {2021})}\BibitemShut {NoStop}%
\bibitem [{\citenamefont {Hardy}(2007)}]{Hardy2007}%
  \BibitemOpen
  \bibfield  {author} {\bibinfo {author} {\bibfnamefont {L.}~\bibnamefont
  {Hardy}},\ }\bibfield  {title} {\bibinfo {title} {Towards quantum gravity: a
  framework for probabilistic theories with non-fixed causal structure},\
  }\href {https://doi.org/10.1088/1751-8113/40/12/s12} {\bibfield  {journal}
  {\bibinfo  {journal} {J. Phys. A: Math. Theor.}\ }\textbf {\bibinfo {volume}
  {40}},\ \bibinfo {pages} {3081} (\bibinfo {year} {2007})}\BibitemShut
  {NoStop}%
\bibitem [{\citenamefont {Chiribella}\ \emph {et~al.}(2013)\citenamefont
  {Chiribella}, \citenamefont {D'Ariano}, \citenamefont {Perinotti},\ and\
  \citenamefont {Valiron}}]{Chiribellaetal2013}%
  \BibitemOpen
  \bibfield  {author} {\bibinfo {author} {\bibfnamefont {G.}~\bibnamefont
  {Chiribella}}, \bibinfo {author} {\bibfnamefont {G.~M.}\ \bibnamefont
  {D'Ariano}}, \bibinfo {author} {\bibfnamefont {P.}~\bibnamefont
  {Perinotti}},\ and\ \bibinfo {author} {\bibfnamefont {B.}~\bibnamefont
  {Valiron}},\ }\bibfield  {title} {\bibinfo {title} {Quantum computations
  without definite causal structure},\ }\href
  {https://doi.org/10.1103/PhysRevA.88.022318} {\bibfield  {journal} {\bibinfo
  {journal} {Phys. Rev. A}\ }\textbf {\bibinfo {volume} {88}},\ \bibinfo
  {pages} {022318} (\bibinfo {year} {2013})}\BibitemShut {NoStop}%
\bibitem [{\citenamefont {Colnaghi}\ \emph {et~al.}(2012)\citenamefont
  {Colnaghi}, \citenamefont {D'Ariano}, \citenamefont {Facchini},\ and\
  \citenamefont {Perinotti}}]{Colnaghietal2012}%
  \BibitemOpen
  \bibfield  {author} {\bibinfo {author} {\bibfnamefont {T.}~\bibnamefont
  {Colnaghi}}, \bibinfo {author} {\bibfnamefont {G.~M.}\ \bibnamefont
  {D'Ariano}}, \bibinfo {author} {\bibfnamefont {S.}~\bibnamefont {Facchini}},\
  and\ \bibinfo {author} {\bibfnamefont {P.}~\bibnamefont {Perinotti}},\
  }\bibfield  {title} {\bibinfo {title} {Quantum computation with programmable
  connections between gates},\ }\href
  {https://doi.org/https://doi.org/10.1016/j.physleta.2012.08.028} {\bibfield
  {journal} {\bibinfo  {journal} {Phys. Lett. A}\ }\textbf {\bibinfo {volume}
  {376}},\ \bibinfo {pages} {2940} (\bibinfo {year} {2012})}\BibitemShut
  {NoStop}%
\bibitem [{\citenamefont {Morimae}(2014)}]{Morimae2014}%
  \BibitemOpen
  \bibfield  {author} {\bibinfo {author} {\bibfnamefont {T.}~\bibnamefont
  {Morimae}},\ }\bibfield  {title} {\bibinfo {title} {Acausal measurement-based
  quantum computing},\ }\href {https://doi.org/10.1103/PhysRevA.90.010101}
  {\bibfield  {journal} {\bibinfo  {journal} {Phys. Rev. A}\ }\textbf {\bibinfo
  {volume} {90}},\ \bibinfo {pages} {010101} (\bibinfo {year}
  {2014})}\BibitemShut {NoStop}%
\bibitem [{\citenamefont {Ara\'ujo}\ \emph {et~al.}(2014)\citenamefont
  {Ara\'ujo}, \citenamefont {Costa},\ and\ \citenamefont
  {Brukner}}]{Araujoetal2014}%
  \BibitemOpen
  \bibfield  {author} {\bibinfo {author} {\bibfnamefont {M.}~\bibnamefont
  {Ara\'ujo}}, \bibinfo {author} {\bibfnamefont {F.}~\bibnamefont {Costa}},\
  and\ \bibinfo {author} {\bibfnamefont {{\v{C}}.}~\bibnamefont {Brukner}},\
  }\bibfield  {title} {\bibinfo {title} {Computational advantage from
  quantum-controlled ordering of gates},\ }\href
  {https://doi.org/10.1103/PhysRevLett.113.250402} {\bibfield  {journal}
  {\bibinfo  {journal} {Phys. Rev. Lett.}\ }\textbf {\bibinfo {volume} {113}},\
  \bibinfo {pages} {250402} (\bibinfo {year} {2014})}\BibitemShut {NoStop}%
\bibitem [{\citenamefont {Taddei}\ \emph {et~al.}(2021)\citenamefont {Taddei},
  \citenamefont {Cari\~ne}, \citenamefont {Mart\'{\i}nez}, \citenamefont
  {Garc\'{\i}a}, \citenamefont {Guerrero}, \citenamefont {Abbott},
  \citenamefont {Ara\'ujo}, \citenamefont {Branciard}, \citenamefont {G\'omez},
  \citenamefont {Walborn}, \citenamefont {Aolita},\ and\ \citenamefont
  {Lima}}]{Taddeietal2021}%
  \BibitemOpen
  \bibfield  {author} {\bibinfo {author} {\bibfnamefont {M.~M.}\ \bibnamefont
  {Taddei}}, \bibinfo {author} {\bibfnamefont {J.}~\bibnamefont {Cari\~ne}},
  \bibinfo {author} {\bibfnamefont {D.}~\bibnamefont {Mart\'{\i}nez}}, \bibinfo
  {author} {\bibfnamefont {T.}~\bibnamefont {Garc\'{\i}a}}, \bibinfo {author}
  {\bibfnamefont {N.}~\bibnamefont {Guerrero}}, \bibinfo {author}
  {\bibfnamefont {A.~A.}\ \bibnamefont {Abbott}}, \bibinfo {author}
  {\bibfnamefont {M.}~\bibnamefont {Ara\'ujo}}, \bibinfo {author}
  {\bibfnamefont {C.}~\bibnamefont {Branciard}}, \bibinfo {author}
  {\bibfnamefont {E.~S.}\ \bibnamefont {G\'omez}}, \bibinfo {author}
  {\bibfnamefont {S.~P.}\ \bibnamefont {Walborn}}, \bibinfo {author}
  {\bibfnamefont {L.}~\bibnamefont {Aolita}},\ and\ \bibinfo {author}
  {\bibfnamefont {G.}~\bibnamefont {Lima}},\ }\bibfield  {title} {\bibinfo
  {title} {Computational advantage from the quantum superposition of multiple
  temporal orders of photonic gates},\ }\href
  {https://doi.org/10.1103/PRXQuantum.2.010320} {\bibfield  {journal} {\bibinfo
   {journal} {PRX Quantum}\ }\textbf {\bibinfo {volume} {2}},\ \bibinfo {pages}
  {010320} (\bibinfo {year} {2021})}\BibitemShut {NoStop}%
\bibitem [{\citenamefont {Wechs}\ \emph {et~al.}(2021)\citenamefont {Wechs},
  \citenamefont {Dourdent}, \citenamefont {Abbott},\ and\ \citenamefont
  {Branciard}}]{Wechsetal2021}%
  \BibitemOpen
  \bibfield  {author} {\bibinfo {author} {\bibfnamefont {J.}~\bibnamefont
  {Wechs}}, \bibinfo {author} {\bibfnamefont {H.}~\bibnamefont {Dourdent}},
  \bibinfo {author} {\bibfnamefont {A.~A.}\ \bibnamefont {Abbott}},\ and\
  \bibinfo {author} {\bibfnamefont {C.}~\bibnamefont {Branciard}},\ }\bibfield
  {title} {\bibinfo {title} {Quantum circuits with classical versus quantum
  control of causal order},\ }\href
  {https://doi.org/10.1103/PRXQuantum.2.030335} {\bibfield  {journal} {\bibinfo
   {journal} {PRX Quantum}\ }\textbf {\bibinfo {volume} {2}},\ \bibinfo {pages}
  {030335} (\bibinfo {year} {2021})}\BibitemShut {NoStop}%
\bibitem [{\citenamefont {Chiribella}(2012)}]{Chiribella2012}%
  \BibitemOpen
  \bibfield  {author} {\bibinfo {author} {\bibfnamefont {G.}~\bibnamefont
  {Chiribella}},\ }\bibfield  {title} {\bibinfo {title} {Perfect discrimination
  of no-signalling channels via quantum superposition of causal structures},\
  }\href {https://doi.org/10.1103/PhysRevA.86.040301} {\bibfield  {journal}
  {\bibinfo  {journal} {Phys. Rev. A}\ }\textbf {\bibinfo {volume} {86}},\
  \bibinfo {pages} {040301} (\bibinfo {year} {2012})}\BibitemShut {NoStop}%
\bibitem [{\citenamefont {Feix}\ \emph {et~al.}(2015)\citenamefont {Feix},
  \citenamefont {Ara\'ujo},\ and\ \citenamefont {Brukner}}]{Feixetal2015}%
  \BibitemOpen
  \bibfield  {author} {\bibinfo {author} {\bibfnamefont {A.}~\bibnamefont
  {Feix}}, \bibinfo {author} {\bibfnamefont {M.}~\bibnamefont {Ara\'ujo}},\
  and\ \bibinfo {author} {\bibfnamefont {{\v{C}}.}~\bibnamefont {Brukner}},\
  }\bibfield  {title} {\bibinfo {title} {Quantum superposition of the order of
  parties as a communication resource},\ }\href
  {https://doi.org/10.1103/PhysRevA.92.052326} {\bibfield  {journal} {\bibinfo
  {journal} {Phys. Rev. A}\ }\textbf {\bibinfo {volume} {92}},\ \bibinfo
  {pages} {052326} (\bibinfo {year} {2015})}\BibitemShut {NoStop}%
\bibitem [{\citenamefont {Gu\'erin}\ \emph {et~al.}(2016)\citenamefont
  {Gu\'erin}, \citenamefont {Feix}, \citenamefont {Ara\'ujo},\ and\
  \citenamefont {Brukner}}]{Guerinetal2016}%
  \BibitemOpen
  \bibfield  {author} {\bibinfo {author} {\bibfnamefont {P.~A.}\ \bibnamefont
  {Gu\'erin}}, \bibinfo {author} {\bibfnamefont {A.}~\bibnamefont {Feix}},
  \bibinfo {author} {\bibfnamefont {M.}~\bibnamefont {Ara\'ujo}},\ and\
  \bibinfo {author} {\bibfnamefont {{\v{C}}.}~\bibnamefont {Brukner}},\
  }\bibfield  {title} {\bibinfo {title} {Exponential communication complexity
  advantage from quantum superposition of the direction of communication},\
  }\href {https://doi.org/10.1103/PhysRevLett.117.100502} {\bibfield  {journal}
  {\bibinfo  {journal} {Phys. Rev. Lett.}\ }\textbf {\bibinfo {volume} {117}},\
  \bibinfo {pages} {100502} (\bibinfo {year} {2016})}\BibitemShut {NoStop}%
\bibitem [{\citenamefont {Del~Santo}\ and\ \citenamefont
  {Daki\ifmmode~\acute{c}\else \'{c}\fi{}}(2018)}]{DelSantoDakic2018}%
  \BibitemOpen
  \bibfield  {author} {\bibinfo {author} {\bibfnamefont {F.}~\bibnamefont
  {Del~Santo}}\ and\ \bibinfo {author} {\bibfnamefont {B.}~\bibnamefont
  {Daki\ifmmode~\acute{c}\else \'{c}\fi{}}},\ }\bibfield  {title} {\bibinfo
  {title} {Two-way communication with a single quantum particle},\ }\href
  {https://doi.org/10.1103/PhysRevLett.120.060503} {\bibfield  {journal}
  {\bibinfo  {journal} {Phys. Rev. Lett.}\ }\textbf {\bibinfo {volume} {120}},\
  \bibinfo {pages} {060503} (\bibinfo {year} {2018})}\BibitemShut {NoStop}%
\bibitem [{\citenamefont {Ebler}\ \emph {et~al.}(2018)\citenamefont {Ebler},
  \citenamefont {Salek},\ and\ \citenamefont {Chiribella}}]{Ebleretal2018}%
  \BibitemOpen
  \bibfield  {author} {\bibinfo {author} {\bibfnamefont {D.}~\bibnamefont
  {Ebler}}, \bibinfo {author} {\bibfnamefont {S.}~\bibnamefont {Salek}},\ and\
  \bibinfo {author} {\bibfnamefont {G.}~\bibnamefont {Chiribella}},\ }\bibfield
   {title} {\bibinfo {title} {Enhanced communication with the assistance of
  indefinite causal order},\ }\href
  {https://doi.org/10.1103/PhysRevLett.120.120502} {\bibfield  {journal}
  {\bibinfo  {journal} {Phys. Rev. Lett.}\ }\textbf {\bibinfo {volume} {120}},\
  \bibinfo {pages} {120502} (\bibinfo {year} {2018})}\BibitemShut {NoStop}%
\bibitem [{\citenamefont {Procopio}\ \emph {et~al.}(2019)\citenamefont
  {Procopio}, \citenamefont {Delgado}, \citenamefont {Enr{\'\i}quez},
  \citenamefont {Belabas},\ and\ \citenamefont {Levenson}}]{Procopioetal2019}%
  \BibitemOpen
  \bibfield  {author} {\bibinfo {author} {\bibfnamefont {L.~M.}\ \bibnamefont
  {Procopio}}, \bibinfo {author} {\bibfnamefont {F.}~\bibnamefont {Delgado}},
  \bibinfo {author} {\bibfnamefont {M.}~\bibnamefont {Enr{\'\i}quez}}, \bibinfo
  {author} {\bibfnamefont {N.}~\bibnamefont {Belabas}},\ and\ \bibinfo {author}
  {\bibfnamefont {J.~A.}\ \bibnamefont {Levenson}},\ }\bibfield  {title}
  {\bibinfo {title} {Communication enhancement through quantum coherent control
  of {N} channels in an indefinite causal-order scenario},\ }\href
  {https://doi.org/10.3390/e21101012} {\bibfield  {journal} {\bibinfo
  {journal} {Entropy}\ }\textbf {\bibinfo {volume} {21}},\ \bibinfo {pages}
  {1012} (\bibinfo {year} {2019})}\BibitemShut {NoStop}%
\bibitem [{\citenamefont {Chiribella}\ \emph {et~al.}(2021)\citenamefont
  {Chiribella}, \citenamefont {Banik}, \citenamefont {Bhattacharya},
  \citenamefont {Guha}, \citenamefont {Alimuddin}, \citenamefont {Roy},
  \citenamefont {Saha}, \citenamefont {Agrawal},\ and\ \citenamefont
  {Kar}}]{Chiribellaetal2021}%
  \BibitemOpen
  \bibfield  {author} {\bibinfo {author} {\bibfnamefont {G.}~\bibnamefont
  {Chiribella}}, \bibinfo {author} {\bibfnamefont {M.}~\bibnamefont {Banik}},
  \bibinfo {author} {\bibfnamefont {S.~S.}\ \bibnamefont {Bhattacharya}},
  \bibinfo {author} {\bibfnamefont {T.}~\bibnamefont {Guha}}, \bibinfo {author}
  {\bibfnamefont {M.}~\bibnamefont {Alimuddin}}, \bibinfo {author}
  {\bibfnamefont {A.}~\bibnamefont {Roy}}, \bibinfo {author} {\bibfnamefont
  {S.}~\bibnamefont {Saha}}, \bibinfo {author} {\bibfnamefont {S.}~\bibnamefont
  {Agrawal}},\ and\ \bibinfo {author} {\bibfnamefont {G.}~\bibnamefont {Kar}},\
  }\bibfield  {title} {\bibinfo {title} {Indefinite causal order enables
  perfect quantum communication with zero capacity channels},\ }\href
  {https://doi.org/10.1088/1367-2630/abe7a0} {\bibfield  {journal} {\bibinfo
  {journal} {New J. Phys.}\ }\textbf {\bibinfo {volume} {23}},\ \bibinfo
  {pages} {033039} (\bibinfo {year} {2021})}\BibitemShut {NoStop}%
\bibitem [{\citenamefont {Felce}\ and\ \citenamefont
  {Vedral}(2020)}]{FelceVedral2020}%
  \BibitemOpen
  \bibfield  {author} {\bibinfo {author} {\bibfnamefont {D.}~\bibnamefont
  {Felce}}\ and\ \bibinfo {author} {\bibfnamefont {V.}~\bibnamefont {Vedral}},\
  }\bibfield  {title} {\bibinfo {title} {Quantum refrigeration with indefinite
  causal order},\ }\href {https://doi.org/10.1103/PhysRevLett.125.070603}
  {\bibfield  {journal} {\bibinfo  {journal} {Phys. Rev. Lett.}\ }\textbf
  {\bibinfo {volume} {125}},\ \bibinfo {pages} {070603} (\bibinfo {year}
  {2020})}\BibitemShut {NoStop}%
\bibitem [{\citenamefont {Cao}\ \emph {et~al.}(2022)\citenamefont {Cao},
  \citenamefont {Wang}, \citenamefont {Jia}, \citenamefont {Zhang},
  \citenamefont {Guo}, \citenamefont {Liu}, \citenamefont {Huang},
  \citenamefont {Li},\ and\ \citenamefont {Guo}}]{Caoetal2022}%
  \BibitemOpen
  \bibfield  {author} {\bibinfo {author} {\bibfnamefont {H.}~\bibnamefont
  {Cao}}, \bibinfo {author} {\bibfnamefont {N.-N.}\ \bibnamefont {Wang}},
  \bibinfo {author} {\bibfnamefont {Z.}~\bibnamefont {Jia}}, \bibinfo {author}
  {\bibfnamefont {C.}~\bibnamefont {Zhang}}, \bibinfo {author} {\bibfnamefont
  {Y.}~\bibnamefont {Guo}}, \bibinfo {author} {\bibfnamefont {B.-H.}\
  \bibnamefont {Liu}}, \bibinfo {author} {\bibfnamefont {Y.-F.}\ \bibnamefont
  {Huang}}, \bibinfo {author} {\bibfnamefont {C.-F.}\ \bibnamefont {Li}},\ and\
  \bibinfo {author} {\bibfnamefont {G.-C.}\ \bibnamefont {Guo}},\ }\bibfield
  {title} {\bibinfo {title} {Quantum simulation of indefinite causal order
  induced quantum refrigeration},\ }\href
  {https://doi.org/10.1103/PhysRevResearch.4.L032029} {\bibfield  {journal}
  {\bibinfo  {journal} {Phys. Rev. Res.}\ }\textbf {\bibinfo {volume} {4}},\
  \bibinfo {pages} {L032029} (\bibinfo {year} {2022})}\BibitemShut {NoStop}%
\bibitem [{\citenamefont {Goldberg}\ and\ \citenamefont
  {Heshami}(2023)}]{GoldbergHeshami2023HBAC}%
  \BibitemOpen
  \bibfield  {author} {\bibinfo {author} {\bibfnamefont {A.~Z.}\ \bibnamefont
  {Goldberg}}\ and\ \bibinfo {author} {\bibfnamefont {K.}~\bibnamefont
  {Heshami}},\ }\bibfield  {title} {\bibinfo {title} {Breaking the limits of
  purification: postselection enhances heat-bath algorithmic cooling},\ }\href
  {https://doi.org/10.1088/2399-6528/acb414} {\bibfield  {journal} {\bibinfo
  {journal} {J. Phys. Commun.}\ }\textbf {\bibinfo {volume} {7}},\ \bibinfo
  {pages} {015003} (\bibinfo {year} {2023})}\BibitemShut {NoStop}%
\bibitem [{\citenamefont {Nie}\ \emph {et~al.}(2022)\citenamefont {Nie},
  \citenamefont {Feng}, \citenamefont {Longden},\ and\ \citenamefont
  {Vedral}}]{Nieetal2022arxiv}%
  \BibitemOpen
  \bibfield  {author} {\bibinfo {author} {\bibfnamefont {H.}~\bibnamefont
  {Nie}}, \bibinfo {author} {\bibfnamefont {T.}~\bibnamefont {Feng}}, \bibinfo
  {author} {\bibfnamefont {S.}~\bibnamefont {Longden}},\ and\ \bibinfo {author}
  {\bibfnamefont {V.}~\bibnamefont {Vedral}},\ }\bibfield  {title} {\bibinfo
  {title} {Quantum cooling activated by coherent-controlled thermalisation},\
  }\href@noop {} {\  (\bibinfo {year} {2022})},\ \Eprint
  {https://arxiv.org/abs/2201.06954} {arXiv:2201.06954 [quant-ph]} \BibitemShut
  {NoStop}%
\bibitem [{\citenamefont {Guha}\ \emph {et~al.}(2020)\citenamefont {Guha},
  \citenamefont {Alimuddin},\ and\ \citenamefont {Parashar}}]{Guhaetal2020}%
  \BibitemOpen
  \bibfield  {author} {\bibinfo {author} {\bibfnamefont {T.}~\bibnamefont
  {Guha}}, \bibinfo {author} {\bibfnamefont {M.}~\bibnamefont {Alimuddin}},\
  and\ \bibinfo {author} {\bibfnamefont {P.}~\bibnamefont {Parashar}},\
  }\bibfield  {title} {\bibinfo {title} {Thermodynamic advancement in the
  causally inseparable occurrence of thermal maps},\ }\href
  {https://doi.org/10.1103/PhysRevA.102.032215} {\bibfield  {journal} {\bibinfo
   {journal} {Phys. Rev. A}\ }\textbf {\bibinfo {volume} {102}},\ \bibinfo
  {pages} {032215} (\bibinfo {year} {2020})}\BibitemShut {NoStop}%
\bibitem [{\citenamefont {Guha}\ \emph {et~al.}(2022)\citenamefont {Guha},
  \citenamefont {Roy}, \citenamefont {Simonov},\ and\ \citenamefont
  {Zimbor{\'a}s}}]{Guhaetal2022arxiv}%
  \BibitemOpen
  \bibfield  {author} {\bibinfo {author} {\bibfnamefont {T.}~\bibnamefont
  {Guha}}, \bibinfo {author} {\bibfnamefont {S.}~\bibnamefont {Roy}}, \bibinfo
  {author} {\bibfnamefont {K.}~\bibnamefont {Simonov}},\ and\ \bibinfo {author}
  {\bibfnamefont {Z.}~\bibnamefont {Zimbor{\'a}s}},\ }\href
  {https://doi.org/10.48550/ARXIV.2208.04034} {\bibinfo {title} {Activation of
  thermal states by quantum switch-driven thermalization and its limits}}
  (\bibinfo {year} {2022})\BibitemShut {NoStop}%
\bibitem [{\citenamefont {Simonov}\ \emph {et~al.}(2022)\citenamefont
  {Simonov}, \citenamefont {Francica}, \citenamefont {Guarnieri},\ and\
  \citenamefont {Paternostro}}]{Simonovetal2022}%
  \BibitemOpen
  \bibfield  {author} {\bibinfo {author} {\bibfnamefont {K.}~\bibnamefont
  {Simonov}}, \bibinfo {author} {\bibfnamefont {G.}~\bibnamefont {Francica}},
  \bibinfo {author} {\bibfnamefont {G.}~\bibnamefont {Guarnieri}},\ and\
  \bibinfo {author} {\bibfnamefont {M.}~\bibnamefont {Paternostro}},\
  }\bibfield  {title} {\bibinfo {title} {Work extraction from coherently
  activated maps via quantum switch},\ }\href
  {https://doi.org/10.1103/PhysRevA.105.032217} {\bibfield  {journal} {\bibinfo
   {journal} {Phys. Rev. A}\ }\textbf {\bibinfo {volume} {105}},\ \bibinfo
  {pages} {032217} (\bibinfo {year} {2022})}\BibitemShut {NoStop}%
\bibitem [{\citenamefont {{Mukhopadhyay}}\ \emph {et~al.}()\citenamefont
  {{Mukhopadhyay}}, \citenamefont {{Gupta}},\ and\ \citenamefont
  {{Pati}}}]{Mukhopadhyayetal2018arxiv}%
  \BibitemOpen
  \bibfield  {author} {\bibinfo {author} {\bibfnamefont {C.}~\bibnamefont
  {{Mukhopadhyay}}}, \bibinfo {author} {\bibfnamefont {M.~K.}\ \bibnamefont
  {{Gupta}}},\ and\ \bibinfo {author} {\bibfnamefont {A.~K.}\ \bibnamefont
  {{Pati}}},\ }\bibfield  {title} {\bibinfo {title} {{Superposition of causal
  order as a metrological resource for quantum thermometry}},\ }\href@noop {}
  {\ }\Eprint {https://arxiv.org/abs/1812.07508} {1812.07508 [quant-ph]}
  \BibitemShut {NoStop}%
\bibitem [{\citenamefont {Frey}(2019)}]{Frey2019}%
  \BibitemOpen
  \bibfield  {author} {\bibinfo {author} {\bibfnamefont {M.}~\bibnamefont
  {Frey}},\ }\bibfield  {title} {\bibinfo {title} {Indefinite causal order aids
  quantum depolarizing channel identification},\ }\href
  {https://doi.org/10.1007/s11128-019-2186-9} {\bibfield  {journal} {\bibinfo
  {journal} {Quantum Inf. Process.}\ }\textbf {\bibinfo {volume} {18}},\
  \bibinfo {pages} {96} (\bibinfo {year} {2019})}\BibitemShut {NoStop}%
\bibitem [{\citenamefont {Zhao}\ \emph {et~al.}(2020)\citenamefont {Zhao},
  \citenamefont {Yang},\ and\ \citenamefont {Chiribella}}]{Zhaoetal2020}%
  \BibitemOpen
  \bibfield  {author} {\bibinfo {author} {\bibfnamefont {X.}~\bibnamefont
  {Zhao}}, \bibinfo {author} {\bibfnamefont {Y.}~\bibnamefont {Yang}},\ and\
  \bibinfo {author} {\bibfnamefont {G.}~\bibnamefont {Chiribella}},\ }\bibfield
   {title} {\bibinfo {title} {Quantum metrology with indefinite causal order},\
  }\href {https://doi.org/10.1103/PhysRevLett.124.190503} {\bibfield  {journal}
  {\bibinfo  {journal} {Phys. Rev. Lett.}\ }\textbf {\bibinfo {volume} {124}},\
  \bibinfo {pages} {190503} (\bibinfo {year} {2020})}\BibitemShut {NoStop}%
\bibitem [{\citenamefont
  {Chapeau-Blondeau}(2021{\natexlab{a}})}]{ChapeauBlondeau2021ICO}%
  \BibitemOpen
  \bibfield  {author} {\bibinfo {author} {\bibfnamefont {F.}~\bibnamefont
  {Chapeau-Blondeau}},\ }\bibfield  {title} {\bibinfo {title} {Noisy quantum
  metrology with the assistance of indefinite causal order},\ }\href
  {https://doi.org/10.1103/PhysRevA.103.032615} {\bibfield  {journal} {\bibinfo
   {journal} {Phys. Rev. A}\ }\textbf {\bibinfo {volume} {103}},\ \bibinfo
  {pages} {032615} (\bibinfo {year} {2021}{\natexlab{a}})}\BibitemShut
  {NoStop}%
\bibitem [{\citenamefont {Liu}\ \emph {et~al.}(2023)\citenamefont {Liu},
  \citenamefont {Hu}, \citenamefont {Yuan},\ and\ \citenamefont
  {Yang}}]{Liuetal2023}%
  \BibitemOpen
  \bibfield  {author} {\bibinfo {author} {\bibfnamefont {Q.}~\bibnamefont
  {Liu}}, \bibinfo {author} {\bibfnamefont {Z.}~\bibnamefont {Hu}}, \bibinfo
  {author} {\bibfnamefont {H.}~\bibnamefont {Yuan}},\ and\ \bibinfo {author}
  {\bibfnamefont {Y.}~\bibnamefont {Yang}},\ }\bibfield  {title} {\bibinfo
  {title} {Optimal strategies of quantum metrology with a strict hierarchy},\
  }\href {https://doi.org/10.1103/PhysRevLett.130.070803} {\bibfield  {journal}
  {\bibinfo  {journal} {Phys. Rev. Lett.}\ }\textbf {\bibinfo {volume} {130}},\
  \bibinfo {pages} {070803} (\bibinfo {year} {2023})}\BibitemShut {NoStop}%
\bibitem [{\citenamefont {Procopio}\ \emph {et~al.}(2015)\citenamefont
  {Procopio}, \citenamefont {Moqanaki}, \citenamefont {Ara{\'u}jo},
  \citenamefont {Costa}, \citenamefont {Alonso~Calafell}, \citenamefont {Dowd},
  \citenamefont {Hamel}, \citenamefont {Rozema}, \citenamefont {Brukner},\ and\
  \citenamefont {Walther}}]{Procopioetal2015}%
  \BibitemOpen
  \bibfield  {author} {\bibinfo {author} {\bibfnamefont {L.~M.}\ \bibnamefont
  {Procopio}}, \bibinfo {author} {\bibfnamefont {A.}~\bibnamefont {Moqanaki}},
  \bibinfo {author} {\bibfnamefont {M.}~\bibnamefont {Ara{\'u}jo}}, \bibinfo
  {author} {\bibfnamefont {F.}~\bibnamefont {Costa}}, \bibinfo {author}
  {\bibfnamefont {I.}~\bibnamefont {Alonso~Calafell}}, \bibinfo {author}
  {\bibfnamefont {E.~G.}\ \bibnamefont {Dowd}}, \bibinfo {author}
  {\bibfnamefont {D.~R.}\ \bibnamefont {Hamel}}, \bibinfo {author}
  {\bibfnamefont {L.~A.}\ \bibnamefont {Rozema}}, \bibinfo {author}
  {\bibfnamefont {{\v{C}}.}~\bibnamefont {Brukner}},\ and\ \bibinfo {author}
  {\bibfnamefont {P.}~\bibnamefont {Walther}},\ }\bibfield  {title} {\bibinfo
  {title} {Experimental superposition of orders of quantum gates},\ }\href
  {https://doi.org/10.1038/ncomms8913} {\bibfield  {journal} {\bibinfo
  {journal} {Nature Commun.}\ }\textbf {\bibinfo {volume} {6}},\ \bibinfo
  {pages} {7913} (\bibinfo {year} {2015})}\BibitemShut {NoStop}%
\bibitem [{\citenamefont {Rubino}\ \emph {et~al.}(2017)\citenamefont {Rubino},
  \citenamefont {Rozema}, \citenamefont {Feix}, \citenamefont {Ara{\'u}jo},
  \citenamefont {Zeuner}, \citenamefont {Procopio}, \citenamefont {Brukner},\
  and\ \citenamefont {Walther}}]{Rubinoetal2017}%
  \BibitemOpen
  \bibfield  {author} {\bibinfo {author} {\bibfnamefont {G.}~\bibnamefont
  {Rubino}}, \bibinfo {author} {\bibfnamefont {L.~A.}\ \bibnamefont {Rozema}},
  \bibinfo {author} {\bibfnamefont {A.}~\bibnamefont {Feix}}, \bibinfo {author}
  {\bibfnamefont {M.}~\bibnamefont {Ara{\'u}jo}}, \bibinfo {author}
  {\bibfnamefont {J.~M.}\ \bibnamefont {Zeuner}}, \bibinfo {author}
  {\bibfnamefont {L.~M.}\ \bibnamefont {Procopio}}, \bibinfo {author}
  {\bibfnamefont {{\v {C}}.}~\bibnamefont {Brukner}},\ and\ \bibinfo {author}
  {\bibfnamefont {P.}~\bibnamefont {Walther}},\ }\bibfield  {title} {\bibinfo
  {title} {Experimental verification of an indefinite causal order},\ }\href
  {https://advances.sciencemag.org/content/3/3/e1602589} {\bibfield  {journal}
  {\bibinfo  {journal} {Sci. Adv.}\ }\textbf {\bibinfo {volume} {3}} (\bibinfo
  {year} {2017})}\BibitemShut {NoStop}%
\bibitem [{\citenamefont {Rubino}\ \emph {et~al.}(2022)\citenamefont {Rubino},
  \citenamefont {Rozema}, \citenamefont {Massa}, \citenamefont {Ara{\'{u}}jo},
  \citenamefont {Zych}, \citenamefont {Brukner},\ and\ \citenamefont
  {Walther}}]{Rubinoetal2022}%
  \BibitemOpen
  \bibfield  {author} {\bibinfo {author} {\bibfnamefont {G.}~\bibnamefont
  {Rubino}}, \bibinfo {author} {\bibfnamefont {L.~A.}\ \bibnamefont {Rozema}},
  \bibinfo {author} {\bibfnamefont {F.}~\bibnamefont {Massa}}, \bibinfo
  {author} {\bibfnamefont {M.}~\bibnamefont {Ara{\'{u}}jo}}, \bibinfo {author}
  {\bibfnamefont {M.}~\bibnamefont {Zych}}, \bibinfo {author} {\bibfnamefont
  {{\v{C}}.}~\bibnamefont {Brukner}},\ and\ \bibinfo {author} {\bibfnamefont
  {P.}~\bibnamefont {Walther}},\ }\bibfield  {title} {\bibinfo {title}
  {Experimental entanglement of temporal order},\ }\href
  {https://doi.org/10.22331/q-2022-01-11-621} {\bibfield  {journal} {\bibinfo
  {journal} {{Quantum}}\ }\textbf {\bibinfo {volume} {6}},\ \bibinfo {pages}
  {621} (\bibinfo {year} {2022})}\BibitemShut {NoStop}%
\bibitem [{\citenamefont {Goswami}\ \emph {et~al.}(2018)\citenamefont
  {Goswami}, \citenamefont {Giarmatzi}, \citenamefont {Kewming}, \citenamefont
  {Costa}, \citenamefont {Branciard}, \citenamefont {Romero},\ and\
  \citenamefont {White}}]{Goswamietal2018}%
  \BibitemOpen
  \bibfield  {author} {\bibinfo {author} {\bibfnamefont {K.}~\bibnamefont
  {Goswami}}, \bibinfo {author} {\bibfnamefont {C.}~\bibnamefont {Giarmatzi}},
  \bibinfo {author} {\bibfnamefont {M.}~\bibnamefont {Kewming}}, \bibinfo
  {author} {\bibfnamefont {F.}~\bibnamefont {Costa}}, \bibinfo {author}
  {\bibfnamefont {C.}~\bibnamefont {Branciard}}, \bibinfo {author}
  {\bibfnamefont {J.}~\bibnamefont {Romero}},\ and\ \bibinfo {author}
  {\bibfnamefont {A.~G.}\ \bibnamefont {White}},\ }\bibfield  {title} {\bibinfo
  {title} {Indefinite causal order in a quantum switch},\ }\href
  {https://doi.org/10.1103/PhysRevLett.121.090503} {\bibfield  {journal}
  {\bibinfo  {journal} {Phys. Rev. Lett.}\ }\textbf {\bibinfo {volume} {121}},\
  \bibinfo {pages} {090503} (\bibinfo {year} {2018})}\BibitemShut {NoStop}%
\bibitem [{\citenamefont {Wei}\ \emph {et~al.}(2019)\citenamefont {Wei},
  \citenamefont {Tischler}, \citenamefont {Zhao}, \citenamefont {Li},
  \citenamefont {Arrazola}, \citenamefont {Liu}, \citenamefont {Zhang},
  \citenamefont {Li}, \citenamefont {You}, \citenamefont {Wang}, \citenamefont
  {Chen}, \citenamefont {Sanders}, \citenamefont {Zhang}, \citenamefont
  {Pryde}, \citenamefont {Xu},\ and\ \citenamefont {Pan}}]{Weietal2019}%
  \BibitemOpen
  \bibfield  {author} {\bibinfo {author} {\bibfnamefont {K.}~\bibnamefont
  {Wei}}, \bibinfo {author} {\bibfnamefont {N.}~\bibnamefont {Tischler}},
  \bibinfo {author} {\bibfnamefont {S.-R.}\ \bibnamefont {Zhao}}, \bibinfo
  {author} {\bibfnamefont {Y.-H.}\ \bibnamefont {Li}}, \bibinfo {author}
  {\bibfnamefont {J.~M.}\ \bibnamefont {Arrazola}}, \bibinfo {author}
  {\bibfnamefont {Y.}~\bibnamefont {Liu}}, \bibinfo {author} {\bibfnamefont
  {W.}~\bibnamefont {Zhang}}, \bibinfo {author} {\bibfnamefont
  {H.}~\bibnamefont {Li}}, \bibinfo {author} {\bibfnamefont {L.}~\bibnamefont
  {You}}, \bibinfo {author} {\bibfnamefont {Z.}~\bibnamefont {Wang}}, \bibinfo
  {author} {\bibfnamefont {Y.-A.}\ \bibnamefont {Chen}}, \bibinfo {author}
  {\bibfnamefont {B.~C.}\ \bibnamefont {Sanders}}, \bibinfo {author}
  {\bibfnamefont {Q.}~\bibnamefont {Zhang}}, \bibinfo {author} {\bibfnamefont
  {G.~J.}\ \bibnamefont {Pryde}}, \bibinfo {author} {\bibfnamefont
  {F.}~\bibnamefont {Xu}},\ and\ \bibinfo {author} {\bibfnamefont {J.-W.}\
  \bibnamefont {Pan}},\ }\bibfield  {title} {\bibinfo {title} {Experimental
  quantum switching for exponentially superior quantum communication
  complexity},\ }\href {https://doi.org/10.1103/PhysRevLett.122.120504}
  {\bibfield  {journal} {\bibinfo  {journal} {Phys. Rev. Lett.}\ }\textbf
  {\bibinfo {volume} {122}},\ \bibinfo {pages} {120504} (\bibinfo {year}
  {2019})}\BibitemShut {NoStop}%
\bibitem [{\citenamefont {Massa}\ \emph {et~al.}(2019)\citenamefont {Massa},
  \citenamefont {Moqanaki}, \citenamefont {Baumeler}, \citenamefont
  {Del~Santo}, \citenamefont {Kettlewell}, \citenamefont {Daki{\'c}},\ and\
  \citenamefont {Walther}}]{Massaetal2019}%
  \BibitemOpen
  \bibfield  {author} {\bibinfo {author} {\bibfnamefont {F.}~\bibnamefont
  {Massa}}, \bibinfo {author} {\bibfnamefont {A.}~\bibnamefont {Moqanaki}},
  \bibinfo {author} {\bibfnamefont {{\"A}.}~\bibnamefont {Baumeler}}, \bibinfo
  {author} {\bibfnamefont {F.}~\bibnamefont {Del~Santo}}, \bibinfo {author}
  {\bibfnamefont {J.~A.}\ \bibnamefont {Kettlewell}}, \bibinfo {author}
  {\bibfnamefont {B.}~\bibnamefont {Daki{\'c}}},\ and\ \bibinfo {author}
  {\bibfnamefont {P.}~\bibnamefont {Walther}},\ }\bibfield  {title} {\bibinfo
  {title} {Experimental two-way communication with one photon},\ }\href
  {https://doi.org/https://doi.org/10.1002/qute.201900050} {\bibfield
  {journal} {\bibinfo  {journal} {Adv. Quantum Technol.}\ }\textbf {\bibinfo
  {volume} {2}},\ \bibinfo {pages} {1900050} (\bibinfo {year}
  {2019})}\BibitemShut {NoStop}%
\bibitem [{\citenamefont {Goswami}\ \emph {et~al.}(2020)\citenamefont
  {Goswami}, \citenamefont {Cao}, \citenamefont {Paz-Silva}, \citenamefont
  {Romero},\ and\ \citenamefont {White}}]{Goswamietal2020}%
  \BibitemOpen
  \bibfield  {author} {\bibinfo {author} {\bibfnamefont {K.}~\bibnamefont
  {Goswami}}, \bibinfo {author} {\bibfnamefont {Y.}~\bibnamefont {Cao}},
  \bibinfo {author} {\bibfnamefont {G.~A.}\ \bibnamefont {Paz-Silva}}, \bibinfo
  {author} {\bibfnamefont {J.}~\bibnamefont {Romero}},\ and\ \bibinfo {author}
  {\bibfnamefont {A.~G.}\ \bibnamefont {White}},\ }\bibfield  {title} {\bibinfo
  {title} {Increasing communication capacity via superposition of order},\
  }\href {https://doi.org/10.1103/PhysRevResearch.2.033292} {\bibfield
  {journal} {\bibinfo  {journal} {Phys. Rev. Research}\ }\textbf {\bibinfo
  {volume} {2}},\ \bibinfo {pages} {033292} (\bibinfo {year}
  {2020})}\BibitemShut {NoStop}%
\bibitem [{\citenamefont {Guo}\ \emph {et~al.}(2020)\citenamefont {Guo},
  \citenamefont {Hu}, \citenamefont {Hou}, \citenamefont {Cao}, \citenamefont
  {Cui}, \citenamefont {Liu}, \citenamefont {Huang}, \citenamefont {Li},
  \citenamefont {Guo},\ and\ \citenamefont {Chiribella}}]{Guoetal2020}%
  \BibitemOpen
  \bibfield  {author} {\bibinfo {author} {\bibfnamefont {Y.}~\bibnamefont
  {Guo}}, \bibinfo {author} {\bibfnamefont {X.-M.}\ \bibnamefont {Hu}},
  \bibinfo {author} {\bibfnamefont {Z.-B.}\ \bibnamefont {Hou}}, \bibinfo
  {author} {\bibfnamefont {H.}~\bibnamefont {Cao}}, \bibinfo {author}
  {\bibfnamefont {J.-M.}\ \bibnamefont {Cui}}, \bibinfo {author} {\bibfnamefont
  {B.-H.}\ \bibnamefont {Liu}}, \bibinfo {author} {\bibfnamefont {Y.-F.}\
  \bibnamefont {Huang}}, \bibinfo {author} {\bibfnamefont {C.-F.}\ \bibnamefont
  {Li}}, \bibinfo {author} {\bibfnamefont {G.-C.}\ \bibnamefont {Guo}},\ and\
  \bibinfo {author} {\bibfnamefont {G.}~\bibnamefont {Chiribella}},\ }\bibfield
   {title} {\bibinfo {title} {Experimental transmission of quantum information
  using a superposition of causal orders},\ }\href
  {https://doi.org/10.1103/PhysRevLett.124.030502} {\bibfield  {journal}
  {\bibinfo  {journal} {Phys. Rev. Lett.}\ }\textbf {\bibinfo {volume} {124}},\
  \bibinfo {pages} {030502} (\bibinfo {year} {2020})}\BibitemShut {NoStop}%
\bibitem [{\citenamefont {{Felce}}\ \emph {et~al.}()\citenamefont {{Felce}},
  \citenamefont {{Vedral}},\ and\ \citenamefont
  {{Tennie}}}]{Felceetal2021arxiv}%
  \BibitemOpen
  \bibfield  {author} {\bibinfo {author} {\bibfnamefont {D.}~\bibnamefont
  {{Felce}}}, \bibinfo {author} {\bibfnamefont {V.}~\bibnamefont {{Vedral}}},\
  and\ \bibinfo {author} {\bibfnamefont {F.}~\bibnamefont {{Tennie}}},\
  }\bibfield  {title} {\bibinfo {title} {{Refrigeration with Indefinite Causal
  Orders on a Cloud Quantum Computer}},\ }\href@noop {} {\ ,\ \bibinfo {eid}
  {arXiv:2107.12413}}\BibitemShut {NoStop}%
\bibitem [{\citenamefont {Rubino}\ \emph {et~al.}(2021)\citenamefont {Rubino},
  \citenamefont {Rozema}, \citenamefont {Ebler}, \citenamefont
  {Kristj\'ansson}, \citenamefont {Salek}, \citenamefont {Allard~Gu\'erin},
  \citenamefont {Abbott}, \citenamefont {Branciard}, \citenamefont {Brukner},
  \citenamefont {Chiribella},\ and\ \citenamefont {Walther}}]{Rubinoetal2021}%
  \BibitemOpen
  \bibfield  {author} {\bibinfo {author} {\bibfnamefont {G.}~\bibnamefont
  {Rubino}}, \bibinfo {author} {\bibfnamefont {L.~A.}\ \bibnamefont {Rozema}},
  \bibinfo {author} {\bibfnamefont {D.}~\bibnamefont {Ebler}}, \bibinfo
  {author} {\bibfnamefont {H.}~\bibnamefont {Kristj\'ansson}}, \bibinfo
  {author} {\bibfnamefont {S.}~\bibnamefont {Salek}}, \bibinfo {author}
  {\bibfnamefont {P.}~\bibnamefont {Allard~Gu\'erin}}, \bibinfo {author}
  {\bibfnamefont {A.~A.}\ \bibnamefont {Abbott}}, \bibinfo {author}
  {\bibfnamefont {C.}~\bibnamefont {Branciard}}, \bibinfo {author}
  {\bibfnamefont {{\v{C}}.}~\bibnamefont {Brukner}}, \bibinfo {author}
  {\bibfnamefont {G.}~\bibnamefont {Chiribella}},\ and\ \bibinfo {author}
  {\bibfnamefont {P.}~\bibnamefont {Walther}},\ }\bibfield  {title} {\bibinfo
  {title} {Experimental quantum communication enhancement by superposing
  trajectories},\ }\href {https://doi.org/10.1103/PhysRevResearch.3.013093}
  {\bibfield  {journal} {\bibinfo  {journal} {Phys. Rev. Research}\ }\textbf
  {\bibinfo {volume} {3}},\ \bibinfo {pages} {013093} (\bibinfo {year}
  {2021})}\BibitemShut {NoStop}%
\bibitem [{\citenamefont {Yin}\ \emph {et~al.}(2022)\citenamefont {Yin},
  \citenamefont {Zhao}, \citenamefont {Yang}, \citenamefont {Guo},
  \citenamefont {Zhang}, \citenamefont {Li}, \citenamefont {Han}, \citenamefont
  {Liu}, \citenamefont {Xu}, \citenamefont {Chiribella} \emph
  {et~al.}}]{Yinetal2022researchsquare}%
  \BibitemOpen
  \bibfield  {author} {\bibinfo {author} {\bibfnamefont {P.}~\bibnamefont
  {Yin}}, \bibinfo {author} {\bibfnamefont {X.}~\bibnamefont {Zhao}}, \bibinfo
  {author} {\bibfnamefont {Y.}~\bibnamefont {Yang}}, \bibinfo {author}
  {\bibfnamefont {Y.}~\bibnamefont {Guo}}, \bibinfo {author} {\bibfnamefont
  {W.-H.}\ \bibnamefont {Zhang}}, \bibinfo {author} {\bibfnamefont {G.-C.}\
  \bibnamefont {Li}}, \bibinfo {author} {\bibfnamefont {Y.-J.}\ \bibnamefont
  {Han}}, \bibinfo {author} {\bibfnamefont {B.-H.}\ \bibnamefont {Liu}},
  \bibinfo {author} {\bibfnamefont {J.-S.}\ \bibnamefont {Xu}}, \bibinfo
  {author} {\bibfnamefont {G.}~\bibnamefont {Chiribella}}, \emph {et~al.},\
  }\bibfield  {title} {\bibinfo {title} {Experimental super-{Heisenberg}
  quantum metrology with indefinite gate order},\ } \href{
  https://doi.org/10.1038/s41567-023-02046-y} {\bibfield
  {journal} {\bibinfo  {journal} {Nat. Phys.}\ } \textbf
  {\bibinfo {volume} {19}}, \ \bibinfo {pages} {1122--1127} (\bibinfo {year}
  {2023})}\BibitemShut {NoStop}%
\bibitem [{\citenamefont {Oreshkov}\ \emph {et~al.}(2012)\citenamefont
  {Oreshkov}, \citenamefont {Costa},\ and\ \citenamefont
  {Brukner}}]{Oreshkovetal2012}%
  \BibitemOpen
  \bibfield  {author} {\bibinfo {author} {\bibfnamefont {O.}~\bibnamefont
  {Oreshkov}}, \bibinfo {author} {\bibfnamefont {F.}~\bibnamefont {Costa}},\
  and\ \bibinfo {author} {\bibfnamefont {{\v{C}}.}~\bibnamefont {Brukner}},\
  }\bibfield  {title} {\bibinfo {title} {Quantum correlations with no causal
  order},\ }\href {https://doi.org/10.1038/ncomms2076} {\bibfield  {journal}
  {\bibinfo  {journal} {Nature Commun.}\ }\textbf {\bibinfo {volume} {3}},\
  \bibinfo {pages} {1092} (\bibinfo {year} {2012})}\BibitemShut {NoStop}%
\bibitem [{\citenamefont {Ara{\'{u}}jo}\ \emph {et~al.}(2015)\citenamefont
  {Ara{\'{u}}jo}, \citenamefont {Branciard}, \citenamefont {Costa},
  \citenamefont {Feix}, \citenamefont {Giarmatzi},\ and\ \citenamefont
  {Brukner}}]{Araujoetal2015}%
  \BibitemOpen
  \bibfield  {author} {\bibinfo {author} {\bibfnamefont {M.}~\bibnamefont
  {Ara{\'{u}}jo}}, \bibinfo {author} {\bibfnamefont {C.}~\bibnamefont
  {Branciard}}, \bibinfo {author} {\bibfnamefont {F.}~\bibnamefont {Costa}},
  \bibinfo {author} {\bibfnamefont {A.}~\bibnamefont {Feix}}, \bibinfo {author}
  {\bibfnamefont {C.}~\bibnamefont {Giarmatzi}},\ and\ \bibinfo {author}
  {\bibfnamefont {{\v{C}}.}~\bibnamefont {Brukner}},\ }\bibfield  {title}
  {\bibinfo {title} {Witnessing causal nonseparability},\ }\href
  {https://doi.org/10.1088/1367-2630/17/10/102001} {\bibfield  {journal}
  {\bibinfo  {journal} {New J. Phys.}\ }\textbf {\bibinfo {volume} {17}},\
  \bibinfo {pages} {102001} (\bibinfo {year} {2015})}\BibitemShut {NoStop}%
\bibitem [{\citenamefont {Baumeler}\ and\ \citenamefont
  {Wolf}(2016)}]{BaumelerWolf2016}%
  \BibitemOpen
  \bibfield  {author} {\bibinfo {author} {\bibfnamefont {{\"A}.}~\bibnamefont
  {Baumeler}}\ and\ \bibinfo {author} {\bibfnamefont {S.}~\bibnamefont
  {Wolf}},\ }\bibfield  {title} {\bibinfo {title} {The space of logically
  consistent classical processes without causal order},\ }\href
  {https://doi.org/10.1088/1367-2630/18/1/013036} {\bibfield  {journal}
  {\bibinfo  {journal} {New J. Phys.}\ }\textbf {\bibinfo {volume} {18}},\
  \bibinfo {pages} {013036} (\bibinfo {year} {2016})}\BibitemShut {NoStop}%
\bibitem [{\citenamefont {Oreshkov}\ and\ \citenamefont
  {Giarmatzi}(2016)}]{OreshkovGiarmatzi2016}%
  \BibitemOpen
  \bibfield  {author} {\bibinfo {author} {\bibfnamefont {O.}~\bibnamefont
  {Oreshkov}}\ and\ \bibinfo {author} {\bibfnamefont {C.}~\bibnamefont
  {Giarmatzi}},\ }\bibfield  {title} {\bibinfo {title} {Causal and causally
  separable processes},\ }\href {https://doi.org/10.1088/1367-2630/18/9/093020}
  {\bibfield  {journal} {\bibinfo  {journal} {New Journal of Physics}\ }\textbf
  {\bibinfo {volume} {18}},\ \bibinfo {pages} {093020} (\bibinfo {year}
  {2016})}\BibitemShut {NoStop}%
\bibitem [{\citenamefont {Zych}\ \emph {et~al.}(2019)\citenamefont {Zych},
  \citenamefont {Costa}, \citenamefont {Pikovski},\ and\ \citenamefont
  {Brukner}}]{Zychetal2019}%
  \BibitemOpen
  \bibfield  {author} {\bibinfo {author} {\bibfnamefont {M.}~\bibnamefont
  {Zych}}, \bibinfo {author} {\bibfnamefont {F.}~\bibnamefont {Costa}},
  \bibinfo {author} {\bibfnamefont {I.}~\bibnamefont {Pikovski}},\ and\
  \bibinfo {author} {\bibfnamefont {{\v{C}}.}~\bibnamefont {Brukner}},\
  }\bibfield  {title} {\bibinfo {title} {Bell's theorem for temporal order},\
  }\href {https://doi.org/10.1038/s41467-019-11579-x} {\bibfield  {journal}
  {\bibinfo  {journal} {Nature Communications}\ }\textbf {\bibinfo {volume}
  {10}},\ \bibinfo {pages} {3772} (\bibinfo {year} {2019})}\BibitemShut
  {NoStop}%
\bibitem [{\citenamefont {Oreshkov}(2019)}]{Oreshkov2019}%
  \BibitemOpen
  \bibfield  {author} {\bibinfo {author} {\bibfnamefont {O.}~\bibnamefont
  {Oreshkov}},\ }\bibfield  {title} {\bibinfo {title} {Time-delocalized quantum
  subsystems and operations: on the existence of processes with indefinite
  causal structure in quantum mechanics},\ }\href
  {https://doi.org/10.22331/q-2019-12-02-206} {\bibfield  {journal} {\bibinfo
  {journal} {{Quantum}}\ }\textbf {\bibinfo {volume} {3}},\ \bibinfo {pages}
  {206} (\bibinfo {year} {2019})}\BibitemShut {NoStop}%
\bibitem [{\citenamefont {Dimi{\'c}}\ \emph {et~al.}(2020)\citenamefont
  {Dimi{\'c}}, \citenamefont {Milivojevi{\'c}}, \citenamefont {Go{\v{C}}anin},
  \citenamefont {M{\'o}ller},\ and\ \citenamefont {Brukner}}]{Dimicetal2020}%
  \BibitemOpen
  \bibfield  {author} {\bibinfo {author} {\bibfnamefont {A.}~\bibnamefont
  {Dimi{\'c}}}, \bibinfo {author} {\bibfnamefont {M.}~\bibnamefont
  {Milivojevi{\'c}}}, \bibinfo {author} {\bibfnamefont {D.}~\bibnamefont
  {Go{\v{C}}anin}}, \bibinfo {author} {\bibfnamefont {N.~S.}\ \bibnamefont
  {M{\'o}ller}},\ and\ \bibinfo {author} {\bibfnamefont {{\v{C}}.}~\bibnamefont
  {Brukner}},\ }\bibfield  {title} {\bibinfo {title} {Simulating indefinite
  causal order with {R}indler observers},\ }\href
  {https://doi.org/10.3389/fphy.2020.525333} {\bibfield  {journal} {\bibinfo
  {journal} {Front. Phys.}\ }\textbf {\bibinfo {volume} {8}},\ \bibinfo {pages}
  {470} (\bibinfo {year} {2020})}\BibitemShut {NoStop}%
\bibitem [{\citenamefont {Milz}\ \emph {et~al.}(2021)\citenamefont {Milz},
  \citenamefont {Jurkschat}, \citenamefont {Pollock},\ and\ \citenamefont
  {Modi}}]{Milzetal2021}%
  \BibitemOpen
  \bibfield  {author} {\bibinfo {author} {\bibfnamefont {S.}~\bibnamefont
  {Milz}}, \bibinfo {author} {\bibfnamefont {D.}~\bibnamefont {Jurkschat}},
  \bibinfo {author} {\bibfnamefont {F.~A.}\ \bibnamefont {Pollock}},\ and\
  \bibinfo {author} {\bibfnamefont {K.}~\bibnamefont {Modi}},\ }\bibfield
  {title} {\bibinfo {title} {Delayed-choice causal order and nonclassical
  correlations},\ }\href {https://doi.org/10.1103/PhysRevResearch.3.023028}
  {\bibfield  {journal} {\bibinfo  {journal} {Phys. Rev. Research}\ }\textbf
  {\bibinfo {volume} {3}},\ \bibinfo {pages} {023028} (\bibinfo {year}
  {2021})}\BibitemShut {NoStop}%
\bibitem [{\citenamefont {Barrett}\ \emph {et~al.}(2021)\citenamefont
  {Barrett}, \citenamefont {Lorenz},\ and\ \citenamefont
  {Oreshkov}}]{Barrettetal2021}%
  \BibitemOpen
  \bibfield  {author} {\bibinfo {author} {\bibfnamefont {J.}~\bibnamefont
  {Barrett}}, \bibinfo {author} {\bibfnamefont {R.}~\bibnamefont {Lorenz}},\
  and\ \bibinfo {author} {\bibfnamefont {O.}~\bibnamefont {Oreshkov}},\
  }\bibfield  {title} {\bibinfo {title} {Cyclic quantum causal models},\ }\href
  {https://doi.org/10.1038/s41467-020-20456-x} {\bibfield  {journal} {\bibinfo
  {journal} {Nature Commun.}\ }\textbf {\bibinfo {volume} {12}},\ \bibinfo
  {pages} {885} (\bibinfo {year} {2021})}\BibitemShut {NoStop}%
\bibitem [{\citenamefont {Purves}\ and\ \citenamefont
  {Short}(2021)}]{PurvesShort2021}%
  \BibitemOpen
  \bibfield  {author} {\bibinfo {author} {\bibfnamefont {T.}~\bibnamefont
  {Purves}}\ and\ \bibinfo {author} {\bibfnamefont {A.~J.}\ \bibnamefont
  {Short}},\ }\bibfield  {title} {\bibinfo {title} {Quantum theory cannot
  violate a causal inequality},\ }\href
  {https://doi.org/10.1103/PhysRevLett.127.110402} {\bibfield  {journal}
  {\bibinfo  {journal} {Phys. Rev. Lett.}\ }\textbf {\bibinfo {volume} {127}},\
  \bibinfo {pages} {110402} (\bibinfo {year} {2021})}\BibitemShut {NoStop}%
\bibitem [{\citenamefont {Vilasini}\ and\ \citenamefont
  {Renner}()}]{VilasiniRenato2022arxiv}%
  \BibitemOpen
  \bibfield  {author} {\bibinfo {author} {\bibfnamefont {V.}~\bibnamefont
  {Vilasini}}\ and\ \bibinfo {author} {\bibfnamefont {R.}~\bibnamefont
  {Renner}},\ }\bibfield  {title} {\bibinfo {title} {Embedding cyclic causal
  structures in acyclic spacetimes: no-go results for process matrices}\ }\href
  {https://doi.org/10.48550/arxiv.2203.11245}
  {10.48550/arxiv.2203.11245}\BibitemShut {NoStop}%
\bibitem [{\citenamefont
  {Chapeau-Blondeau}(2021{\natexlab{b}})}]{ChapeauBlondeau2021ICOinspired}%
  \BibitemOpen
  \bibfield  {author} {\bibinfo {author} {\bibfnamefont {F.}~\bibnamefont
  {Chapeau-Blondeau}},\ }\bibfield  {title} {\bibinfo {title} {Quantum
  parameter estimation on coherently superposed noisy channels},\ }\href
  {https://doi.org/10.1103/PhysRevA.104.032214} {\bibfield  {journal} {\bibinfo
   {journal} {Phys. Rev. A}\ }\textbf {\bibinfo {volume} {104}},\ \bibinfo
  {pages} {032214} (\bibinfo {year} {2021}{\natexlab{b}})}\BibitemShut
  {NoStop}%
\bibitem [{\citenamefont {Goldberg}\ \emph {et~al.}(2023)\citenamefont
  {Goldberg}, \citenamefont {S\'{a}nchez-Soto},\ and\ \citenamefont
  {Heshami}}]{companion2023depol}%
  \BibitemOpen
  \bibfield  {author} {\bibinfo {author} {\bibfnamefont {A.}~\bibnamefont
  {Goldberg}}, \bibinfo {author} {\bibfnamefont {L.~L.}\ \bibnamefont
  {S\'{a}nchez-Soto}},\ and\ \bibinfo {author} {\bibfnamefont {K.}~\bibnamefont
  {Heshami}},\ }\bibfield  {title} {\bibinfo {title} {Measuring impossible
  parameters with indefinite causal order},\ }\href@noop {} {\bibfield
  {journal} {\bibinfo  {journal} {under review}\ } (\bibinfo {year}
  {2023})}\BibitemShut {NoStop}%
\bibitem [{\citenamefont {Demkowicz-Dobrzanski}\ \emph
  {et~al.}(2009)\citenamefont {Demkowicz-Dobrzanski}, \citenamefont {Dorner},
  \citenamefont {Smith}, \citenamefont {Lundeen}, \citenamefont {Wasilewski},
  \citenamefont {Banaszek},\ and\ \citenamefont
  {Walmsley}}]{DemkowiczDobrzanskietal2009}%
  \BibitemOpen
  \bibfield  {author} {\bibinfo {author} {\bibfnamefont {R.}~\bibnamefont
  {Demkowicz-Dobrzanski}}, \bibinfo {author} {\bibfnamefont {U.}~\bibnamefont
  {Dorner}}, \bibinfo {author} {\bibfnamefont {B.~J.}\ \bibnamefont {Smith}},
  \bibinfo {author} {\bibfnamefont {J.~S.}\ \bibnamefont {Lundeen}}, \bibinfo
  {author} {\bibfnamefont {W.}~\bibnamefont {Wasilewski}}, \bibinfo {author}
  {\bibfnamefont {K.}~\bibnamefont {Banaszek}},\ and\ \bibinfo {author}
  {\bibfnamefont {I.~A.}\ \bibnamefont {Walmsley}},\ }\bibfield  {title}
  {\bibinfo {title} {Quantum phase estimation with lossy interferometers},\
  }\href {https://doi.org/10.1103/PhysRevA.80.013825} {\bibfield  {journal}
  {\bibinfo  {journal} {Phys. Rev. A}\ }\textbf {\bibinfo {volume} {80}},\
  \bibinfo {pages} {013825} (\bibinfo {year} {2009})}\BibitemShut {NoStop}%
\bibitem [{\citenamefont {Demkowicz-Dobrza{\'n}ski}\ \emph
  {et~al.}(2012)\citenamefont {Demkowicz-Dobrza{\'n}ski}, \citenamefont
  {Ko{\l}ody{\'n}ski},\ and\ \citenamefont {Gu{\c t}{\u
  a}}}]{DemkowiczDobrzanskietal2012}%
  \BibitemOpen
  \bibfield  {author} {\bibinfo {author} {\bibfnamefont {R.}~\bibnamefont
  {Demkowicz-Dobrza{\'n}ski}}, \bibinfo {author} {\bibfnamefont
  {J.}~\bibnamefont {Ko{\l}ody{\'n}ski}},\ and\ \bibinfo {author}
  {\bibfnamefont {M.}~\bibnamefont {Gu{\c t}{\u a}}},\ }\bibfield  {title}
  {\bibinfo {title} {The elusive {Heisenberg} limit in quantum-enhanced
  metrology},\ }\href {https://doi.org/10.1038/ncomms2067} {\bibfield
  {journal} {\bibinfo  {journal} {Nature Commun.}\ }\textbf {\bibinfo {volume}
  {3}},\ \bibinfo {pages} {1063} (\bibinfo {year} {2012})}\BibitemShut
  {NoStop}%
\bibitem [{\citenamefont {Crowley}\ \emph {et~al.}(2014)\citenamefont
  {Crowley}, \citenamefont {Datta}, \citenamefont {Barbieri},\ and\
  \citenamefont {Walmsley}}]{Crowleyetal2014}%
  \BibitemOpen
  \bibfield  {author} {\bibinfo {author} {\bibfnamefont {P.~J.~D.}\
  \bibnamefont {Crowley}}, \bibinfo {author} {\bibfnamefont {A.}~\bibnamefont
  {Datta}}, \bibinfo {author} {\bibfnamefont {M.}~\bibnamefont {Barbieri}},\
  and\ \bibinfo {author} {\bibfnamefont {I.~A.}\ \bibnamefont {Walmsley}},\
  }\bibfield  {title} {\bibinfo {title} {Tradeoff in simultaneous
  quantum-limited phase and loss estimation in interferometry},\ }\href
  {https://doi.org/10.1103/PhysRevA.89.023845} {\bibfield  {journal} {\bibinfo
  {journal} {Phys. Rev. A}\ }\textbf {\bibinfo {volume} {89}},\ \bibinfo
  {pages} {023845} (\bibinfo {year} {2014})}\BibitemShut {NoStop}%
\bibitem [{\citenamefont {Gianani}\ \emph {et~al.}(2021)\citenamefont
  {Gianani}, \citenamefont {Albarelli}, \citenamefont {Verna}, \citenamefont
  {Cimini}, \citenamefont {Demkowicz-Dobrzanski},\ and\ \citenamefont
  {Barbieri}}]{Giananietal2021}%
  \BibitemOpen
  \bibfield  {author} {\bibinfo {author} {\bibfnamefont {I.}~\bibnamefont
  {Gianani}}, \bibinfo {author} {\bibfnamefont {F.}~\bibnamefont {Albarelli}},
  \bibinfo {author} {\bibfnamefont {A.}~\bibnamefont {Verna}}, \bibinfo
  {author} {\bibfnamefont {V.}~\bibnamefont {Cimini}}, \bibinfo {author}
  {\bibfnamefont {R.}~\bibnamefont {Demkowicz-Dobrzanski}},\ and\ \bibinfo
  {author} {\bibfnamefont {M.}~\bibnamefont {Barbieri}},\ }\bibfield  {title}
  {\bibinfo {title} {Kramers--Kronig relations and precision limits in quantum
  phase estimation},\ }\href {https://doi.org/10.1364/OPTICA.440438} {\bibfield
   {journal} {\bibinfo  {journal} {Optica}\ }\textbf {\bibinfo {volume} {8}},\
  \bibinfo {pages} {1642} (\bibinfo {year} {2021})}\BibitemShut {NoStop}%
\bibitem [{\citenamefont {{Bai}}\ and\ \citenamefont
  {{An}}()}]{BaiAn2023arxiv}%
  \BibitemOpen
  \bibfield  {author} {\bibinfo {author} {\bibfnamefont {S.-Y.}\ \bibnamefont
  {{Bai}}}\ and\ \bibinfo {author} {\bibfnamefont {J.-H.}\ \bibnamefont
  {{An}}},\ }\bibfield  {title} {\bibinfo {title} {{Floquet engineering to
  overcome no-go theorem of noisy quantum metrology}}\ }\href
  {https://doi.org/10.48550/arXiv.2303.00392}
  {10.48550/arXiv.2303.00392}\BibitemShut {NoStop}%
\bibitem [{\citenamefont {{Chiribella}}\ and\ \citenamefont
  {{Zhao}}()}]{ChiribellaZhao2022arxiv}%
  \BibitemOpen
  \bibfield  {author} {\bibinfo {author} {\bibfnamefont {G.}~\bibnamefont
  {{Chiribella}}}\ and\ \bibinfo {author} {\bibfnamefont {X.}~\bibnamefont
  {{Zhao}}},\ }\bibfield  {title} {\bibinfo {title} {{Heisenberg-limited
  metrology with coherent control on the probes' configuration}}\ }\href
  {https://doi.org/10.48550/arXiv.2206.03052}
  {10.48550/arXiv.2206.03052}\BibitemShut {NoStop}%
\bibitem [{\citenamefont {Chapeau-Blondeau}(2022)}]{ChapeauBlondeau2022}%
  \BibitemOpen
  \bibfield  {author} {\bibinfo {author} {\bibfnamefont {F.}~\bibnamefont
  {Chapeau-Blondeau}},\ }\bibfield  {title} {\bibinfo {title} {Indefinite
  causal order for quantum metrology with quantum thermal noise},\ }\href
  {https://doi.org/https://doi.org/10.1016/j.physleta.2022.128300} {\bibfield
  {journal} {\bibinfo  {journal} {Phys. Lett. A}\ }\textbf {\bibinfo {volume}
  {447}},\ \bibinfo {pages} {128300} (\bibinfo {year} {2022})}\BibitemShut
  {NoStop}%
\bibitem [{\citenamefont {Delgado}(2022)}]{Delgado2022}%
  \BibitemOpen
  \bibfield  {author} {\bibinfo {author} {\bibfnamefont {F.}~\bibnamefont
  {Delgado}},\ }\bibfield  {title} {\bibinfo {title} {Symmetries of quantum
  {F}isher information as parameter estimator for {P}auli channels under
  indefinite causal order},\ }\href {https://www.mdpi.com/2073-8994/14/9/1813}
  {\bibfield  {journal} {\bibinfo  {journal} {Symmetry}\ }\textbf {\bibinfo
  {volume} {14}},\ \bibinfo {pages} {1813} (\bibinfo {year}
  {2022})}\BibitemShut {NoStop}%
\bibitem [{Note1()}]{Note1}%
  \BibitemOpen
  \bibinfo {note} {If there are fewer channels, some of the Kraus operators can
  be set to an appropriate multiple of the identity and, if there are more
  channels, some of them can be concatenated into one channel [$ABC$ and $CAB$
  to $(AB)C$ and $C(AB)$] or applied with a definite causal order ($ABC$ and
  $CBA$ always have operation $B$ at the same time).}\BibitemShut {Stop}%
\bibitem [{\citenamefont {Carollo}\ \emph {et~al.}(2019)\citenamefont
  {Carollo}, \citenamefont {Spagnolo}, \citenamefont {Dubkov},\ and\
  \citenamefont {Valenti}}]{Carolloetal2019}%
  \BibitemOpen
  \bibfield  {author} {\bibinfo {author} {\bibfnamefont {A.}~\bibnamefont
  {Carollo}}, \bibinfo {author} {\bibfnamefont {B.}~\bibnamefont {Spagnolo}},
  \bibinfo {author} {\bibfnamefont {A.~A.}\ \bibnamefont {Dubkov}},\ and\
  \bibinfo {author} {\bibfnamefont {D.}~\bibnamefont {Valenti}},\ }\bibfield
  {title} {\bibinfo {title} {On quantumness in multi-parameter quantum
  estimation},\ }\href {https://doi.org/10.1088/1742-5468/ab3ccb} {\bibfield
  {journal} {\bibinfo  {journal} {J. Stat. Mech.}\ }\textbf {\bibinfo {volume}
  {2019}},\ \bibinfo {pages} {094010} (\bibinfo {year} {2019})}\BibitemShut
  {NoStop}%
\bibitem [{\citenamefont {Tsang}\ \emph {et~al.}(2020)\citenamefont {Tsang},
  \citenamefont {Albarelli},\ and\ \citenamefont {Datta}}]{Tsangetal2020}%
  \BibitemOpen
  \bibfield  {author} {\bibinfo {author} {\bibfnamefont {M.}~\bibnamefont
  {Tsang}}, \bibinfo {author} {\bibfnamefont {F.}~\bibnamefont {Albarelli}},\
  and\ \bibinfo {author} {\bibfnamefont {A.}~\bibnamefont {Datta}},\ }\bibfield
   {title} {\bibinfo {title} {Quantum semiparametric estimation},\ }\href
  {https://doi.org/10.1103/PhysRevX.10.031023} {\bibfield  {journal} {\bibinfo
  {journal} {Phys. Rev. X}\ }\textbf {\bibinfo {volume} {10}},\ \bibinfo
  {pages} {031023} (\bibinfo {year} {2020})}\BibitemShut {NoStop}%
\bibitem [{\citenamefont {Giovannetti}\ \emph {et~al.}(2006)\citenamefont
  {Giovannetti}, \citenamefont {Lloyd},\ and\ \citenamefont
  {Maccone}}]{Giovannettietal2006}%
  \BibitemOpen
  \bibfield  {author} {\bibinfo {author} {\bibfnamefont {V.}~\bibnamefont
  {Giovannetti}}, \bibinfo {author} {\bibfnamefont {S.}~\bibnamefont {Lloyd}},\
  and\ \bibinfo {author} {\bibfnamefont {L.}~\bibnamefont {Maccone}},\
  }\bibfield  {title} {\bibinfo {title} {Quantum metrology},\ }\href
  {https://doi.org/10.1103/PhysRevLett.96.010401} {\bibfield  {journal}
  {\bibinfo  {journal} {Phys. Rev. Lett.}\ }\textbf {\bibinfo {volume} {96}},\
  \bibinfo {pages} {010401} (\bibinfo {year} {2006})}\BibitemShut {NoStop}%
\bibitem [{\citenamefont {Dowling}(2008)}]{Dowling2008}%
  \BibitemOpen
  \bibfield  {author} {\bibinfo {author} {\bibfnamefont {J.~P.}\ \bibnamefont
  {Dowling}},\ }\bibfield  {title} {\bibinfo {title} {Quantum optical metrology
  -- the lowdown on {high-N00N} states},\ }\href
  {https://doi.org/10.1080/00107510802091298} {\bibfield  {journal} {\bibinfo
  {journal} {Contemp. Phys.}\ }\textbf {\bibinfo {volume} {49}},\ \bibinfo
  {pages} {125} (\bibinfo {year} {2008})}\BibitemShut {NoStop}%
\bibitem [{\citenamefont {Greenberger}\ \emph {et~al.}(1990)\citenamefont
  {Greenberger}, \citenamefont {Horne}, \citenamefont {Shimony},\ and\
  \citenamefont {Zeilinger}}]{Greenbergeretal1990}%
  \BibitemOpen
  \bibfield  {author} {\bibinfo {author} {\bibfnamefont {D.~M.}\ \bibnamefont
  {Greenberger}}, \bibinfo {author} {\bibfnamefont {M.~A.}\ \bibnamefont
  {Horne}}, \bibinfo {author} {\bibfnamefont {A.}~\bibnamefont {Shimony}},\
  and\ \bibinfo {author} {\bibfnamefont {A.}~\bibnamefont {Zeilinger}},\
  }\bibfield  {title} {\bibinfo {title} {Bell's theorem without inequalities},\
  }\href {https://doi.org/10.1119/1.16243} {\bibfield  {journal} {\bibinfo
  {journal} {Am. J. Phys.}\ }\textbf {\bibinfo {volume} {58}},\ \bibinfo
  {pages} {1131} (\bibinfo {year} {1990})} \BibitemShut {NoStop}%
\bibitem [{Note2()}]{Note2}%
  \BibitemOpen
  \bibinfo {note} {We take this opportunity to correct a typo in the factor of
  two that does not appear in the earlier sources \cite
  {Toth2012,Hyllusetal2012} but propagates in review articles~\cite
  {TothApellaniz2014,SidhuKok2020}.}\BibitemShut {Stop}%
\bibitem [{\citenamefont {Yunger~Halpern}\ \emph {et~al.}(2018)\citenamefont
  {Yunger~Halpern}, \citenamefont {Swingle},\ and\ \citenamefont
  {Dressel}}]{YungerHalpernetal2018}%
  \BibitemOpen
  \bibfield  {author} {\bibinfo {author} {\bibfnamefont {N.}~\bibnamefont
  {Yunger~Halpern}}, \bibinfo {author} {\bibfnamefont {B.}~\bibnamefont
  {Swingle}},\ and\ \bibinfo {author} {\bibfnamefont {J.}~\bibnamefont
  {Dressel}},\ }\bibfield  {title} {\bibinfo {title} {Quasiprobability behind
  the out-of-time-ordered correlator},\ }\href
  {https://doi.org/10.1103/PhysRevA.97.042105} {\bibfield  {journal} {\bibinfo
  {journal} {Phys. Rev. A}\ }\textbf {\bibinfo {volume} {97}},\ \bibinfo
  {pages} {042105} (\bibinfo {year} {2018})}\BibitemShut {NoStop}%
\bibitem [{\citenamefont {{Gao}}\ \emph {et~al.}()\citenamefont {{Gao}},
  \citenamefont {{Li}}, \citenamefont {{Mishra}}, \citenamefont {{Yan}},
  \citenamefont {{Simonov}},\ and\ \citenamefont
  {{Chiribella}}}]{Gaoetal2022arxiv}%
  \BibitemOpen
  \bibfield  {author} {\bibinfo {author} {\bibfnamefont {N.}~\bibnamefont
  {{Gao}}}, \bibinfo {author} {\bibfnamefont {D.}~\bibnamefont {{Li}}},
  \bibinfo {author} {\bibfnamefont {A.}~\bibnamefont {{Mishra}}}, \bibinfo
  {author} {\bibfnamefont {J.}~\bibnamefont {{Yan}}}, \bibinfo {author}
  {\bibfnamefont {K.}~\bibnamefont {{Simonov}}},\ and\ \bibinfo {author}
  {\bibfnamefont {G.}~\bibnamefont {{Chiribella}}},\ }\bibfield  {title}
  {\bibinfo {title} {{Measuring incompatibility and clustering quantum
  observables with a quantum switch}}\ }\href
  {https://doi.org/10.48550/arXiv.2208.06210}
  {10.48550/arXiv.2208.06210}\BibitemShut {NoStop}%
\bibitem [{\citenamefont {Nielsen}\ and\ \citenamefont
  {Chuang}(2000)}]{NielsenChuang2000}%
  \BibitemOpen
  \bibfield  {author} {\bibinfo {author} {\bibfnamefont {M.~A.}\ \bibnamefont
  {Nielsen}}\ and\ \bibinfo {author} {\bibfnamefont {I.~L.}\ \bibnamefont
  {Chuang}},\ }\href@noop {} {\emph {\bibinfo {title} {{Quantum Computation and
  Quantum Information}}}}\ (\bibinfo  {publisher} {Cambridge University
  Press},\ \bibinfo {address} {Cambridge},\ \bibinfo {year} {2000})\BibitemShut
  {NoStop}%
\bibitem [{Note3()}]{Note3}%
  \BibitemOpen
  \bibinfo {note} {See Ref.~\cite {Liuetal2023} for a comparison of ICO
  strategies for the amplitude damping channel when given access to two (or
  more) copies of $U$ each followed by an amplitude damping channel, where one
  can control the orders of the two noisy channels but cannot control the order
  of the noise and the unitary within each channel.}\BibitemShut {Stop}%
\bibitem [{\citenamefont {Kurdzialek}\ \emph {et~al.}()\citenamefont
  {Kurdzialek}, \citenamefont {Gorecki}, \citenamefont {Albarelli},\ and\
  \citenamefont {Demkowicz-Dobrzanski}}]{Kurdzialeketal2022arxiv}%
  \BibitemOpen
  \bibfield  {author} {\bibinfo {author} {\bibfnamefont {S.}~\bibnamefont
  {Kurdzialek}}, \bibinfo {author} {\bibfnamefont {W.}~\bibnamefont {Gorecki}},
  \bibinfo {author} {\bibfnamefont {F.}~\bibnamefont {Albarelli}},\ and\
  \bibinfo {author} {\bibfnamefont {R.}~\bibnamefont {Demkowicz-Dobrzanski}},\
  }\bibfield  {title} {\bibinfo {title} {Using adaptiveness and causal
  superpositions against noise in quantum metrology}\ }\href
  {10.1103/PhysRevLett.131.090801}{\bibfield  {journal} {\bibinfo
  {journal} {Phys. Rev. Lett.}\ }\textbf {\bibinfo {volume} {131}},\ \bibinfo
  {pages} {090801} (\bibinfo {year} {2023})}\BibitemShut {NoStop}%
\bibitem [{\citenamefont {Cubitt}\ \emph {et~al.}(2008)\citenamefont {Cubitt},
  \citenamefont {Ruskai},\ and\ \citenamefont {Smith}}]{Cubittetal2008}%
  \BibitemOpen
  \bibfield  {author} {\bibinfo {author} {\bibfnamefont {T.~S.}\ \bibnamefont
  {Cubitt}}, \bibinfo {author} {\bibfnamefont {M.~B.}\ \bibnamefont {Ruskai}},\
  and\ \bibinfo {author} {\bibfnamefont {G.}~\bibnamefont {Smith}},\ }\bibfield
   {title} {\bibinfo {title} {The structure of degradable quantum channels},\
  }\href {https://doi.org/10.1063/1.2953685} {\bibfield  {journal} {\bibinfo
  {journal} {J. Math. Phys.}\ }\textbf {\bibinfo {volume} {49}},\ \bibinfo
  {pages} {102104} (\bibinfo {year} {2008})}\BibitemShut {NoStop}%
\bibitem [{\citenamefont {Leung}\ and\ \citenamefont
  {Watrous}(2017)}]{LeungWatrous2017}%
  \BibitemOpen
  \bibfield  {author} {\bibinfo {author} {\bibfnamefont {D.}~\bibnamefont
  {Leung}}\ and\ \bibinfo {author} {\bibfnamefont {J.}~\bibnamefont
  {Watrous}},\ }\bibfield  {title} {\bibinfo {title} {On the complementary
  quantum capacity of the depolarizing channel},\ }\href
  {https://doi.org/10.22331/q-2017-09-19-28} {\bibfield  {journal} {\bibinfo
  {journal} {{Quantum}}\ }\textbf {\bibinfo {volume} {1}},\ \bibinfo {pages}
  {28} (\bibinfo {year} {2017})}\BibitemShut {NoStop}%
\bibitem [{\citenamefont {T\'oth}(2012)}]{Toth2012}%
  \BibitemOpen
  \bibfield  {author} {\bibinfo {author} {\bibfnamefont {G.}~\bibnamefont
  {T\'oth}},\ }\bibfield  {title} {\bibinfo {title} {Multipartite entanglement
  and high-precision metrology},\ }\href
  {https://doi.org/10.1103/PhysRevA.85.022322} {\bibfield  {journal} {\bibinfo
  {journal} {Phys. Rev. A}\ }\textbf {\bibinfo {volume} {85}},\ \bibinfo
  {pages} {022322} (\bibinfo {year} {2012})}\BibitemShut {NoStop}%
\bibitem [{\citenamefont {Hyllus}\ \emph {et~al.}(2012)\citenamefont {Hyllus},
  \citenamefont {Laskowski}, \citenamefont {Krischek}, \citenamefont
  {Schwemmer}, \citenamefont {Wieczorek}, \citenamefont {Weinfurter},
  \citenamefont {Pezz\'e},\ and\ \citenamefont {Smerzi}}]{Hyllusetal2012}%
  \BibitemOpen
  \bibfield  {author} {\bibinfo {author} {\bibfnamefont {P.}~\bibnamefont
  {Hyllus}}, \bibinfo {author} {\bibfnamefont {W.}~\bibnamefont {Laskowski}},
  \bibinfo {author} {\bibfnamefont {R.}~\bibnamefont {Krischek}}, \bibinfo
  {author} {\bibfnamefont {C.}~\bibnamefont {Schwemmer}}, \bibinfo {author}
  {\bibfnamefont {W.}~\bibnamefont {Wieczorek}}, \bibinfo {author}
  {\bibfnamefont {H.}~\bibnamefont {Weinfurter}}, \bibinfo {author}
  {\bibfnamefont {L.}~\bibnamefont {Pezz\'e}},\ and\ \bibinfo {author}
  {\bibfnamefont {A.}~\bibnamefont {Smerzi}},\ }\bibfield  {title} {\bibinfo
  {title} {Fisher information and multiparticle entanglement},\ }\href
  {https://doi.org/10.1103/PhysRevA.85.022321} {\bibfield  {journal} {\bibinfo
  {journal} {Phys. Rev. A}\ }\textbf {\bibinfo {volume} {85}},\ \bibinfo
  {pages} {022321} (\bibinfo {year} {2012})}\BibitemShut {NoStop}%
\end{thebibliography}

%

\newpage

\end{document}